\documentclass[a4paper]{article}
\usepackage[margin=1in]{geometry}

\usepackage[english]{babel}
\usepackage[utf8x]{inputenc}
\usepackage{amsmath,amsthm, amssymb}
\usepackage{enumerate}
\usepackage{bm}
\usepackage{graphicx}
\usepackage{verbatim}
\usepackage{booktabs}
\usepackage{bbm}
\usepackage{hyperref}
\usepackage{natbib}
\setcitestyle{authoryear,open={(},close={)}}
\usepackage[usenames,dvipsnames]{color}
\usepackage{algorithm}
\usepackage{enumitem}
\usepackage{longtable}
\usepackage{framed}
\usepackage{multirow}
\RequirePackage{rotating}
\usepackage{tablefootnote}

\title{A Bayesian Selection Model for Correcting Outcome Reporting Bias With Application to a Meta-analysis on Heart Failure Interventions
}
\author{Ray Bai\thanks{Department of Statistics, University of South Carolina, Columbia, SC, USA. Co-first author. E-mail: \href{mailto:rbai@mailbox.sc.edu}{\tt rbai@mailbox.sc.edu}}, Xiaokang Liu\thanks{Department of Biostatistics, Epidemiology, and Informatics, University of Pennsylvania, Philadelphia, PA, USA. Co-first author. Email: \href{mailto:xiaokang.liu@pennmedicine.upenn.edu}{\tt xiaokang.liu@pennmedicine.upenn.edu}}, Lifeng Lin\thanks{Department of Statistics, Florida State University, Tallahassee, FL, USA}, 
	Yulun Liu\thanks{Department of Population and Data Sciences, University of Texas Southwestern Medical Center, Dallas, TX, USA}, \\
	Stephen E. Kimmel\thanks{Department of Epidemiology, University of Florida, Gainesville, FL, USA}, Haitao Chu\thanks{Department of Biostatistics, University of Minnesota, Minneapolis, MN, USA}, Yong Chen\thanks{Department of Biostatistics, Epidemiology, and Informatics, University of Pennsylvania, Philadelphia, PA, USA. Co-corresponding author. Email: \href{mailto:ychen123@pennmedicine.upenn.edu}{\tt ychen123@pennmedicine.upenn.edu}}}
\date{\today}

\numberwithin{equation}{section}

\theoremstyle{definition}

\theoremstyle{plain}

\begin{document}
\maketitle
\begin{abstract}
Multivariate meta-analysis (MMA) is a powerful tool for jointly estimating multiple outcomes' treatment effects. However, the validity of results from MMA is potentially compromised by outcome reporting bias (ORB), or the tendency for studies to selectively report outcomes. Until recently, ORB has been understudied. Since ORB can lead to biased conclusions, it is crucial to correct the estimates of effect sizes \textit{and} quantify their uncertainty in the presence of ORB. With this goal, we develop a Bayesian selection model to adjust for ORB in MMA. We further propose a measure for quantifying the \textit{impact} of ORB on the results from MMA. We evaluate our approaches through a meta-evaluation of 748 bivariate meta-analyses from the Cochrane Database of Systematic Reviews. Our model is motivated by and applied to a meta-analysis of interventions on hospital readmission and quality of life for heart failure patients. In our analysis, the relative risk (RR) of hospital readmission for the intervention group changes from a significant decrease (RR: 0.931, 95\% confidence interval [CI]: 0.862-0.993) to a statistically \textit{nonsignificant} effect (RR: 0.955, 95\% CI: 0.876--1.051) after adjusting for ORB. This study demonstrates that failing to account for ORB can lead to different conclusions in a meta-analysis.
\end{abstract}

\section{Introduction} \label{Introduction}
Evidence-based medicine intends to optimize healthcare decision-making by using evidence from well-designed and conducted research \citep{guyatt2002users, moher2006systematic, egger2008systematic}. It classifies evidence by its epistemological strength and recommends using evidence from randomized controlled trials (RCTs), systematic reviews, and meta-analyses when available, to inform guidelines and policies. When conducted properly, systematic reviews and meta-analyses provide the most reliable evidence for synthesizing the benefits and harms associated with various treatment options, and can provide patients, caregivers, and doctors with integrated information for healthcare decision-making \citep{moher1999improving,bossuyt2003towards, moher2009preferred,stewart2015preferred}. 

Almost all RCTs measure and report more than one outcome, and often these outcomes are correlated with each other. For example, in a cardiovascular trial, the reduction in lipids level may be correlated with risk of clinical events such as stroke and myocardial infarction. In many RCTs, there is a balance between safety and efficacy; an experimental treatment may have greater efficacy than the placebo or standard therapy, but it may also have higher risk of adverse side effects such as transient toxicity or death. In practice, clinical decision-making relies on both efficacy and safety, so these outcomes must be considered simultaneously. Multivariate meta-analysis (MMA) is one technique proposed to jointly analyze multiple outcomes. MMA can borrow information from the potential correlation among the outcomes to improve the estimation of the pooled effect sizes \citep{riley2007bivariate, riley2007evaluation,jackson2011multivariate}.

On the other hand, since multiple outcomes are simultaneously considered in medical decision-making, biases in some of the outcomes can affect the overall decision of treatment. Recently, empirical studies have provided convincing evidence of the existence of selective reporting. \citet{chan2004empirical} compared the protocols of 102 trials with 122 published reports. Their investigation showed that, on average, 50\% of efficacy outcomes and 65\% of safety outcomes in each trial were incompletely reported; 62\% of the 82 trials had major inconsistencies between outcomes stated in the trial protocols and those reported in publications. They also found that, compared with nonsignificant outcomes, statistically significant outcomes are more likely to have higher odds of being reported for both efficacy outcomes (odds ratio = 2.4) and safety outcomes (odds ratio = 4.7).  Other studies have found similar results, such as in the reporting of toxicity in seven different medical areas \citep{hemminki1980study, chan2004outcome, hazell2006under, al2008selective, chowers2009reporting, mathieu2009comparison}, and in the reporting of safety outcomes for breast cancer treatments \citep{vera2013bias}. In the present article's motivating study for the effects of interventions on hospital readmission and quality of life for heart failure patients, 11 studies out of 45 do not report readmission, while 30 studies do not report quality of life. 

As such, outcome reporting bias (ORB), defined as ``{\it{the selective reporting of some outcomes but not others, depending on the nature and direction of the results}}'' \citep{sterne2016chapter}, may lead to biased inference in the pooled estimates of the outcomes and negatively affect patient outcomes. In addition to biased inference, ORB can also invalidate results from meta-analyses. For example, in this article's case study, the significant decrease in relative risk (RR) of hospital readmission for heart failure patients in the intervention group (95\% confidence interval [CI] 0.862--0.993) is no longer present \textit{after} we adjust for ORB (95\% CI 0.876--1.051). In our meta-evaluation of 748 bivariate meta-analyses from the Cochrane Database of Systematic Reviews in Section \ref{meta-meta}, we also found that 157 reviews experienced a change in statistical significance for at least one outcome \textit{after} correcting for ORB.

Until recently, ORB has been understudied, especially compared to the well-studied publication bias (PB) problem, defined as ``{\it{the publication or nonpublication of research findings, depending on the nature and direction of the results}}'' \citep{sterne2016chapter}. In the presence of PB, the published studies form a biased selection of the research in certain areas, which then leads to biased estimates \citep{jackson2007assessing}. The ORB problem is different from the PB problem in that, although outcomes of a study have been selectively reported under ORB, the remaining outcomes are still available. For PB, on the other hand, studies are completely missing, and we do not even know the number of studies that have been conducted but not published. Thus, the strategy for addressing ORB differs from that for PB, especially when leveraging the partially observed outcomes to infer the unreported outcomes. 

Since part of the outcomes are available, the current strategy for MMA with missing outcomes has focused on joint modeling of multiple outcomes that ``borrow strength'' across correlated outcomes \citep{riley2009multivariate,kirkham2012multivariate,frosi2015multivariate}. The idea is that the set of studies with outcomes reported can inform the correlations among multiple outcomes, which can be used to ``impute'' the missing outcomes from the reported outcomes. Unfortunately, this joint modeling strategy alone is insufficient as an approach to account for ORB because it relies on the {\emph{missing at random}} (MAR) assumption. This assumption is often not true in RCTs, since evidence suggests that the majority of missing outcomes are \textit{selectively} unreported \citep{chan2004empirical, vera2013bias}. It is also unclear if joint modeling alone can lead to less biased estimates in the presence of ORB. 

The evaluation of ORB has been included as a key component by the Cochrane risk of bias tool  \citep{higgins2011cochrane}, which is becoming a standard procedure in conducting a systematic review. However, the \textit{Cochrane Handbook for Systematic Reviews of Interventions} (Chapter 8.14.2, version 5.1.0, 2011), has acknowledged that ``statistical methods to detect within-study selective reporting (i.e., outcome-reporting bias) are, as yet, not well developed'' \citep{higgins2011chapter}. The \textit{Journal of Clinical Epidemiology} has also stressed that ``guidance is needed for using multiple outcomes and results in systematic reviews'' \citep{mayo2017multiple}. 

Motivated by the critical need for statistical models that can adjust for and evaluate the impact of ORB, we develop {\bf{A}} {\bf{B}}aysian {\bf{S}}election model for correcting {\bf{ORB}} (abbreviated as ABSORB henceforth) in this article. Specifically, we rely on selection models where multivariate latent variables are used to model the process of selective reporting of multiple outcomes in a flexible way. We then use a Bayesian approach to conduct estimation by placing appropriate priors on the unknown parameters. {\emph{From a modeling point of view}}, the distributions of the latent variables that govern the reporting processes are allowed to be correlated with not only the significance of the outcomes but also the characteristics of the study. 
{\emph{From a statistical inference point of view}}, the Bayesian approach allows the implementation of the model straightforwardly using Markov chain Monte Carlo (MCMC) and naturally provides uncertainty quantification for the model parameters through their posterior distributions. While there have been several approaches proposed for quantifying PB in \textit{univariate} meta-analysis \citep{lin2018quantifying, BaiLinBolandChen2020}, we are not aware of any existing approaches to quantify the impact of \textit{ORB} in \textit{multivariate} meta-analyses. By taking the Hellinger distance between the bias-corrected and non-bias corrected posterior densities for model parameters, we propose a measure to quantify the impact of outcome reporting bias.

The rest of the article is structured as follows. Section~\ref{MotivatingData} describes the motivating case study of the effects of interventions on quality of life and hospital readmission for heart failure patients. Section~\ref{ABSORB} introduces our proposed ABSORB model and our measure for quantifying the impact of ORB using our model. Section~\ref{meta-meta} empirically evaluates these approaches through a meta-evaluation of bivariate meta-analyses from the Cochrane Database of Systematic Reviews. Section~\ref{Application} applies our approaches to the case study of heart failure patients. Section~\ref{Discussion} concludes the article with a discussion of our findings and potential extensions for future work.

\section{A Motivating Meta-Analysis on Interventions for Heart Failure Patients} \label{MotivatingData}

For heart failure (HF) patients, readmission (ReAd) after discharging from the hospital is not rare, which places substantial burdens on both the patients and the health system. According to Medicare, the median risk-standardized 30-day readmission rate for HF was 23.0\% \citep{ZiaeianFonarow2016}. Due to the high cost of HF, preventing ReAd for HF patients has received particular attention from clinicians, researchers, and policymakers. For example, the Affordable Care Act has instituted a financial penalty for excessive readmissions for hospitals that is capped at 3\% of a hospital's total Medicare payments for 2015 and beyond \citep{ZiaeianFonarow2016}. On the other hand, quality of life (QoL) is an outcome that attracts more attention from patients, and the factors that affect the QoL of HF patients include anxiety, depression, and physical disability. A literature review by \citet{Celano2018} found that these adverse QoL outcomes were associated with poor function, reduced adherence to treatment, and elevated mortality in HF patients.  

Telemonitoring (TM) and structured telephone support (STS) are two common interventions and are demonstrated to be effective in reducing HF-specific readmission \citep{inglis2015structured}. A series of RCTs measuring both the all-cause ReAd and QoL provides a good opportunity to systematically evaluate the effects of interventions (TM or STS) on these two outcomes for patients with heart failure. Moreover, \citet{Celano2018} found that QoL was significantly associated with rehospitalization rates for HF patients. Therefore, it is of practical interest to \textit{jointly} model the effects of interventions on both ReAd and QoL in order to capture the inherent correlations between these two outcomes.

After a systematic search of scientific literature, 45 intervention studies were included in our analysis. For ReAd, we calculated the RR in order to quantify the change in risk of readmission due to the interventions compared to the usual care. Since the quantitative measure of QoL differed across studies, we calculated the standardized mean difference (SMD) in order to quantify the change of QoL between the intervention group and the group with usual care. 

For multiple studies in our meta-analysis, either ReAd or QoL was missing. For each of the studies, there were three possible scenarios: 1) the study reported both ReAd \textit{and} QoL, 2) the study reported \textit{only} ReAd, and 3) the study reported \textit{only} QoL. Among the 45 studies, only 8 studies published the results for both ReAd and QoL, 33 studies published only one of the two outcomes, and four studies did not publish either ReAd or QoL. Among the 41 studies with at least one outcome reported, 34 studies published the effect size of interventions on ReAd, and 15 studies published the effect size of interventions on QoL.

\begin{table}[!htbp]
	\centering
	\caption{Number of studies in the meta-analysis of interventions for HF patients, summarized by outcomes (columns) and by missingness scenarios (rows).  \checkmark : reported, \text{\sffamily X}: missing.}
	\medskip 
	
	\begin{tabular}{ccccccc}
		\hline
		& \multicolumn{3}{c}{Published studies} & \multicolumn{3}{c}{Updated studies}\\
		\cline{2-7}
		\multirow{2}{*}{\parbox{2cm}{\centering Scenario}} & \multicolumn{2}{c}{Outcome} & \multirow{2}{*}{\parbox{1.5cm}{\centering No.\ of studies}} &\multicolumn{2}{c}{Outcome}&\multirow{2}{*}{\parbox{1.5cm}{\centering No.\ of studies}} \\
		\cline{2-3}\cline{5-6} 
		& ReAd & QoL &&  ReAd & QoL \\
		\hline
		1 & \checkmark & \checkmark & 8   &\checkmark &\checkmark & 11\\
		2 & \checkmark & \text{\sffamily X} & 26 &\checkmark &\text{\sffamily X} & 23 \\
		3 & \text{\sffamily X} & \checkmark & 7 &\text{\sffamily X} &\checkmark & 10 \\
		No.\ of studies & 34 & 15 & 41 & 34 & 21 & 44 \\
		\hline
	\end{tabular} \label{ReAdQoLTable}
\end{table}

We queried the corresponding authors for the studies that did not report either QoL or ReAd, and only half of the authors we contacted replied. Specifically, we obtained QoL results for six of the 30 studies that did not publish results on QoL.
These new results gave us an updated sample with 11 studies that reported both ReAd and QoL and three \textit{new} studies that \textit{only} reported QoL. Table~\ref{ReAdQoLTable} summarizes the outcome reporting in our initial dataset (i.e., published studies) and in our new dataset \textit{after} obtaining six unpublished results on QoL from corresponding authors (i.e., updated studies).

\begin{figure}[!htbp]
	\centering
	\includegraphics[width=.9\linewidth]{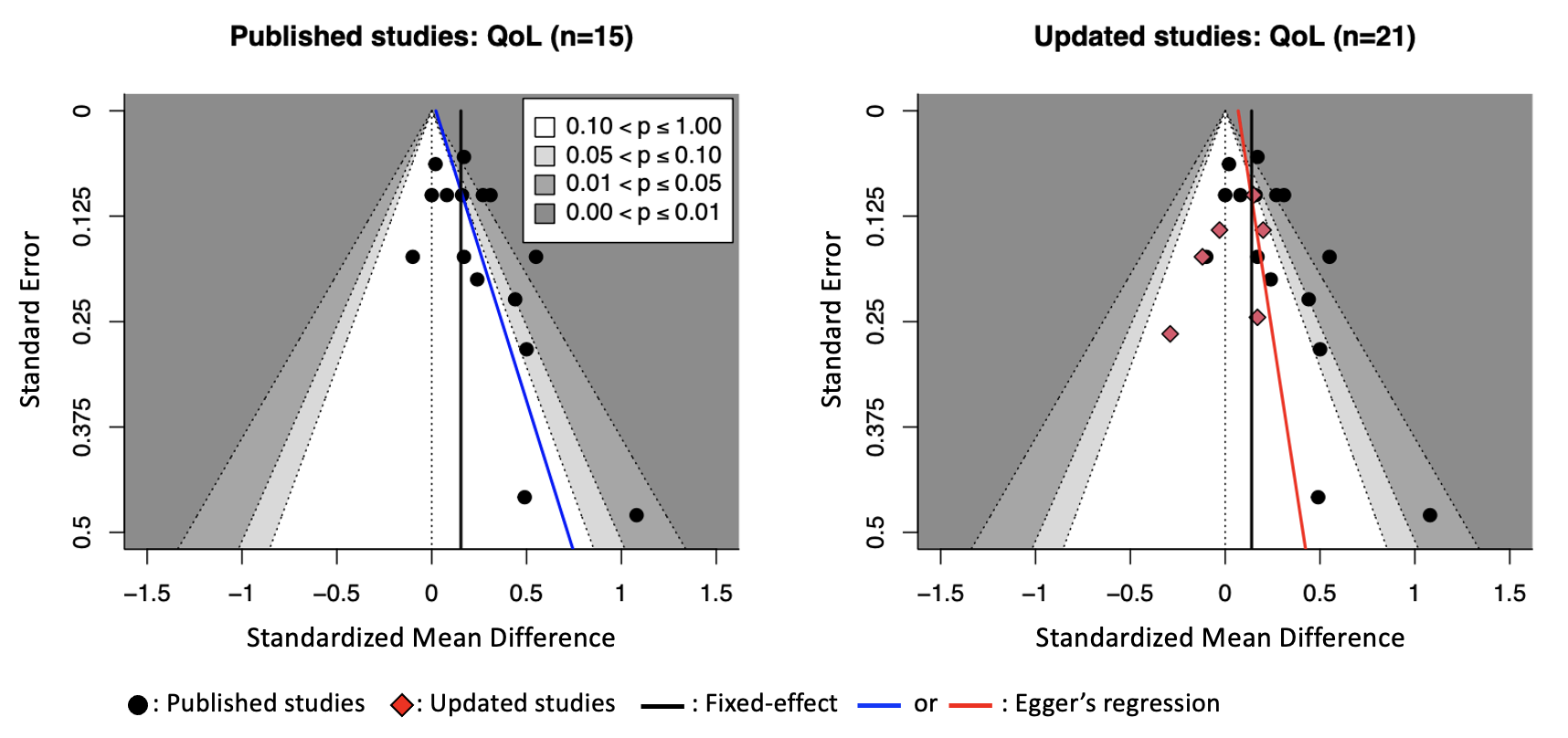}
	\caption{Contour-enhanced funnel plots for QoL in the published studies (left panel) and the updated studies (right panel). There is slightly less asymmetry in the funnel plot for the updated studies, suggesting the existence of selective reporting for QoL.}  \label{funnelplots}
\end{figure}

As a preliminary investigation, we conducted Begg's test \citep{Begg1994} and Egger's test \citep{Egger1997} for PB on ReAd and QoL separately. These tests suggested strong evidence of publication bias for ReAd (Begg's test: p-value = 0.02, Egger's test: p-value = 0.01). For QoL, there was moderate evidence of publication bias (Egger's test: p-value=0.10). However, by plotting the funnel plot for QoL, the evidence of selective reporting became more pronounced. As depicted in the left panel of Figure~\ref{funnelplots}, there was a moderate degree of asymmetry in the funnel plot for the published studies, as evidenced by a missing chunk out of the funnel on the left hand side. In the updated studies (right panel of Figure~\ref{funnelplots}), we found that \textit{all} six missing studies' QoL (represented by diamonds) were statistically \textit{nonsignificant}. This strongly suggested the existence of selective reporting of QoL. Even though the six missing studies were updated, there was still evidence of outcome reporting bias, as shown by the Egger's regression \citep{Egger1997}, i.e., the intercept was found to deviate from zero. 

While our initial investigation analyzed the ReAd and QoL outcomes separately, ReAd and QoL are likely to be correlated \citep{Celano2018} in practice, and biased estimation in one outcome can affect estimation in the other. In addition, given the evidence that many missing outcomes in RCTs are selectively unreported rather than missing at random \citep{chan2004empirical} (including in our case study), we were motivated to: 1) \textit{jointly} model ReAd and QoL in such a way that \textit{adjusts} for potential ORB, and 2) \textit{quantify} the impact of ORB on our MMA. We detail our novel modeling approaches in Section~\ref{ABSORB}.

\section{Statistical Methods} \label{ABSORB}

Based on our motivating case study, we focus on meta-analyses where two outcomes are of interest (or \textit{bivariate} meta-analysis). In practice, a bivariate meta-analysis of studies of diagnostic test accuracy is the most common medical application of MMA \citep{jackson2011multivariate, reitsma2005bivariate, chu2006bivariate}. In studies for drugs and other medical treatments, clinical efficacy and safety are also typically the two outcomes of greatest interest \citep{chan2004empirical}. However, the extension of ABSORB to meta-analyses with more than two outcomes is relatively straightforward and is discussed in Section~\ref{Discussion}. 

In a bivariate meta-analysis, our main parameter of interest is an unknown vector of two population treatment effects $\bm{\mu} = (\mu_1, \mu_2)'$. For example, the first endpoint $\mu_1$ could be a quantitative measure of the efficacy of a treatment, while the second endpoint $\mu_2$ is a quantitative measure for the treatment's safety. In our case study, $\mu_1$ is the RR of readmission, and $\mu_2$ is the SMD of the quality of life for heart failure patients. We let $\bm{y} = ( y_{1}, y_{2})'$ denote the reported effects for $\bm{\mu}$.

\subsection{The ABSORB Model} \label{ABSORBModel}

As discussed in Section~\ref{Introduction}, a common difficulty with conducting MMA is that in practice, outcomes are frequently unreported \citep{jackson2011multivariate}. Selective reporting of $y_1$ or $y_2$ might lead to biased estimation and misleading inference about $\bm{\mu}$. With the ABSORB model, we aim to adjust for this ORB.

\subsubsection{Model Specification and Assumptions} \label{ModelSpecification}

Building upon the selection model literature for correcting PB in meta-analysis \citep{copas1999works, copas2000meta, copas2001sensitivity, BaiLinBolandChen2020}, our goal is to explicitly model the selective reporting mechanism for partially reported outcomes. We assume that for each outcome $y_j, j = 1, 2$, there is a latent variable $z_j$ which determines the likelihood of $y_j$ being reported. 

Let $n$ denote the number of studies in our MMA. We assume that
\begin{align} \label{YgivenZ}
y_{ij} \mid ( z_{ij} > 0 ) = \mu_j + \tau_j u_{ij} + s_{ij} \epsilon_{ij},  \hspace{.5cm} i = 1, \ldots, n, \hspace{.2cm} j = 1, 2,
\end{align}
where $y_{ij}$ is the reported outcome for the $j$th endpoint for the $i$th study, $\mu_j$ is the mean effect for the $j$th endpoint, and $s_{ij}$ is the reported standard error for $y_{ij}$. We assume that $u_{ij}$ and $\epsilon_{ij}$ are marginally distributed as $\mathcal{N}(0,1)$ and that $\textrm{corr}(u_{ij}, \epsilon_{ij}) = 0$. The $u_{ij}$'s are random effects that capture the between-study heterogeneity for the $j$th endpoint, while $\tau_j > 0$ quantifies the amount of between-study heterogeneity. Meanwhile, the within-study random error is captured by $\epsilon_{ij}$. Under \eqref{YgivenZ}, we assume that $y_{ij}$ is only reported if the associated latent variable $z_{ij}$ is greater than zero. We further assume that the $z_{ij}$'s are generated according to
\begin{align} \label{latentZ}
	z_{ij} = \gamma_{0j} + \gamma_{1j} / s_{ij} + \delta_{ij},
\end{align}
where $\delta_{ij} \sim \mathcal{N}(0, 1)$. In \eqref{latentZ}, the parameter $\gamma_{0j}$ determines the overall probability of reporting $y_{ij}$, while $\gamma_{1j}$ determines how the likelihood of reporting depends on sample size. In general, $\gamma_{1j} \geq 0$, so that studies with larger sample sizes are
more likely to report their outcomes. We assume that
\begin{align} \label{EpsilonDeltaCorrelation}
\textrm{corr} (\epsilon_{ij}, \delta_{ij}) = \rho_j,
\end{align}
that is, the reported outcome $y_{ij}$ and the latent variables $z_{ij}$ are correlated through $\rho_j$. The correlation parameters $\rho_1$ and $\rho_2$ in \eqref{EpsilonDeltaCorrelation} control how the probability of reporting for the first and second endpoint respectively is influenced by the effect size of the study. When ORB for both endpoints is present, then $\rho_1 \neq 0$ and $\rho_2 \neq 0$. In this case, standard meta-analyses may lead to \textit{biased} estimation of $\bm{\mu}$. 

In line with standard bivariate meta-analysis \citep{jackson2011multivariate}, we further assume that there is both within-study correlation between the $\epsilon_{ij}$'s in \eqref{YgivenZ}, as well as between-study correlation for the two endpoints. To model the within-study correlation, we assume that
\begin{align} \label{WithinStudyCorrelation}
	\textrm{corr} ( \epsilon_{i1}, \epsilon_{i2} ) = \rho_\text{W}.
\end{align}
Although the assumption that the within-study correlation is a constant $\rho_\text{W}$ across all the studies may be strong, this approach is commonly adopted in practice for MMA \citep{RileyThompsonAbrams2007, LinChu2018} in order to keep the model parsimonious.

To model the between-study correlation, we assume that the random effects $(u_{i1}, u_{i2})'$ for the two endpoints in \eqref{YgivenZ} are also correlated. That is, we assume that
\begin{align} \label{BetweenStudyCorrelation}
	\textrm{corr} ( u_{i1}, u_{i2} ) = \rho_\text{B}.
\end{align}
Finally, we assume that
\begin{align} \label{ZeroCorrelations}
	\textrm{corr} ( \epsilon_{i1}, \delta_{i2}) = \textrm{corr} ( \epsilon_{i2}, \delta_{i1} ) = \textrm{corr}(\delta_{i1}, \delta_{i2}) = 0.
\end{align} 
Assumption \eqref{ZeroCorrelations} implies that $y_{i1} \mid (z_{i1} > 0, z_{i2}) = y_{i1} \mid (z_{i1} > 0)$ and $y_{i2} \mid ( z_{i1}, z_{i2} > 0) = y_{i2} \mid (z_{i2} > 0)$. In other words, $y_{i1}$ is reported only if $z_{i1} > 0$ and does not depend on the value of $z_{i2}$. Similarly, $y_{i2}$ does not depend on $z_{i1}$, and $z_{i1}$ does not depend on $z_{i2}$. We stress that the outcomes $y_{1}$ and $y_{2}$ themselves are likely to be correlated, and this is captured in our model through the within-study correlation $\rho_\text{W}$ \eqref{WithinStudyCorrelation} and the between-study correlation $\rho_\text{B}$ \eqref{BetweenStudyCorrelation}. However, the probability of \textit{reporting} each individual outcome should depend only on the associated latent variable.

\subsection{Estimation in the ABSORB Model} \label{PriorSpecification}
The basic ABSORB model is given in \eqref{YgivenZ}--\eqref{latentZ}, while additional assumptions about the correlation structure of different parameters are encoded in \eqref{EpsilonDeltaCorrelation}--\eqref{ZeroCorrelations}. In summary, we have a total of 12 unknown parameters $(\mu_1, \mu_2, \tau_1, \tau_2, \gamma_{01}, \gamma_{02}, \gamma_{11}, \gamma_{12}, \rho_1, \rho_2, \rho_\text{W}, \rho_\text{B})'$ under the ABSORB model. We propose a Bayesian approach to estimating all these parameters by placing appropriate priors on them. 

For the mean treatment effects $\bm{\mu}$, we place the vague priors,
\begin{align} \label{muprior}
\mu_j \sim \mathcal{N}(0, 10^4),
\end{align}
and for the heterogeneity parameters $(\tau_1, \tau_2)'$, we place vague half-Cauchy priors,
\begin{align} \label{tauprior}
\tau_j \sim \mathcal{C}^{+} (0, 1).
\end{align}
Next, we consider priors for $(\gamma_{01}, \gamma_{02}, \gamma_{11}, \gamma_{12})'$, the parameters that control the overall likelihood of reporting for the first and second endpoint respectively. To induce weakly informative priors on these parameters, we follow \cite{BaiLinBolandChen2020} and specify the priors as,
\begin{align} \label{gamma0prior}
\gamma_{0j} \sim \mathcal{U} (-2, 2);
\end{align}
\begin{align} \label{gamma1prior}
\gamma_{1j} \sim \mathcal{U} (0, \max_i s_{ij} ).
\end{align}
The priors \eqref{gamma0prior}--\eqref{gamma1prior} ensure that most of the mass for each of the latent variables $z_{ij}$ lies in the interval $(-2, 3)$, leading to selection probabilities between 2.5\% and 99.7\%. Finally, in order to complete the prior specification, we place noninformative uniform priors on each of the correlation parameters,
\begin{align} \label{rhoprior}
\rho_1, \rho_2, \rho_\text{W}, \rho_\text{B} \sim \mathcal{U}(-1, 1).
\end{align}
The Bayesian approach is especially appealing for several reasons. First, we can implement the model straightforwardly using MCMC, thus avoiding the difficulties of maximum likelihood estimation (MLE). The main issue with the MLE in selection models is that it can face non-convergence \citep{copas2001sensitivity}. This can arise from poor initializations, a flat plateau in the likelihood, or instability in the computation of a $12 \times 12$ Hessian matrix during the optimization procedure \citep{copas2001sensitivity, ning2017maximum}. MCMC sampling does not encounter such difficulties (provided that we run the MCMC for enough iterations), and we can monitor convergence for the posteriors $p(\mu_1 \mid \bm{y}_1, \ldots, \bm{y}_n)$ and $p(\mu_2 \mid \bm{y}_1, \ldots, \bm{y}_n)$ using trace plots or the effective sample size (ESS). Besides the computational advantages of the Bayesian approach, we can also obtain natural uncertainty quantification for the model parameters through their posterior distributions. These posterior densities will ultimately allow us to quantify the \textit{impact} of outcome reporting bias, as we discuss in Section~\ref{QuantifyingORB}.

\subsection{The ABSORB Likelihood and Implementation} \label{LikelihoodImplementation}

In order to perform Bayesian inference under the ABSORB model \eqref{YgivenZ}--\eqref{ZeroCorrelations}, we need to obtain the likelihood function and then place the priors \eqref{muprior}--\eqref{rhoprior} on the model parameters. In this section, we describe how to derive the ABSORB likelihood for the $n$ studies in our MMA and perform posterior inference under this likelihood. 

Because not all studies report both $y_{1}$ and $y_{2}$, we may not have an equal number of observations for $y_1$ and $y_2$. Consequently, we need to consider three separate cases for the reported outcomes in our meta-analysis: 1) both endpoints are reported, 2) only the first endpoint is reported, or 3) only the second endpoint is reported. Without loss of generality, suppose that the first $m_1$ studies report both endpoints, the next $m_2$ studies report only the first endpoint $y_1$, and the remaining $m_3 = n-(m_1+m_2)$ studies report only the second endpoint $y_2$. 

First note that we can reparameterize the ABSORB model \eqref{YgivenZ}--\eqref{ZeroCorrelations} as a hierarchical model by introducing further latent parameters $(\theta_{i1}, \theta_{i2})'$ for the $m_1$ studies that report both endpoints, $\widetilde{\theta}_{i1}$ for the $m_2$ studies that report only $y_1$, and $\check{\theta}_{i2}$ for the $m_3$ studies that report only $y_2$. The main reason for introducing these additional latent parameters is to ensure that the joint densities in our likelihood can be written explicitly. We denote $\bm{\Xi}$ as the collection of all unknown parameters, including these latent parameters.

For the $m_1$ studies that report both outcomes,  we rewrite \eqref{YgivenZ} as
\begin{equation} \label{ABSORBreparam}
\begin{array}{rl}
y_{i1} \mid ( z_{i1} > 0) \sim & \mathcal{N} ( \theta_{i1}, s_{i1}^2 ); \\
y_{i2} \mid (z _{i2} > 0) \sim & \mathcal{N} ( \theta_{i2}, s_{i2}^2),
\end{array} \hspace{.5cm} i = 1, \ldots, m_1,
\end{equation}
where $(\theta_{i1}, \theta_{i2})$ is jointly distributed as
\begin{align} \label{latentTheta}
\begin{pmatrix} \theta_{i1}\\ \theta_{i2} \end{pmatrix} \sim \mathcal{N} \left( \begin{pmatrix} \mu_1 \\ \mu_2 \end{pmatrix}, \begin{pmatrix} \tau_1^2 & \rho_\text{B} \tau_1 \tau_2 \\ \rho_\text{B} \tau_1 \tau_2 & \tau_2^2 \end{pmatrix} \right), \hspace{.5cm} i = 1, \ldots, m_1.
\end{align}
 For studies $i = 1, \ldots, m_1$ that report both outcomes, the ABSORB model can be represented as
\begin{align} \label{JointYZbothoutcomes}
\begin{pmatrix} y_{i1} \\ y_{i2} \\ z_{i1} \\ z_{i2} \end{pmatrix} \sim \mathcal{N} \left( \begin{pmatrix} \theta_{i1} \\ \theta_{i2} \\ \gamma_{01} + \gamma_{11} / s_{i1} \\ \gamma_{02} + \gamma_{12}/s_{i2} \end{pmatrix}, \begin{pmatrix} s_{i1}^2 & \rho_\text{W} s_{i1} s_{i2} & \rho_1 s_{i1} & 0 \\ \rho_\text{W} s_{i1} s_{i2} & s_{i2}^2 & 0 & \rho_2 s_{i2} \\ \rho_1 s_{i1} & 0 & 1 & 0 \\ 0 & \rho_2 s_{i2} & 0 & 1 \end{pmatrix} \right) \mathbb{I}_{ [ z_{i1} > 0 \cap z_{i2} > 0]}.
\end{align}
Namely, for each of these $m_1$ studies, the joint density of $(y_{i1}, y_{i2}, z_{i1}, z_{i2})$ is a truncated normal density that contains both endpoints $(y_{i1}, y_{i2})'$ only because \textit{both} associated latent variables $z_{1}$ and $z_{2}$ are greater than zero. The off-diagonal entries in the covariance matrix in \eqref{JointYZbothoutcomes} capture the correlations between $y_{i1}$, $y_{i2}$, $z_{i1}$, and $z_{i2}$. The likelihood function for $\bm{\Xi}$ in these $m_1$ studies is easily seen to be
\begin{equation} \label{LikelihoodBothYs}
	L_1 (\bm{\Xi} ) = \prod_{i=1}^{m_1} f( y_{i1}, y_{i2}, z_{i1}, z_{i2} \mid \theta_{i1}, \theta_{i2}, \gamma_{01}, \gamma_{11}, \gamma_{02}, \gamma_{12}, \rho_1, \rho_2, \rho_\text{W} ),
\end{equation}
where $f(y_{i1}, y_{i2}, z_{i1}, z_{i2} \mid \cdot )$ is the probability density function (pdf) for the truncated normal density in \eqref{JointYZbothoutcomes}. 

For the $m_2$ studies that only report the first endpoint $y_{1}$ but not $y_{2}$, we can also represent the model with a truncated normal density. However, since we do not observe $y_{2}$ for these studies, we can only write the joint density of $(y_{i1}, z_{i1}, z_{i2})'$ for studies $i = m_1+1, \ldots, m_1+m_2$ as follows: 
\begin{align} \label{jointYZfirstoutcomeonly}
\begin{pmatrix} y_{i1} \\ z_{i1} \\ z_{i2} \end{pmatrix} \sim \mathcal{N} \left( \begin{pmatrix} \widetilde{\theta}_{i1} \\ \gamma_{01} + \gamma_{11} / s_{i1} \\ \gamma_{02} + \gamma_{12} / s_{i2} \end{pmatrix}, \begin{pmatrix} s_{i1}^2 & \rho_1 s_{i1} & 0 \\ \rho_1 s_{i1} & 1 & 0 \\ 0 & 0 & 1 \end{pmatrix} \right) \mathbb{I}_{[ z_{i1} > 0 \cap z_{i2} < 0 ]},
\end{align} 
where $\widetilde{\theta}_{i1}$ is marginally distributed as
\begin{equation} \label{latenttheta1}
    \widetilde{\theta}_{i1} \sim \mathcal{N} (\mu_1, \tau_1^2), \hspace{.5cm} i = m_1+1, \ldots, m_1+m_2.
\end{equation}
The representation in \eqref{jointYZfirstoutcomeonly} ensures that the first endpoint $y_1$ is only reported because the corresponding latent variable $z_1$ is greater than zero, while the second endpoint $y_2$ is \textit{not} reported because the corresponding latent variable $z_2$ is \textit{less} than zero. The main issue that we encounter with \eqref{jointYZfirstoutcomeonly} is that the standard errors $s_{i2}$'s are not available for these $m_2$ studies (since $y_2$ was not reported for any of them), and our model requires these standard errors in order to parameterize the mean of $z_{i2}$. Nevertheless, we can estimate the missing $s_{i2}$'s using the approach given in Section~3.5 of \cite{copas2014model}. Specifically, we use the relationship that $1 / s_{i2}^2 = k_2 n_i$, where $k_2$ is a constant and $n_i$ is the sample size of the $i$th study. Based on the other $n-m_2$ studies that reported $s_{i2}$ (i.e., the $m_1$ studies where both endpoints were reported and the $m_3$ studies that only reported the second endpoint $y_2$) and the corresponding sample sizes for these studies, we then estimate $k_2$ as
\begin{align*}
	\widehat{k}_2 = \frac{\sum_{i \in R_2} 1 / s_{i2}^2}{\sum_{i \in R_2} n_i },
	\end{align*}
where $R_2$ is the index set of the $n-m_2$ studies that have reported $s_{i2}$. The missing $s_{i2}$'s for the $m_2$ studies in \eqref{jointYZfirstoutcomeonly} can then be estimated as $\widehat{s}_{i2} = \sqrt{1 / ( \widehat{k}_2 n_i ) }$ \citep{copas2014model}. Substituting the $s_{i2}$'s with their estimates $\widehat{s}_{i2}$'s, the likelihood function for the $m_2$ studies that only report $y_1$ but not $y_2$ can be written as
\begin{equation} \label{LikelihoodY1Only}
	L_2 (\bm{\Xi}) = \prod_{i=m_1+1}^{m_1+m_2} f( y_{i1}, z_{i1}, z_{i2} \mid \widetilde{\theta}_{i1}, \gamma_{01}, \gamma_{11}, \gamma_{02}, \gamma_{12}, \rho_1) ,
\end{equation}
where $f(y_{i1}, z_{i1}, z_{i2} \mid \cdot )$ is the pdf of the truncated normal density in \eqref{jointYZfirstoutcomeonly}.

Finally, for the remaining $m_3$ studies that only report the second endpoint $y_{2}$ but not $y_1$, we can similarly represent the model as follows. For $i = m_1 + m_2 + 1, \ldots, n$, we have 
\begin{align} \label{jointYZsecondoutcomeonly}
\begin{pmatrix} y_{i2} \\ z_{i1} \\ z_{i2} \end{pmatrix} \sim \mathcal{N} \left( \begin{pmatrix} \check{\theta}_{i2} \\ \gamma_{01}+ \gamma_{11} / s_{i1} \\ \gamma_{02} + \gamma_{12} / s_{i2} \end{pmatrix}, \begin{pmatrix} s_{i2}^2 & 0 & \rho_2 s_{i2} \\ 0 & 1 & 0 \\ \rho_2 s_{i2} & 0 & 1 \end{pmatrix} \right) \mathbb{I}_{[ z_{i1} < 0 \cap z_{i2} > 0]},
\end{align} 
where $\check{\theta}_{i2}$ is marginally distributed as
\begin{equation}\label{latenttheta2}
\check{\theta}_{i2} \sim \mathcal{N}(\mu_2, \tau_2^2), \hspace{.5cm} i = m_1+m_2+1, \ldots, n.
\end{equation}
The truncated normal density in \eqref{jointYZsecondoutcomeonly} ensures that the second endpoint $y_2$ is only reported because the corresponding latent variable $z_2$ is greater than zero, while $y_1$ is \textit{not} reported because $z_1$ is \textit{less} than zero. For these $m_3$ studies, we do not observe the standard errors $s_{i1}$'s since none of these studies reported $y_1$. As we require these $s_{i1}$'s in order to parameterize the mean of $z_{i1}$, we again follow the approach of \cite{copas2014model} and first estimate
\begin{align*}
	\widehat{k}_1 = \frac{ \sum_{i \in R_1} 1 / s_{i1}^2}{\sum_{i \in R_1} n_i},
\end{align*}
where $R_1$ is the index set for the $m_1+m_2$ studies that have reported $s_{i1}$. The missing $s_{i1}$'s for the $n-(m_1+m_2)$ studies in (3.7) are then estimated as $\widehat{s}_{i1} = \sqrt{1/(\widehat{k}_1 n_i)}$. Similarly as in \eqref{LikelihoodY1Only}, the likelihood for these $n-(m_1+m_2)$ studies after substituting the $s_{i1}$'s with the $\widehat{s}_{i1}$'s is
\begin{equation} \label{LikelihoodY2Only}
	L_3 (\bm{\Xi} ) = \prod_{i=m_1+m_2+1}^{n} f( y_{i2}, z_{i1}, z_{i2} \mid \check{\theta}_{i2}, \gamma_{01}, \gamma_{11}, \gamma_{02}, \gamma_{12}, \rho_2 ),
\end{equation}
where $f(y_{i2}, z_{i1}, z_{i2} \mid \cdot )$ is the pdf of the truncated normal density in \eqref{jointYZsecondoutcomeonly}. Combining \eqref{LikelihoodBothYs}, \eqref{LikelihoodY1Only}, and \eqref{LikelihoodY2Only}, we see that the complete likelihood function for all $n$ studies is
\begin{align} \label{ABSORBLikelihood}
L (  \bm{\Xi} \mid \bm{y}_1, \ldots, \bm{y}_n ) = L_1 (\bm{\Xi} )  L_2 (\bm{\Xi}) L_3 (\bm{\Xi}).
	\end{align}
 Under \eqref{ABSORBLikelihood}, the joint posterior distribution for $\bm{\Xi}$ is then
\begin{align} \label{ABSORBposterior}
p ( \bm{\Xi} \mid  \bm{y}_1, \ldots, \bm{y}_n ) \propto L ( \bm{\Xi} \mid  \bm{y}_1, \ldots, \bm{y}_n ) p ( \bm{\Xi} ),
\end{align}
where $p (\bm{\Xi})$ is the product of the priors \eqref{muprior}--\eqref{rhoprior}, \eqref{latentTheta}, \eqref{latenttheta1}, and \eqref{latenttheta2} on the model parameters. The main challenge with the ABSORB model is sampling from the truncated densities \eqref{JointYZbothoutcomes}, \eqref{jointYZfirstoutcomeonly}, and \eqref{jointYZsecondoutcomeonly} in the full likelihood \eqref{ABSORBLikelihood}. In Appendix~\ref{Sampling}, we describe how to approximately sample from these truncated densities. With the prior for $\bm{\Xi}$ specified, the complete ABSORB model can then be implemented in any standard MCMC software to approximate the posterior distributions $p( \mu_1 \mid \bm{y}_1, \ldots, \bm{y}_n )$ and $p ( \mu_2 \mid \bm{y}_1, \ldots, \bm{y}_n )$.  For our implementation, we use the \texttt{JAGS} software. 

Note that it may be the case that only one of the endpoints $y_1$ or $y_2$ in our MMA contains missing values. When there are no missing outcomes for $y_2$, the number of studies that only report $y_1$ but not $y_2$ is $m_2 = 0$, and we replace \eqref{ABSORBLikelihood} with $L(\bm{\Xi} \mid \bm{y}_1, \ldots, \bm{y}_n) = L_1 (\bm{\Xi}) L_3(\bm{\Xi})$. Similarly, if there are no missing outcomes for $y_1$, then $m_3 = 0$ and we replace \eqref{ABSORBLikelihood} with $L(\bm{\Xi} \mid \bm{y}_1, \ldots, \bm{y}_n ) = L_1 (\bm{\Xi}) L_2 (\bm{\Xi})$. In Appendix~\ref{ABSORBISM}, we describe how to further extend the ABSORB model to incorporate studies that are \textit{completely} missing due to publication bias (i.e., studies that do not report \textit{either} $y_1$ or $y_2$). Such an extension of ABSORB to account for PB, in addition to ORB, is possible if we know the \textit{number} of missing studies.

\subsection{Quantifying the Impact of Outcome Reporting Bias} \label{QuantifyingORB}

In addition to correcting the bias in estimation of $\bm{\mu}$, it is also of practical interest to evaluate the \textit{impact} of ORB on MMA. To the best of our knowledge, there are no existing approaches to quantify the impact of ORB, either frequentist or Bayesian. The Bayesian approach has a natural way of doing this through comparing the bias-corrected posteriors for $\mu_1$ and/or $\mu_2$ under the ABSORB model against their \textit{non}-bias corrected posteriors.

\subsubsection{Estimation for the Non-Bias Corrected Model} \label{NonBiasCorrectedModel}

We first describe how to estimate the parameters in MMA with missing outcomes \textit{without} accounting for ORB. The ABSORB model \eqref{YgivenZ}--\eqref{ZeroCorrelations} explicitly models the selective reporting mechanism through the latent variables $z_{1}$ and $z_{2}$. These variables control whether or not the corresponding outcomes $y_{1}$ or $y_{2}$ are reported, and thus, we obtain bias-corrected estimates of $\bm{\mu}$ under ABSORB. The likelihood of reporting $y_1$ and $y_2$ ultimately depends on the correlation parameters $\rho_1$ and $\rho_2$ in \eqref{EpsilonDeltaCorrelation}. However, if $\rho_1 = \rho_2 = 0$, then $\textrm{corr}(y_{ij}, z_{ij}) = 0$ for all $i = 1, \ldots, n, j = 1, 2$, and the model \eqref{YgivenZ}--\eqref{ZeroCorrelations} reduces to
\begin{equation} \label{ABSORBNoCorrelations}
	\begin{array}{lll}
	y_{i1} & = \mu_1 + \tau_1 u_{i1} + s_{i1} \epsilon_{i1}, & \textrm{corr}(\epsilon_{i1}, \epsilon_{i2}) = \rho_\text{W}; \\
	y_{i2} & = \mu_2 + \tau_2 u_{i2} + s_{i2} \epsilon_{i2}, & \textrm{corr}(u_{i1}, u_{i2}) = \rho_\text{B}.
\end{array}
\end{equation}
In other words, when $\rho_1 = \rho_2 = 0$, the dependence of $y_{i1}$ and $y_{i2}$ on $z_{i1}$ and $z_{i2}$ respectively is removed in \eqref{ABSORBNoCorrelations}, and we \textit{only} have the unknown parameters $(\mu_1, \mu_2, \tau_1, \tau_2, \rho_\text{W}, \rho_\text{B})'$. In this case, the ABSORB model reduces to a joint model with a bivariate random effects model for the $m_1$ studies that report both $(y_1, y_2)'$ and univariate random effects models for the $m_2$ studies that report only $y_1$ and the $m_3$ studies that report only $y_2$. We call model \eqref{ABSORBNoCorrelations} the \textit{non}-bias corrected model because we ignore the selection process that was induced through the latent variables $z_1$ and $z_2$.

Similar to the bias-corrected ABSORB model, we introduce the latent parameters $(\theta_{i1}, \theta_{i2})'$ for $i = 1, \ldots, m_1$, $\widetilde{\theta}_{i1}$ for $i = m_1+1, \ldots, m_1+m_2$, and $\check{\theta}_{i2}$ for $i = m_1+m_2+1, \ldots, n$, as in \eqref{latentTheta}, \eqref{latenttheta1}, and \eqref{latenttheta2}. Let $\bm{\Omega}$ denote all the unknown parameters in the non-bias corrected model, including these latent parameters. Note that $\bm{\Omega}$ does not include the parameters $(\rho_1, \rho_2, \gamma_{01}, \gamma_{11}, \gamma_{02}, \gamma_{12})'$, because $\rho_1$ and $\rho_2$ are fixed at zero and we no longer need to condition on the latent variables $(z_1, z_2)'$ in our analysis. In the non-bias corrected model, we model the $m_1$ studies that report both outcomes as
\begin{align} \label{StandardBivariateMetaAnalysis}
	\begin{pmatrix} y_{i1} \\ y_{i2} \end{pmatrix} \sim \mathcal{N} \left( \begin{pmatrix} \theta_{i1} \\ \theta_{i2} \end{pmatrix}, \begin{pmatrix} s_{i1}^2 & \rho_\text{W} s_{i1} s_{i2} \\ \rho_\text{W} s_{i1} s_{i2} & s_{i2}^2 \end{pmatrix} \right), \hspace{.5cm} i = 1, \ldots, m_1,
\end{align} 
where the joint distribution of $(\theta_{i1}, \theta_{i2})'$ is given in \eqref{latentTheta}. The likelihood function for these $m_1$ studies in the non-bias corrected model is
\begin{align} \label{NonBiasCorrectedBothOutcomes}
	L_1 ( \bm{\Omega} ) = \prod_{i=1}^{m_1}  f(y_{i1}, y_{i2} \mid \theta_{i1}, \theta_{i2}, \rho_\text{W} ),
\end{align}
where $f( y_{i1}, y_{i2} \mid \cdot)$ is the pdf of the bivariate normal density in \eqref{StandardBivariateMetaAnalysis}. For the $m_2$ studies that only report $y_{1}$ but not $y_{2}$, the non-bias corrected model reduces to $y_{i1} \sim \mathcal{N} ( \widetilde{\theta}_{i1}, s_{i1}^2)$, where $\widetilde{\theta}_{i1} \sim \mathcal{N}(\mu_1, \tau_1^2)$. The corresponding likelihood function for these $m_2$ studies is
\begin{align} \label{NonBiasCorrectedFirstOutcome}
	L_2 ( \bm{\Omega} )  = \prod_{i=m_1+1}^{m_1+m_2} f(y_{i1} \mid \widetilde{\theta}_{i1} ) ,
\end{align}
where $f(y_{i1} \mid \widetilde{\theta}_{i1})$ is the pdf for $\mathcal{N}( \widetilde{\theta}_{i1}, s_{i1}^2)$. Similarly, for the $m_3$ studies that only report $y_{2}$ but not $y_{1}$, the non-bias corrected model reduces to $y_{i2} \sim \mathcal{N} (\check{\theta}_{i2}, s_{i2}^2)$, where $\check{\theta}_{i2} \sim \mathcal{N}( \mu_2, \tau_2^2)$. The corresponding likelihood for these $m_3$ studies is
\begin{align} \label{NonBiasCorrectedSecondOutcome}
	L_3 (\bm{\Omega})  = \prod_{i=m_1+m_2+1}^{n} f(y_{i2} \mid \check{\theta}_{i2} ),
\end{align}
where $f(y_{i2} \mid \check{\theta}_{i2})$ is the pdf for $\mathcal{N}(\check{\theta}_{i2}, s_{i2}^2)$. Altogether, the joint likelihood for all $n$ studies in the \textit{non}-bias corrected model is the product of the likelihoods in \eqref{NonBiasCorrectedBothOutcomes}--\eqref{NonBiasCorrectedSecondOutcome}:
\begin{equation} \label{NonBiasCorrectedLikelihood}
	L (\bm{\Omega} \mid \bm{y}_1, \ldots, \bm{y}_n ) = L_1(\bm{\Omega})  L_2 ( \bm{\Omega} )  L_3 (\bm{\Omega}).
\end{equation}
From \eqref{NonBiasCorrectedLikelihood}, we conduct posterior inference for $\bm{\Omega}$ by placing the priors \eqref{latentTheta}, \eqref{latenttheta1}, and \eqref{latenttheta2} on the latent variables $(\theta_{i1}, \theta_{i2})'$, $\widetilde{\theta}_{i1}$, and $\check{\theta}_{i2}$ respectively, and the priors \eqref{muprior} on $\bm{\mu}$, \eqref{tauprior} on $(\tau_1, \tau_2)'$, and \eqref{rhoprior} on $(\rho_\text{W}, \rho_\text{B})'$. We thus obtain the posterior for $\bm{\Omega}$ as
\begin{align} \label{NonBiasCorrectedPosterior}
	p( \bm{\Omega} \mid \bm{y}_1, \ldots, \bm{y}_n) \propto L ( \bm{\Omega} \mid \bm{y}_1, \dots, \bm{y}_n) p(\bm{\Omega}).
\end{align}
With the model fully specified, we can approximate the marginal posteriors $p(\mu_1 \mid \bm{y}_1, \ldots, \bm{y}_n)$ and $p(\mu_2 \mid \bm{y}_1, \ldots, \bm{y}_n)$ using MCMC. As before, if $m_2=0$, then we replace \eqref{NonBiasCorrectedLikelihood} with $L(\bm{\Omega} \mid \bm{y}_1, \ldots, \bm{y}_n) = L_1 (\bm{\Omega})  L_3 (\bm{\Omega})$, and if $m_3 = 0$, then we replace \eqref{NonBiasCorrectedLikelihood} with $L(\bm{\Omega} \mid \bm{y}_1, \ldots, \bm{y}_n) = L_1(\bm{\Omega})  L_2 (\bm{\Omega})$.

\subsubsection{The $D$ Measure for Quantifying the Impact of ORB} \label{DMeasure}

To quantify the impact of publication bias in \textit{univariate} meta-analysis, \cite{BaiLinBolandChen2020} proposed the $D$ measure as a way of measuring the difference between a publication bias-corrected and \textit{non}-bias-corrected posterior for a mean treatment effect. Here, we extend the $D$ measure to quantify the impact of ORB bias in MMA. 

Let $p_\text{ABS} ( \mu_j \mid \bm{y}_1, \ldots, \bm{y}_n)$ denote the posterior for $\mu_j, j=1,2$ under the ABSORB model as described in Section~\ref{LikelihoodImplementation}, and let $p_{NBC} ( \mu_j \mid \bm{y}_1, \ldots, \bm{y}_n )$ denote the posterior for $\mu_j$ under the non-bias corrected model described in Section~\ref{NonBiasCorrectedModel}. To quantify the impact of ORB for each individual endpoint $\mu_j$, we propose taking the Hellinger distance between $p_\text{ABS}(\mu_j \mid \bm{y}_1, \ldots, \bm{y}_n)$ and $p_{NBC} (\mu_j \mid \bm{y}_1, \ldots, \bm{y}_n)$. If we are instead interested in quantifying the joint impact from ORB on both endpoints, we can take the Hellinger distance between the joint posteriors $p_\text{ABS}(\mu_1, \mu_2 \mid \bm{y}_1, \ldots, \bm{y}_n)$ and $p_{NBC} ( \mu_1, \mu_2 \mid \bm{y}_1, \ldots, \bm{y}_n)$. 

Let $\bm{x}$ be either a random scalar or a random vector. The Hellinger distance between densities $f$ and $g$ is defined as
\begin{equation}  \label{Hellinger}
	H(f,g) = \left[ 1 - \displaystyle \int \sqrt{f (\bm{x}) g (\bm{x}) } d \bm{x} \right]^{1/2},
\end{equation}
The Hellinger distance is an appealing way to quantify the dissimilarity between two probability densities. Unlike other probability distance measures such as the Kullback-Leibler distance, the Hellinger distance is symmetric \textit{and} always bounded between zero and one. This gives the Hellinger distance a clear interpretation. Values close to zero indicate that $f$ and $g$ are nearly identical distributions, while values close to one indicate that the majority of the probability mass in $f$ does \textit{not} overlap with that of $g$. 

For shorthand, let $p_\text{ABS}$ and $p_{NBC}$ be the posteriors for either $\mu_1$, $\mu_2$, or $\bm{\mu}$. Unfortunately, these posterior distributions are intractable and therefore need to be approximated. In the present context, we approximate the posteriors $p_\text{ABS}$ and $p_{NBC}$ using MCMC samples to obtain kernel density estimates, $\widehat{p}_\text{ABS}$ and $\widehat{p}_{NBC}$. We then use numerical integration to estimate the Hellinger distance \eqref{Hellinger} between $\widehat{p}_\text{ABS}$ and $\widehat{p}_{NBC}$. In short, our measure for the impact of ORB is
\begin{equation} \label{Dmeasure}
	D = H \left( \widehat{p}_\text{ABS}, \widehat{p}_{NBC}  \right),
\end{equation}
The $D$ measure \eqref{Dmeasure} quantifies the degree to which the ABSORB posterior changes from the non-bias corrected posterior. Smaller values of $D$ ($D \approx 0$) indicate that $p_\text{ABS}$ and $p_{NBC}$ are almost identical. Thus, we conclude that there is negligible impact from ORB on the MMA. Meanwhile larger values of $D$ ($D \approx 1$) indicate that ORB has a strong impact on the estimation of $\bm{\mu}$. In this case, the ABSORB posterior differs quite drastically from the non-bias corrected posterior. In the next section, we provide several illustrations of the $D$ measure on real systematic reviews from the Cochrane Database of Systematic Reviews.

\section{Meta-Evaluation with the Cochrane Database of Systematic Reviews}
\label{meta-meta}

To evaluate the performance of our model, we conducted a meta-evaluation of 748 systematic reviews from the Cochrane Database of Systematic Reviews (hereinafter refer to as the ``Cochrane Database''). We describe how we arrived at these 748 eligible reviews in Section A of the Supplementary Material. For dichotomous outcomes, we performed a log transformation to risk ratio and odds ratio outcomes. For each of the reviews in our meta-evaluation, we fit the ABSORB and non-bias corrected models of Section~\ref{ABSORB}.  For both models, we ran three separate chains of the MCMC algorithm for 50,000 iterations, discarding the first 10,000 samples as burn-in. This left us with a total of 120,000 samples from three chains with which to approximate the posteriors and calculate the $D$ measures \eqref{Dmeasure}. We monitored the convergence of the MCMC using ESS; if the ESS was below 100 for $\mu_1$ or $\mu_2$, then we increased the number of iterations to 100,000, 200,000, etc.\ as needed.

\begin{figure}[t!]
\centering
\includegraphics[width=.52\linewidth]{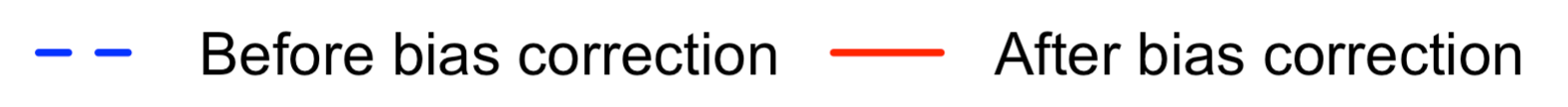} \\
\includegraphics[width=.25\textwidth]{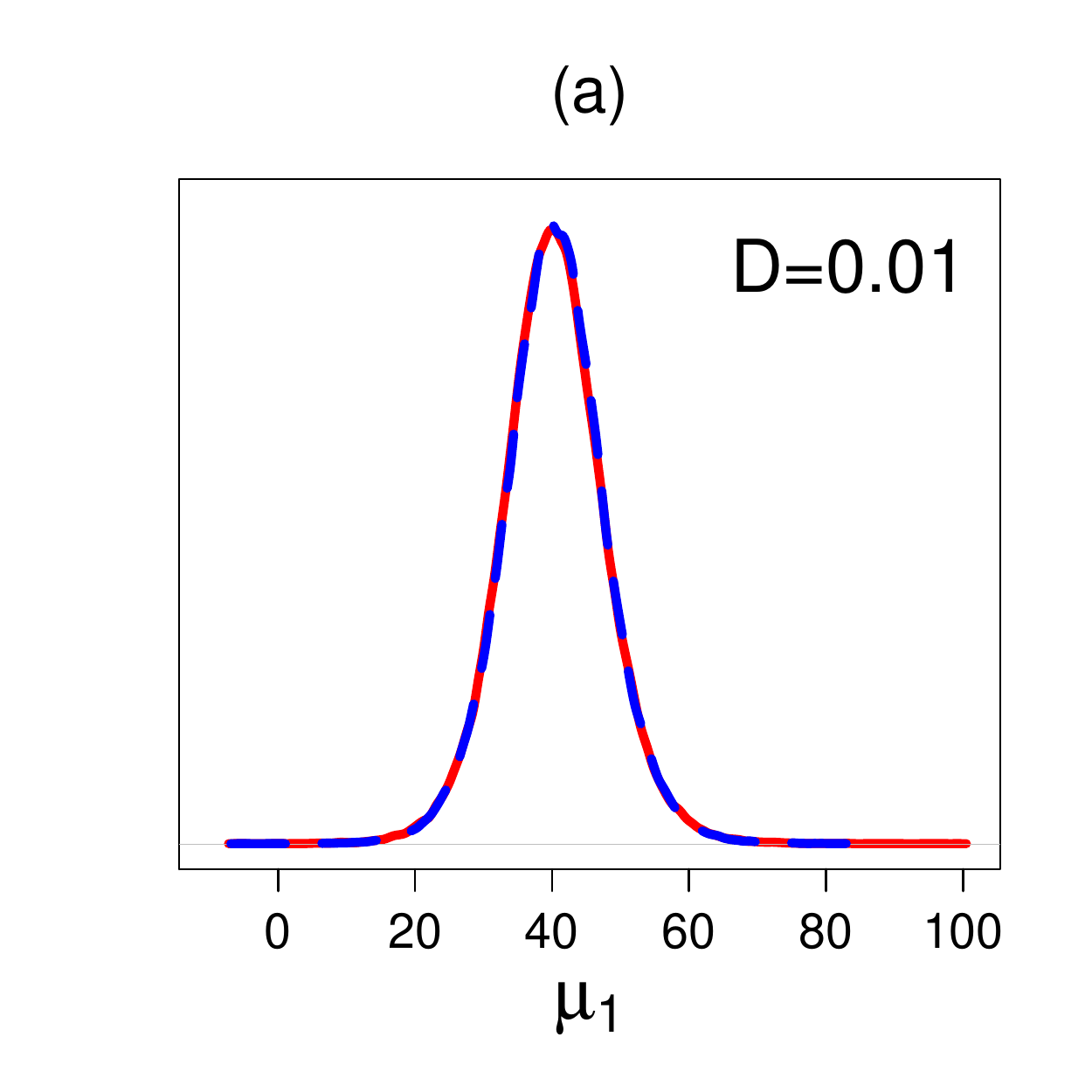}
\includegraphics[width=.25\textwidth]{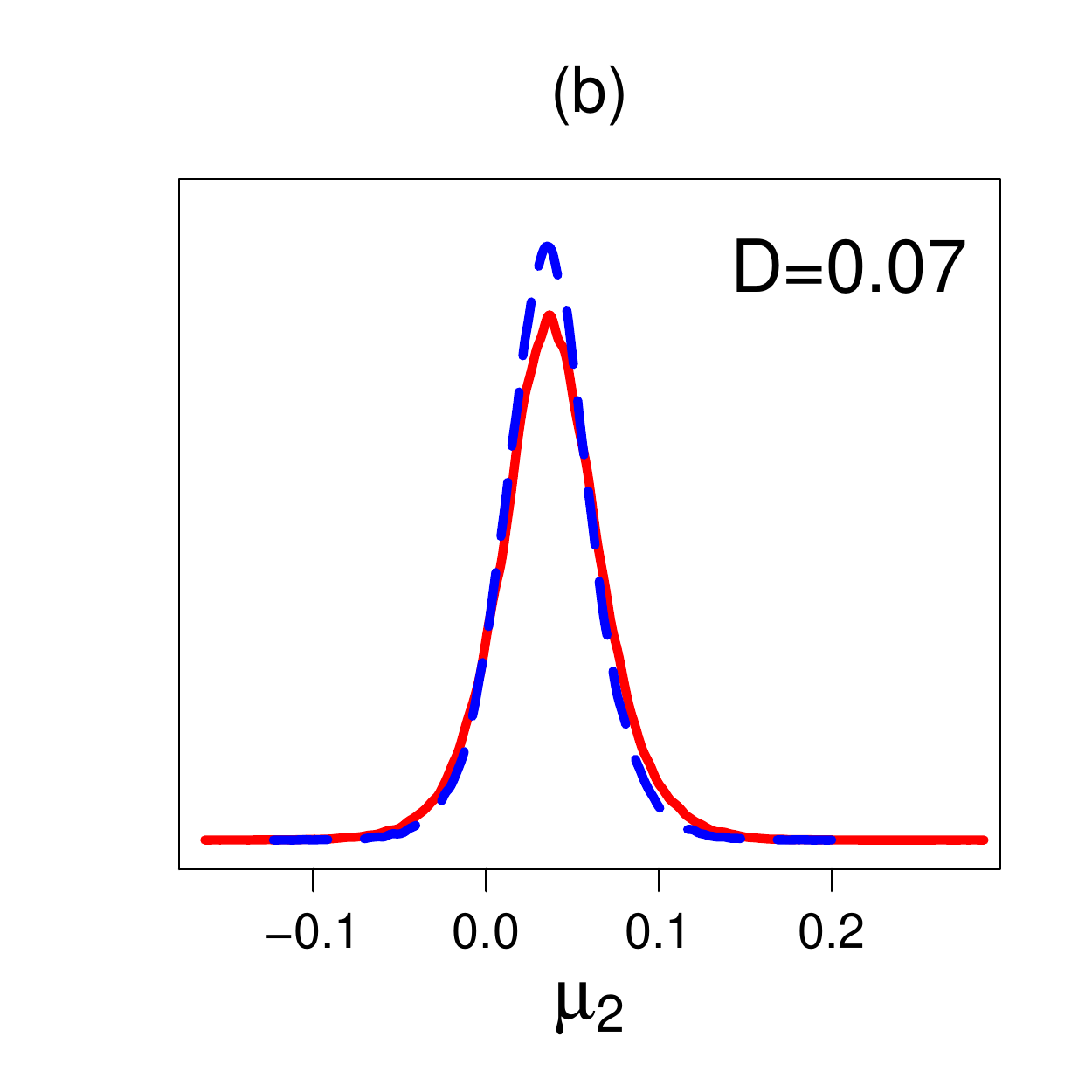}
\includegraphics[width=.25\textwidth]{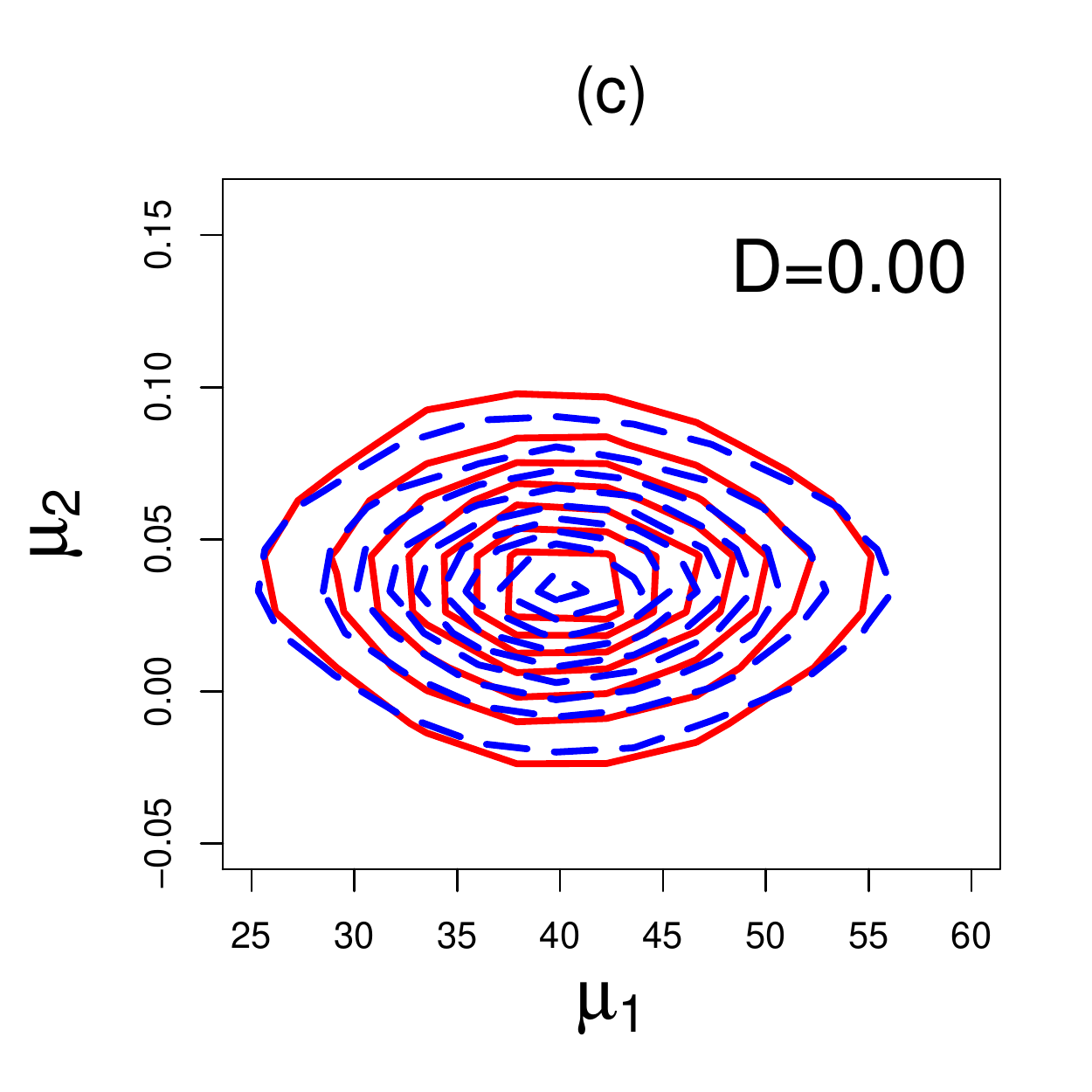}
\includegraphics[width=.25\textwidth]{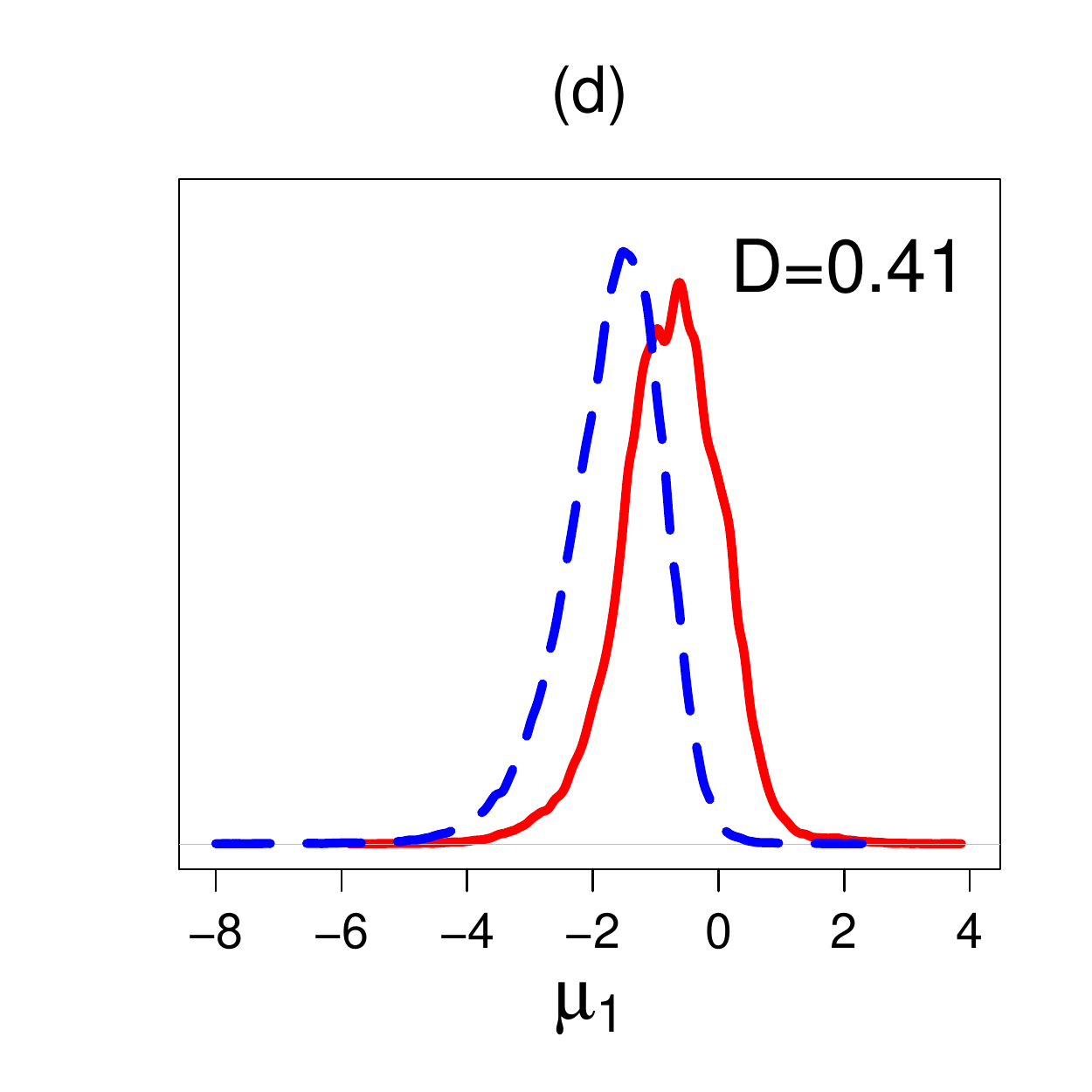}
\includegraphics[width=.25\textwidth]{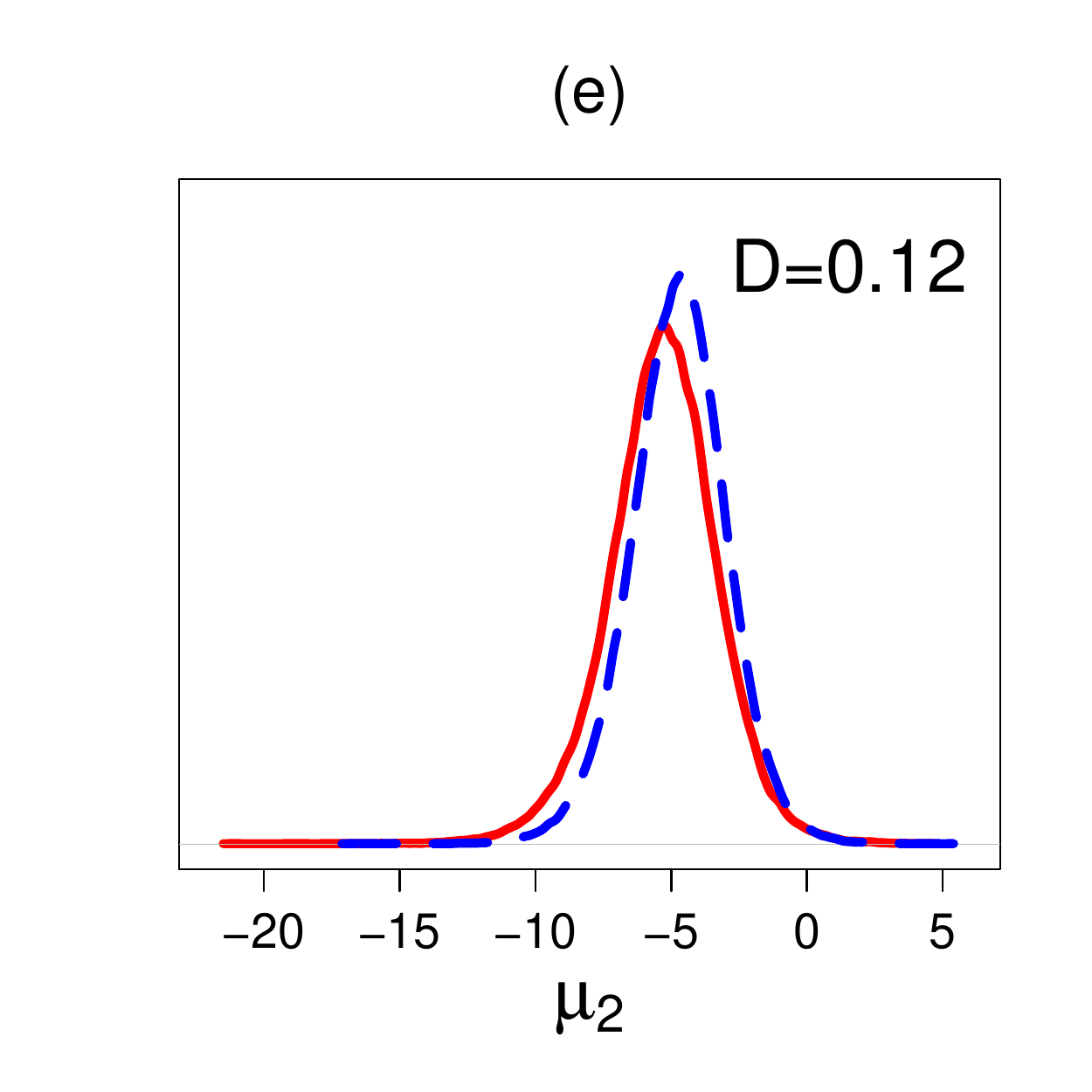}
\includegraphics[width=.25\textwidth]{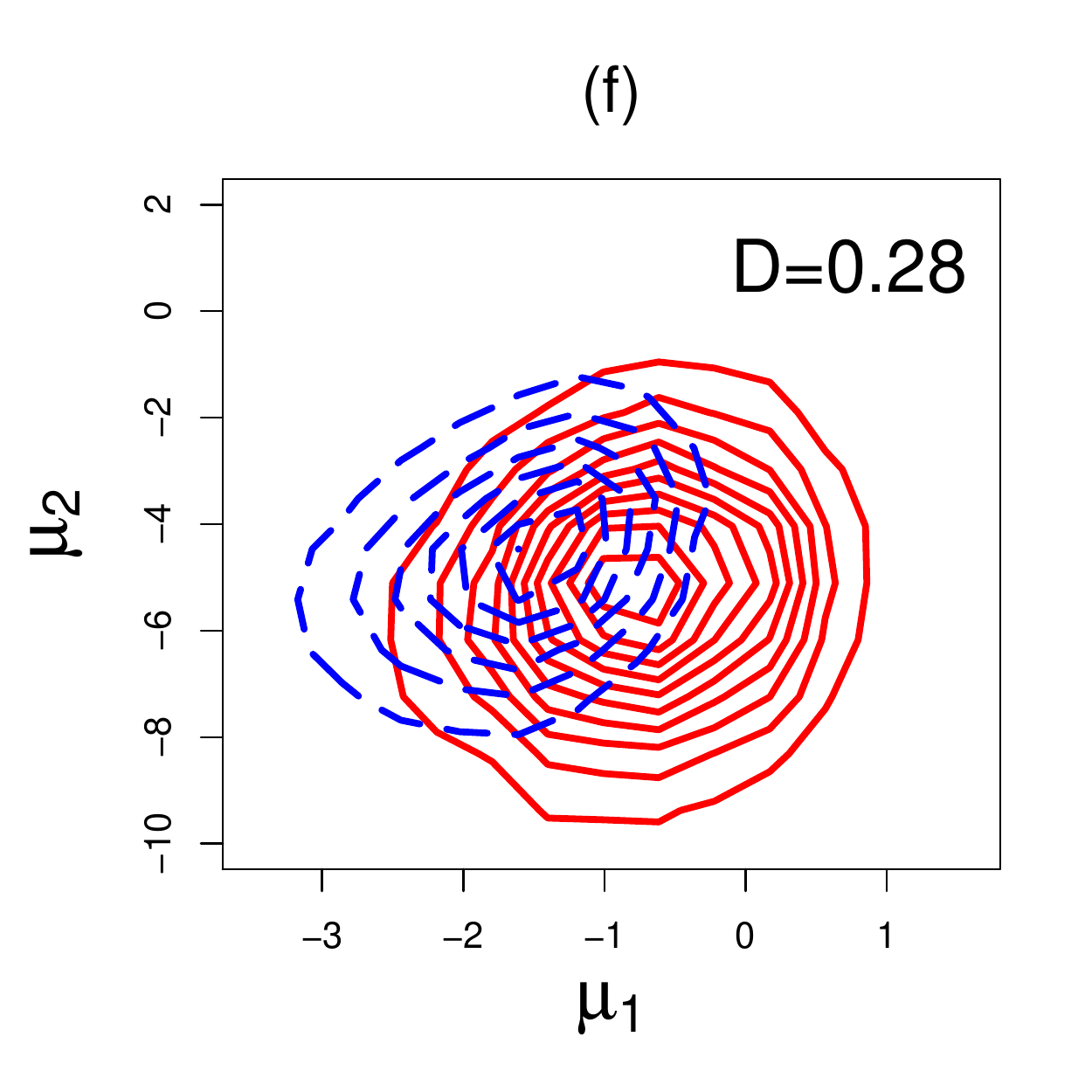}
\includegraphics[width=.25\textwidth]{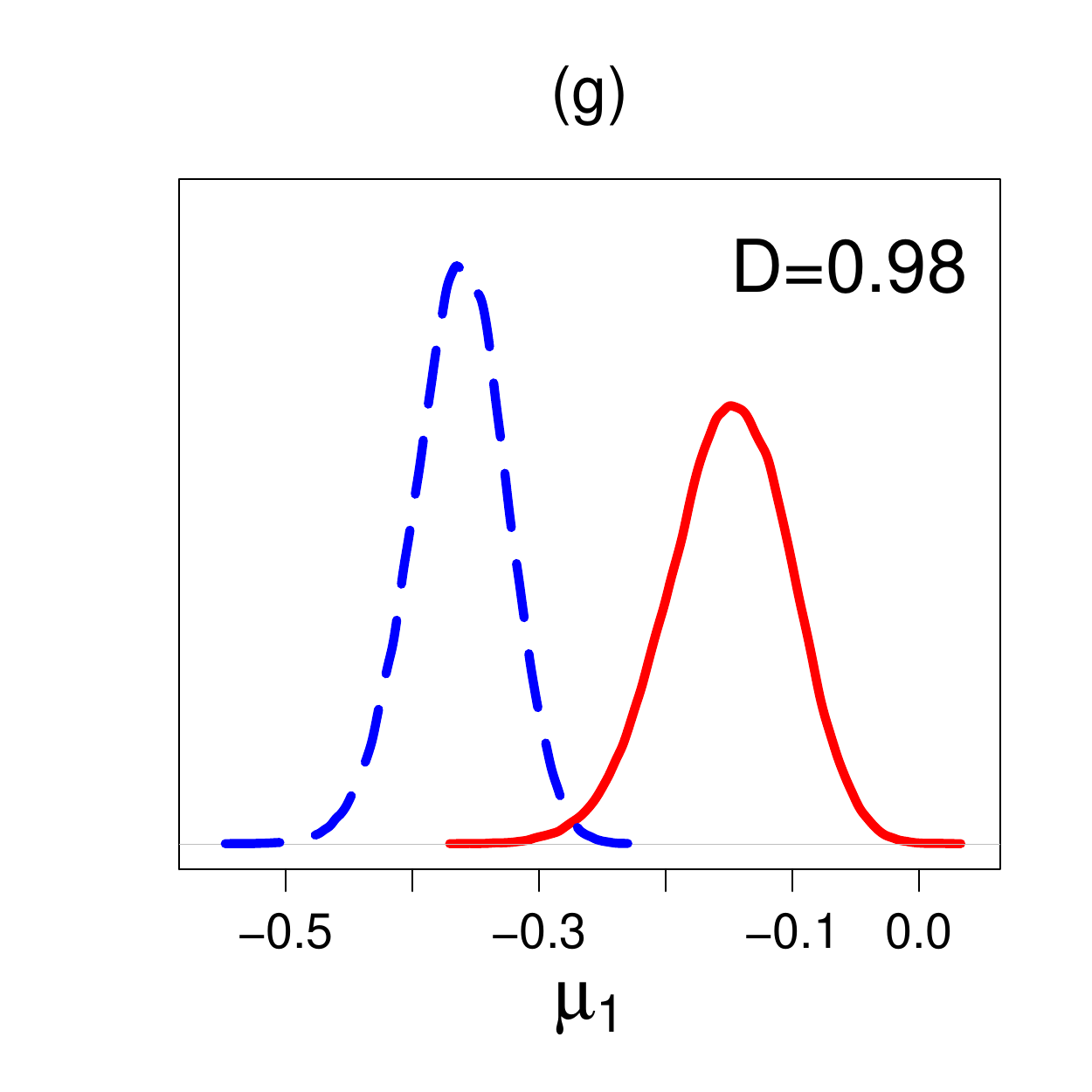}
\includegraphics[width=.25\textwidth]{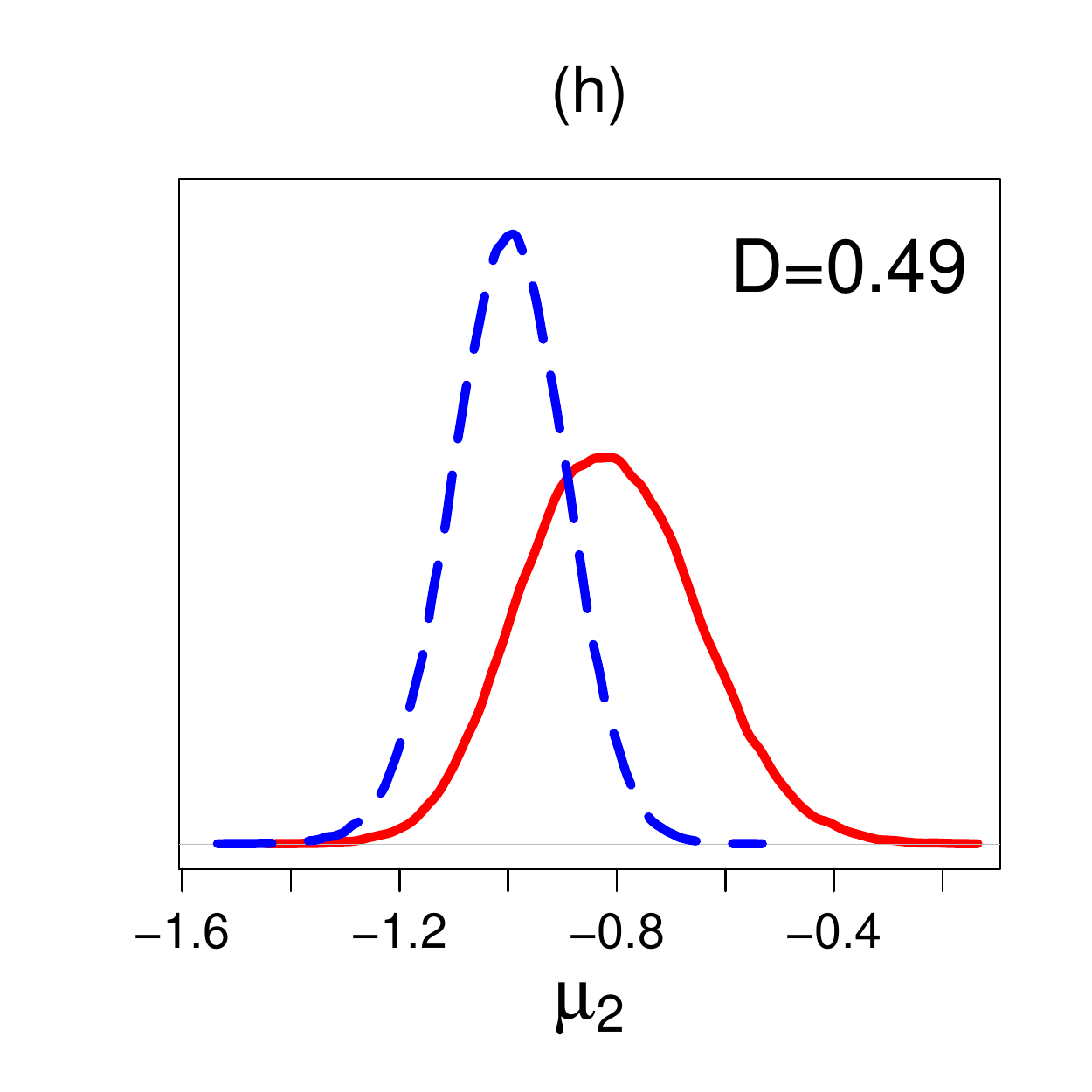}
\includegraphics[width=.25\textwidth]{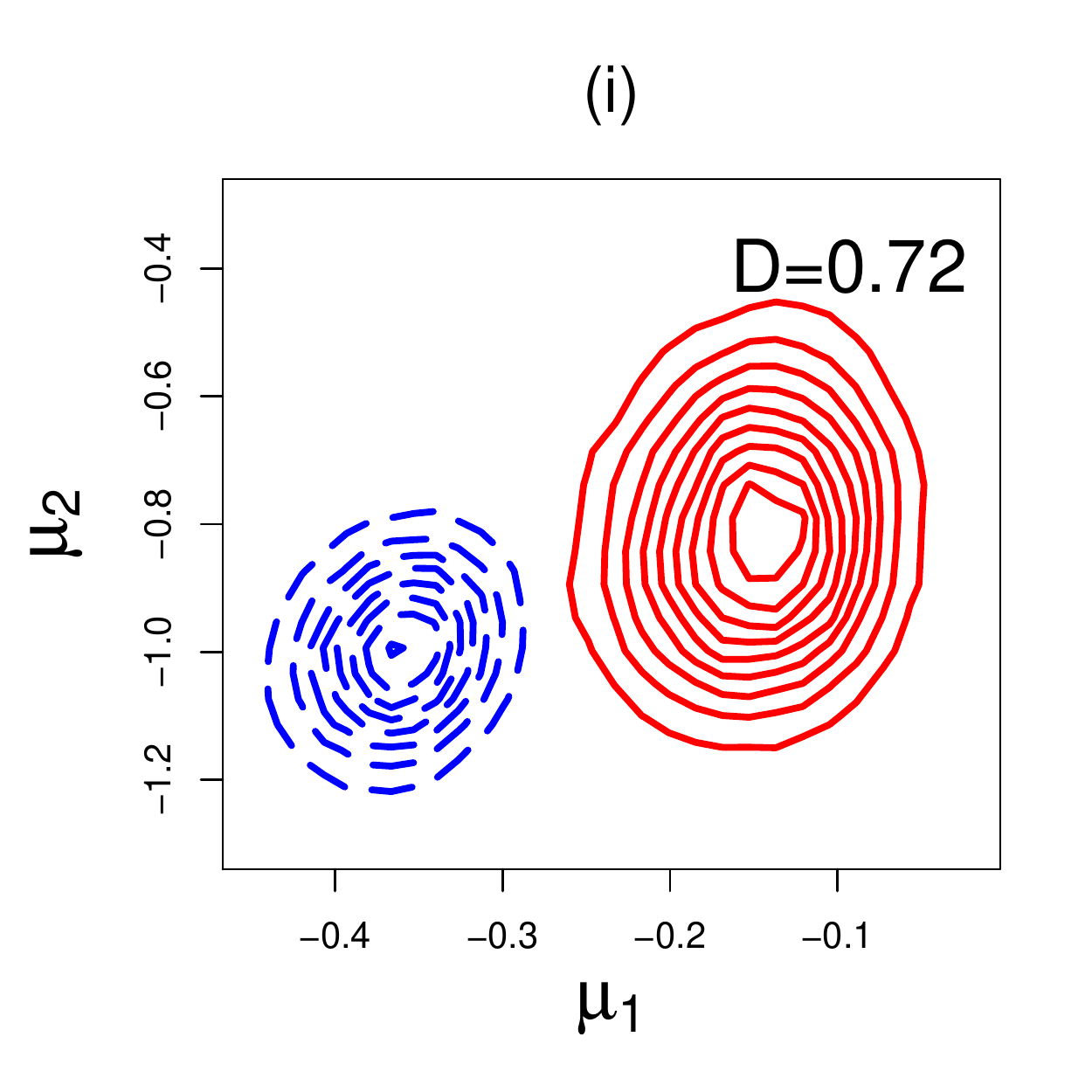}
	\caption{Illustrations of three meta-analyses from the Cochrane Database. Panels~(a)--(c) show the results for MMA1, panels (d)--(f) show the results for MMA2, and panels (g)--(i) show the results for MMA3. In panels (g) and (i), $\mu_1$ is plotted on the log-RR scale. }\label{meta:CochraneExamples}
	\end{figure}

We present three representative meta-analyses from our meta-evaluation, which we denote as MMA1, MMA2, and MMA3. Table~\ref{ThreeMetaAnalyses} provides the details of these meta-analyses, including the review topic, the effect measure, and descriptions of the bivariate treatment effects of interest. The results from these three meta-analyses are depicted in Figure~\ref{meta:CochraneExamples}. In Figure~\ref{meta:CochraneExamples}, we plot the bias-corrected posteriors under the ABSORB model (solid line) against their non-biased corrected posteriors (dashed line) for $\mu_1$ and $\mu_2$, as well as the contour plots for the bias-corrected and non-biased corrected joint posteriors of $\bm{\mu}$. We also report the $D$ measures for $\mu_1$, $\mu_2$, and $\bm{\mu}$. For MMA1 (panels (a)-(c)), we see that there is a negligible impact from ORB for both endpoints, and thus the $D$ measures are all close to zero. In MMA2 (panels (d)-(f)), there is a fairly strong impact from ORB for the first endpoint ($D=0.41$) and a negligible impact ($D=0.12)$ for the second endpoint. In MMA3 (panels (g)-(i)), there is a very strong impact from ORB for the first endpoint ($D=0.98$) and a fairly strong impact ($D=0.49$) for the second endpoint. The bottom left graph in Figure~\ref{meta:CochraneExamples} shows very little overlap between the bias-corrected and non-bias corrected posteriors for $\mu_1$ in MMA3, and hence, we obtained a $D$ measure close to one.

\begin{table}[t!]
    \centering
    \caption{Three representative meta-analyses from the Cochrane Database.}
    \begin{center}
\resizebox{\textwidth}{!}{
\begin{tabular}{lllll}
    \hline 
    & Topic & Outcome & Effect measure & Analysis  \\
    \hline
    \multirow{2}{*}{MMA1\tablefootnote{Cochrane Database ID:CD000990, DOI: 10.1002/14651858.CD000990.pub4.}}& Exercise for intermittent  & $\mu_1$ & Mean difference & Change in maximal walking distance or time  \\
    & claudication & $\mu_2$ & Mean difference & Ankle brachial index  \\
    \hline
     \multirow{2}{*}{MMA2\tablefootnote{Cochrane Database ID:CD000335, DOI: 10.1002/14651858.CD000335.pub2.}} & Exercise therapy for treatment & $\mu_1$ & Mean difference & Function measure  \\ 
    & of non-specific low back pain & $\mu_2$ & Mean difference & Pain measure  \\
    \hline
    \multirow{2}{*}{MMA3\tablefootnote{Cochrane Database ID:CD001886, DOI: 10.1002/14651858.CD001886.pub4.}}& Anti-fibrinolytic use for minimizing & $\mu_1$ & Risk ratio & Number of patients exposed to allogeneic blood \\ 
    & perioperative allogeneic blood transfusion & $\mu_2$ & Mean difference & Units of allogeneic blood transfused \\
    \hline
\end{tabular}}
\end{center} \label{ThreeMetaAnalyses}
\end{table}

In particular, for MMA2, the 95\% posterior credible interval for $\mu_1$ (i.e., the mean change in function measure after exercise therapy for lower back pain) shifted from $(-3.52, -0.38)$ under the non-bias corrected posterior to $(-2.53, 0.69)$ under the ABSORB bias-corrected posterior. This indicates that after adjusting for ORB, the 95\% bias-corrected posterior interval contained zero, and the mean change in function measure after exercise therapy was \textit{no longer} statistically significant. As a consequence of non-negligible ORB, 12.03\% of all 748 meta-analyses in our meta-evaluation (90 reviews) had a change in statistical significance for the first outcome, and 10.56\% (79 reviews) had a change in statistical significance for the second outcome. For 12 reviews, the statistical significance changed for \textit{both} $\mu_1$ and $\mu_2$. These results demonstrate that non-negligible ORB can have a profound effect on the conclusions from MMA.  

 In Appendix~\ref{AdditionalResults}, we provide the specific quantiles of the $D$ measure from our analysis. Based on these quantiles, we determined the following guidelines for interpreting the $D$ measure:
\begin{itemize}
    \item 0.00 to 0.20: probably no impact from ORB;
    \item 0.10 to 0.40: may represent moderate impact from ORB;
    \item 0.30 to 0.60: may represent substantial impact from ORB;
    \item 0.50 to 1.00: may represent severe impact from ORB.
\end{itemize}
Our intervals were inspired by the guidelines given for the $I^2$ statistic  \citep{higgins2002quantifying} in the Cochrane Handbook for Systematic Reviews of Interventions.\footnote{\url{https://handbook-5-1.cochrane.org/chapter_9/9_5_2_identifying_and_measuring_heterogeneity.htm}.} The $I^2$ statistic (for measuring heterogeneity in univariate meta-analyses) also lies between 0 and 1, and the Cochrane Handbook provides overlapping intervals for ``unimportant,'' ``moderate,'' ``substantial,'' and ``considerable'' heterogeneity based on $I^2$, so as to avoid setting hard cutoffs for its interpretation. 

In our experience, a $D$ measure of 0.20 or higher usually suggested non-negligible ORB or the potential to qualitatively change the conclusions from meta-analyses. Meanwhile, a $D$ measure below 0.10 normally ruled out any impact from ORB (as illustrated in panels~(a)--(c) of Figure~\ref{meta:CochraneExamples}). However, there were a few reviews where the statistical significance changed for an outcome even when $D<0.10$. This occurred when one of the CI endpoints was extremely close to zero -- in this case, the 95\% CIs before and after bias correction were very similar to each other, but even a tiny discrepancy near zero changed the conclusion. Thus, the systematic reviewer should also investigate the CIs, not just the $D$ measure.

\begin{figure}[h!]
	\centering
	\includegraphics[width=.32\textwidth]{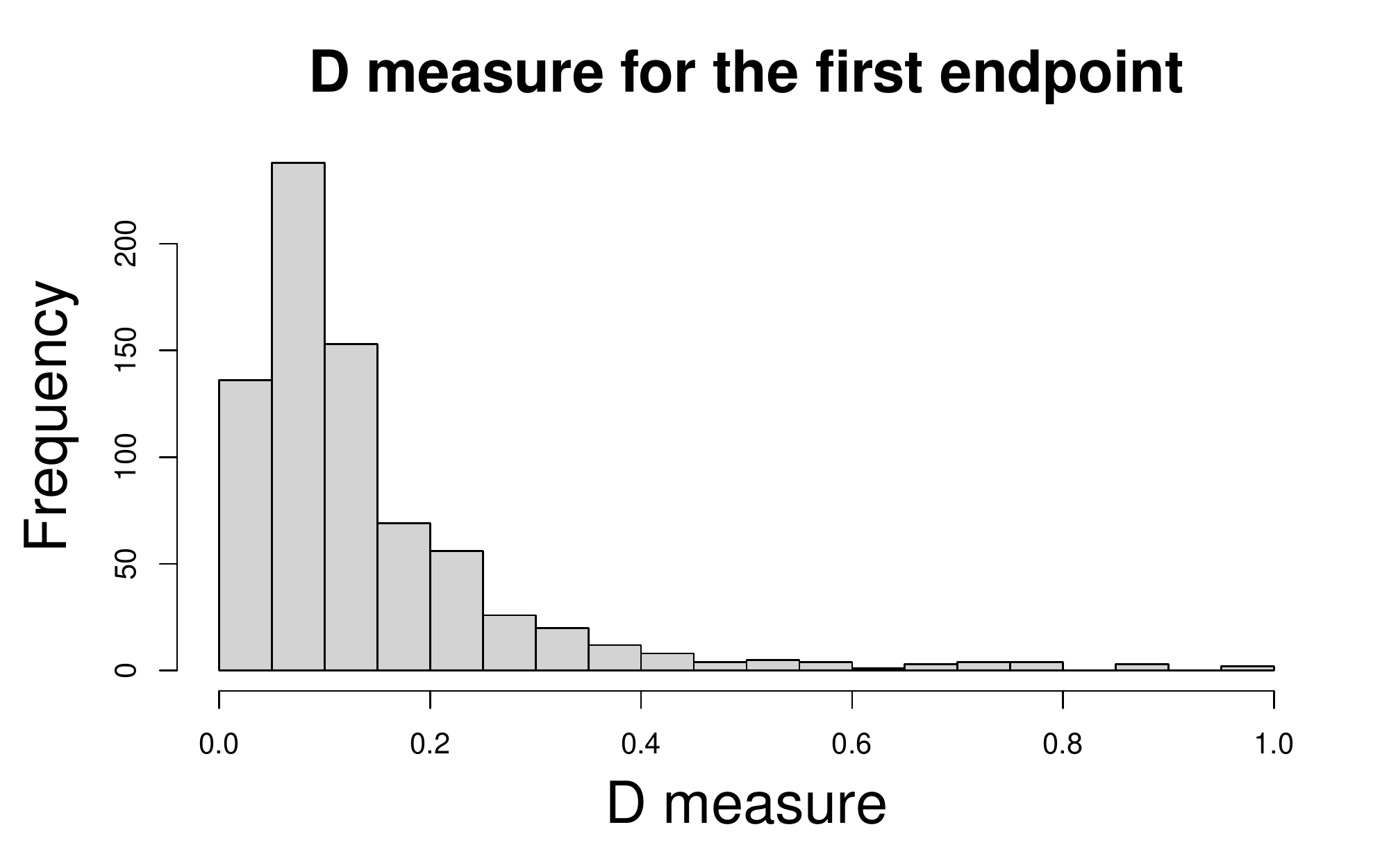}
	\includegraphics[width=.32\textwidth]{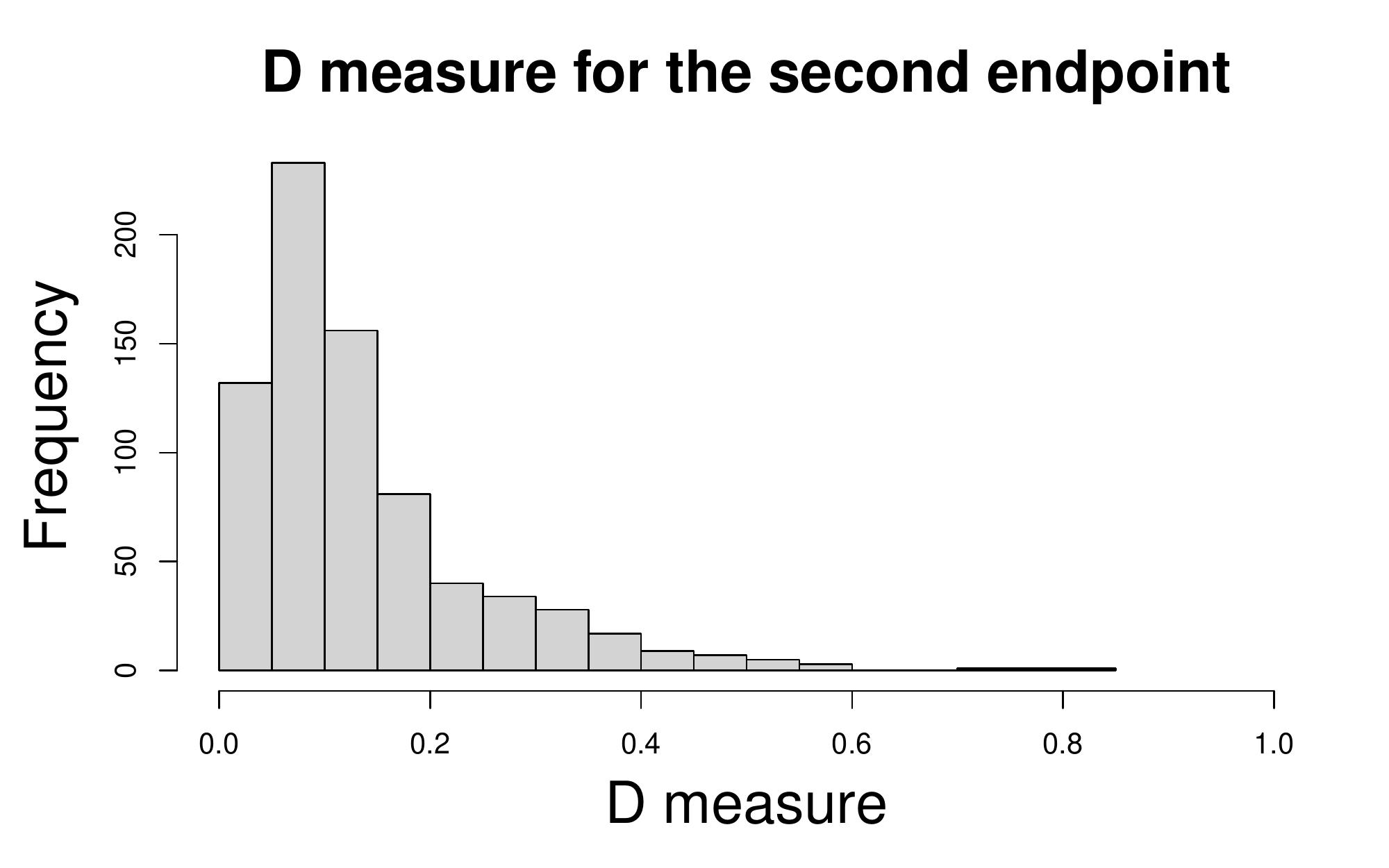}
	\includegraphics[width=.32\textwidth]{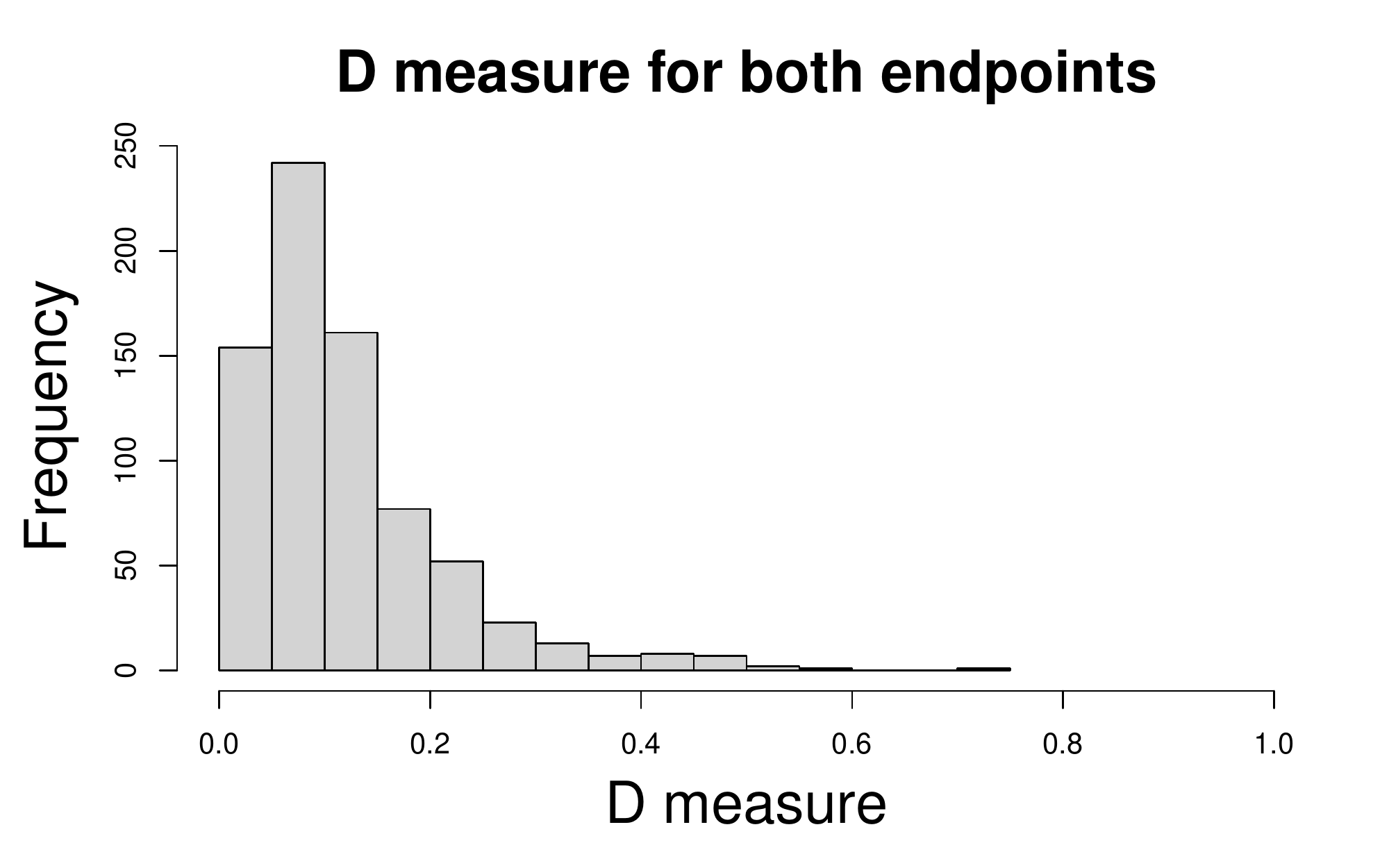}
	\caption{Empirical histograms of the $D$ measure for $\mu_1$ (left panel), $\mu_2$ (middle panel) and $\bm{\mu}$ (right panel) from all 748 reviews in our meta-evaluation.}\label{CochraneHistograms}
\end{figure}

Our meta-evaluation of the Cochrane Database found that 50.00\% of MMAs had $D < 0.10$ for the first endpoint, 48.80\% had $D < 0.10$ for the second endpoint, and 52.94\% had $D < 0.10$ for both endpoints. However, there were also a few reviews where ORB had a very high impact. Namely, 26 reviews had $D$ measures greater than 0.50 for the first endpoint, and 11 reviews had $D$ measures greater than 0.50 for the second endpoint. Figure~\ref{CochraneHistograms} plots the empirical histograms for the $D$ measure from our meta-evaluation. 

Instead of using the provided guidelines, an alternative is to simply use the quantiles from our meta-evaluation for interpretation. The quantiles for the $D$ measures are provided in Table \ref{Dtable1} of Appendix \ref{AdditionalResults}. Using this table, the systematic reviewer can locate the percentile of the $D$ measure obtained from his or her dataset among the $D$ measures from the Cochrane database and conclude that the evidence for ORB in his or her study is in the top, e.g., 20\% of all analyzed datasets. 

Our meta-evaluation of real systematic reviews from the Cochrane Database illustrates the potential of the ABSORB model for adjusting the estimates of effect sizes in the presence of ORB and the $D$ measure for quantifying the impact of ORB. In Appendix~\ref{Simulations}, we further validate our model through simulation studies under a variety of degrees of between-study heterogeneity and missingness. To summarize briefly, our simulation studies demonstrate that when ORB is present, the ABSORB model has lower bias and better empirical coverage than alternative approaches that remove studies with missing data, that impute the missing outcomes, or that ignore the correlations between the two endpoints.

\section{Case Study Results} \label{Application}

We now apply the ABSORB model to the meta-analysis introduced in Section~\ref{MotivatingData} on the effects of intervention on ReAd and QoL for HF patients. As in Section~\ref{ABSORB}, $m_1$ denotes the number of studies that reported both ReAd and QoL, $m_2$ is the number of studies that reported only ReAd, and $m_3$ is the number of studies that reported only QoL. As discussed in Section~\ref{MotivatingData}, our sample contained 45 studies on the effects of interventions on HF patients. We initially had $n=41$ published studies that reported at least one of ReAd or QoL ($m_1 = 8$, $m_2 = 26$, $m_3 = 7$). After querying the corresponding authors, we were able to obtain six additional results for QoL and a total of $n = 44$ studies with results for at least one of ReAd or QoL ($m_1 = 11$, $m_2 = 23$, $m_3 = 10$). 

With our \textit{a priori} knowledge about which studies had QoL results after querying the corresponding authors, we proceeded to conduct a two-stage analysis. In this first stage, we applied the ABSORB model to only the $n=41$ published studies that reported at least one of the ReAd or QoL outcomes (i.e., \textit{before} we had queried the authors). In the second stage, we performed our analysis with the $n=44$ updated studies (i.e., \textit{after} querying the authors). Our purpose for conducting this two-stage analysis was to see how our results changed after we were able to partially mitigate some of the ORB for QoL by querying corresponding authors. For our analysis, we did not include the studies that failed to report either ReAd or QoL. In Appendix~\ref{AdditionalHFResults}, we apply an augmented model (introduced in Appendix~\ref{ABSORBISM}) to \textit{all} 45 intervention studies in both the published and the updated data. 

To quantify the impact of outcome reporting bias in our MMA, we fit the ABSORB model of Section~\ref{ABSORBModel} and the non-bias corrected model of Section~\ref{NonBiasCorrectedModel} and used their posterior samples to compute the $D$ measure \eqref{Dmeasure}. For both models, we ran three MCMC chains of 100,000 iterations, discarding the first 10,000 iterations as burn-in. In Appendix~\ref{AdditionalHFResults}, we provide trace plots for these models, which show that the three chains mixed well and that the number of iterations we used was sufficient to achieve convergence.
 

\begin{figure}[!htbp]
\centering
\hspace{.5cm} \includegraphics[width=.5\linewidth]{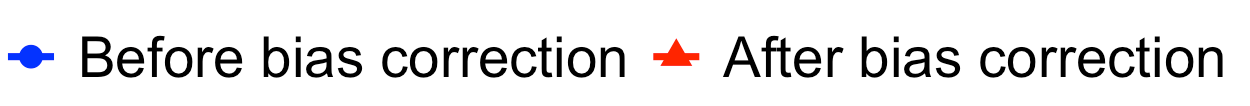} \\
\includegraphics[width=.4\textwidth]{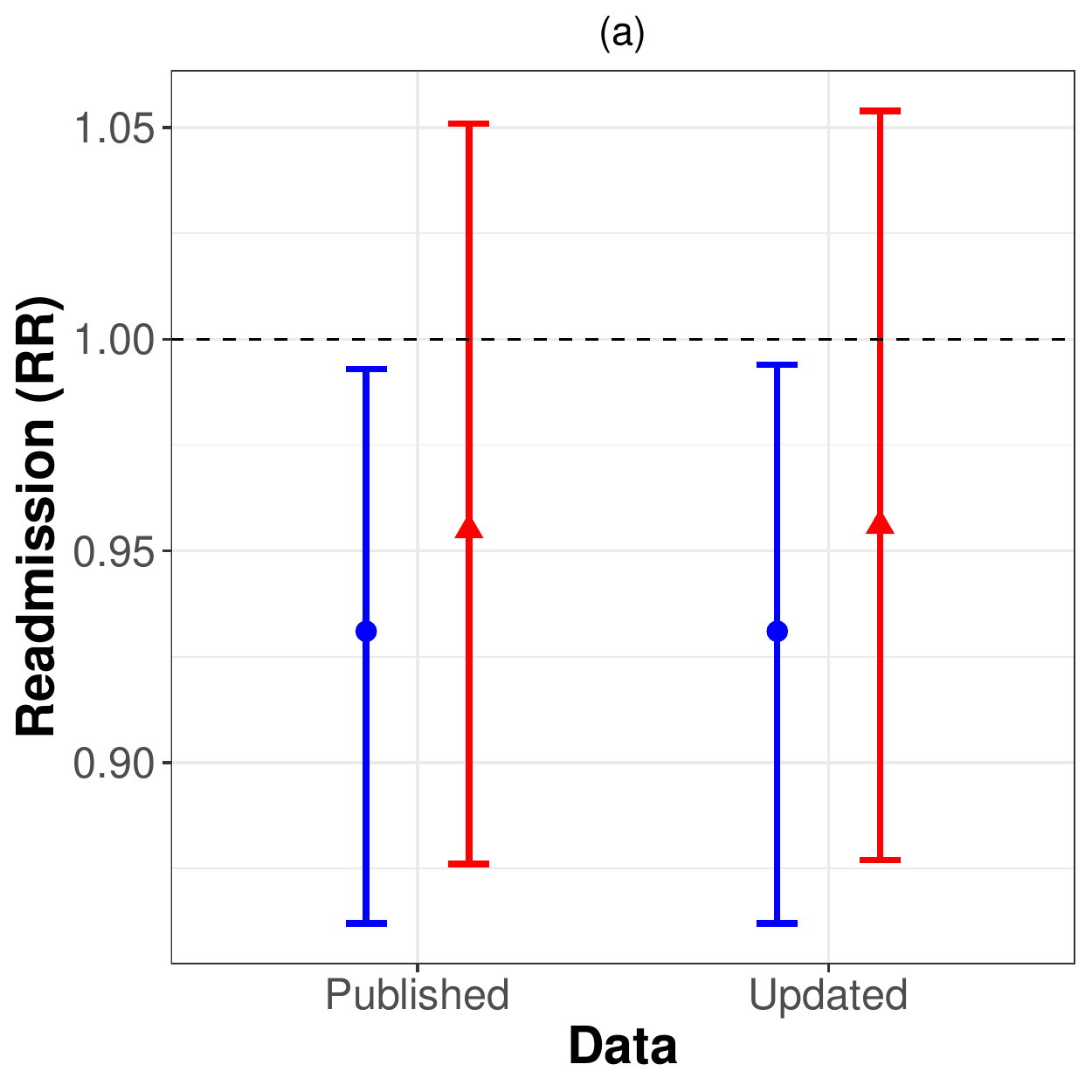}
\includegraphics[width=.4\textwidth]{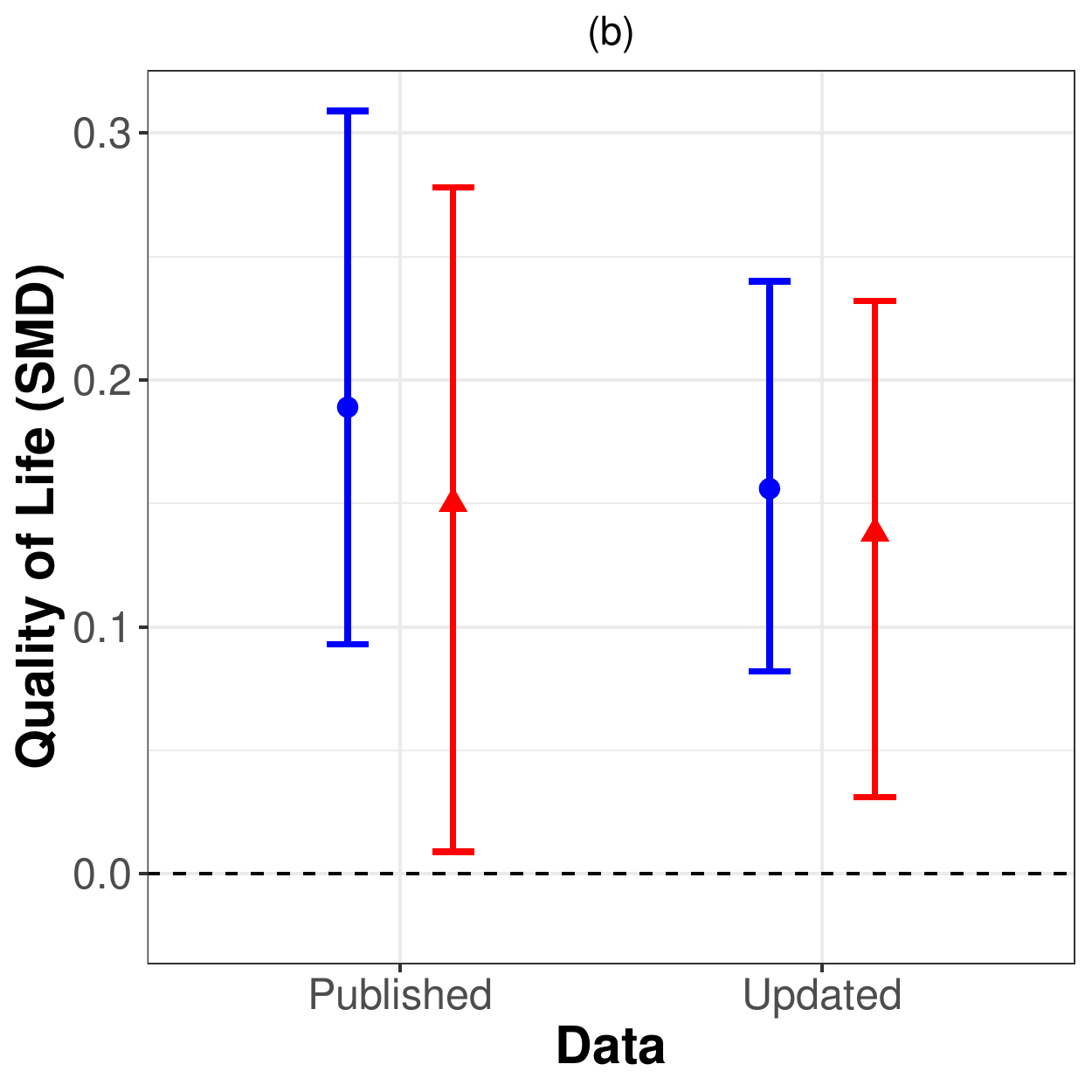}
\caption{Plots of the posterior means and 95\% posterior credible intervals for our case study on interventions for HF patients under the non-bias corrected and ABOSRB models. Panel~(a) plots the results for ReAd and panel~(b) plots the results for QoL.} \label{HeartFailureResults}
\end{figure}

Figure \ref{HeartFailureResults} shows the posterior mean effect sizes and 95\% posterior credible intervals for ReAd and QoL. For ReAd (panel (a) of Figure \ref{HeartFailureResults}), there was little difference between the MMA results obtained from the published and updated datasets. The ABSORB model estimated a mean RR of 0.955 with a 95\% CI of (0.876, 1.051) for the published data, and a mean RR of 0.956 with a 95\% CI of (0.877, 1.054) for the updated data, which indicated \textit{no} significant reduction of risk for hospital readmission for the intervention group. However, there \textit{was} a qualitative difference in the clinical conclusions for ReAd from the \textit{non}-bias corrected models. In addition to slightly lower mean estimates of RR for hospital readmission (0.931 for both the published and the updated data), the non-bias corrected models estimated 95\% CIs of (0.862, 0.993) for the published data and (0.862, 0.994) for the updated data. This indicates that \textit{without} correcting for ORB with the ABSORB model, our meta-analysis would have concluded that there was a \textit{significant} reduction in risk of hospital readmission. 

As for QoL (panel (b) of Figure \ref{HeartFailureResults}), the ABSORB model estimated an SMD of 0.15 for QoL between intervention and control groups in the published data, which was slightly larger than the result obtained from the updated data (0.138). The 95\% CI for the updated data (0.031, 0.232) was also narrower than the interval for the published data (0.009, 0.278). Based on these 95\% CIs, there was a significant improvement in QoL for heart failure patients in the intervention group. Meanwhile, in the \textit{non}-bias corrected model, the point estimates obtained from the published data and the updated data were both higher than their corresponding estimates under the ABSORB model. However, there was no change in the clinical conclusion from our ORB correction, since the non-bias corrected model also showed a significant improvement in QoL for the intervention group.



\begin{figure}[!htbp]
	\centering
	\hspace{.4cm} \includegraphics[width=.55\linewidth]{Figure1_legend.png} \\
	\includegraphics[width=.28\textwidth]{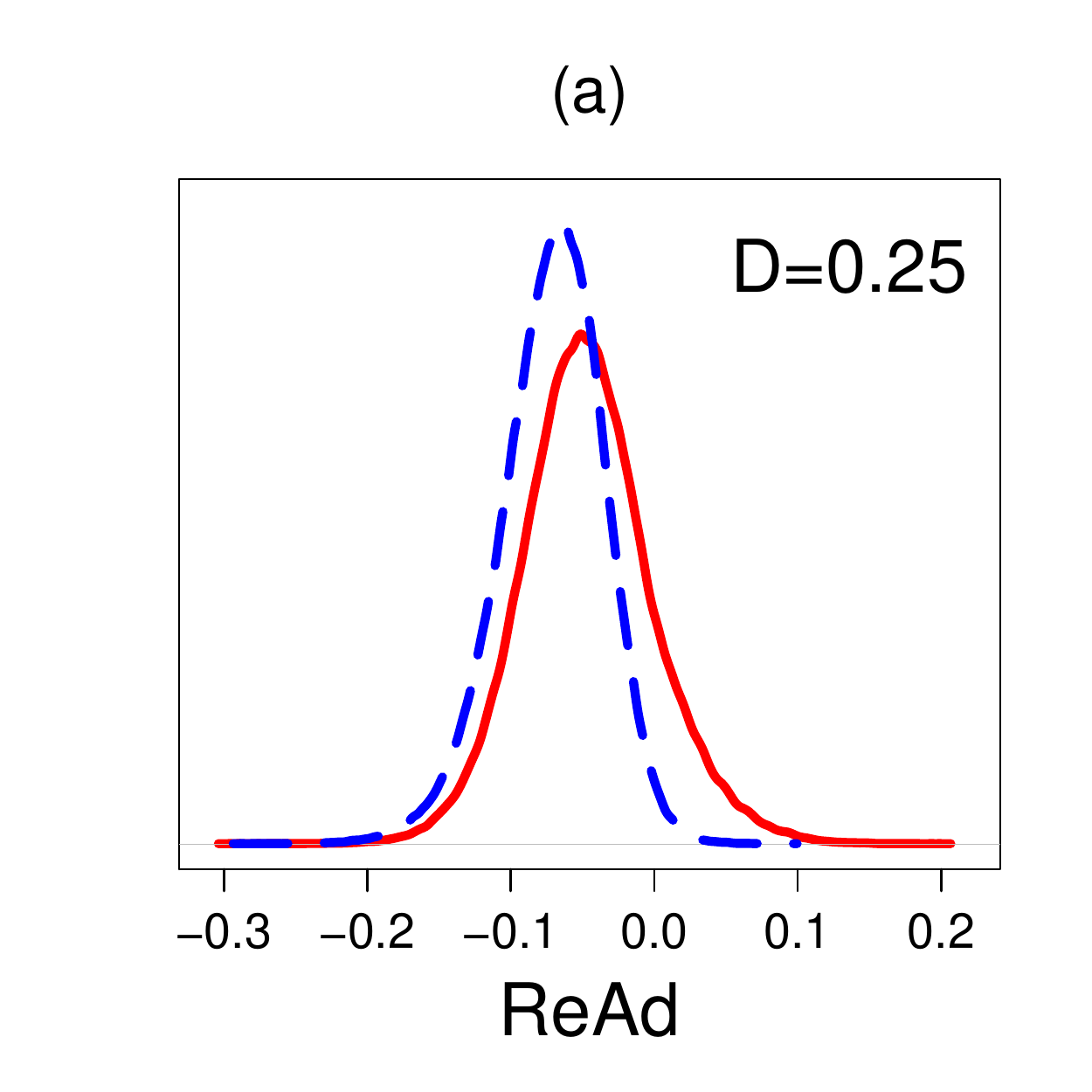}
	\includegraphics[width=.28\textwidth]{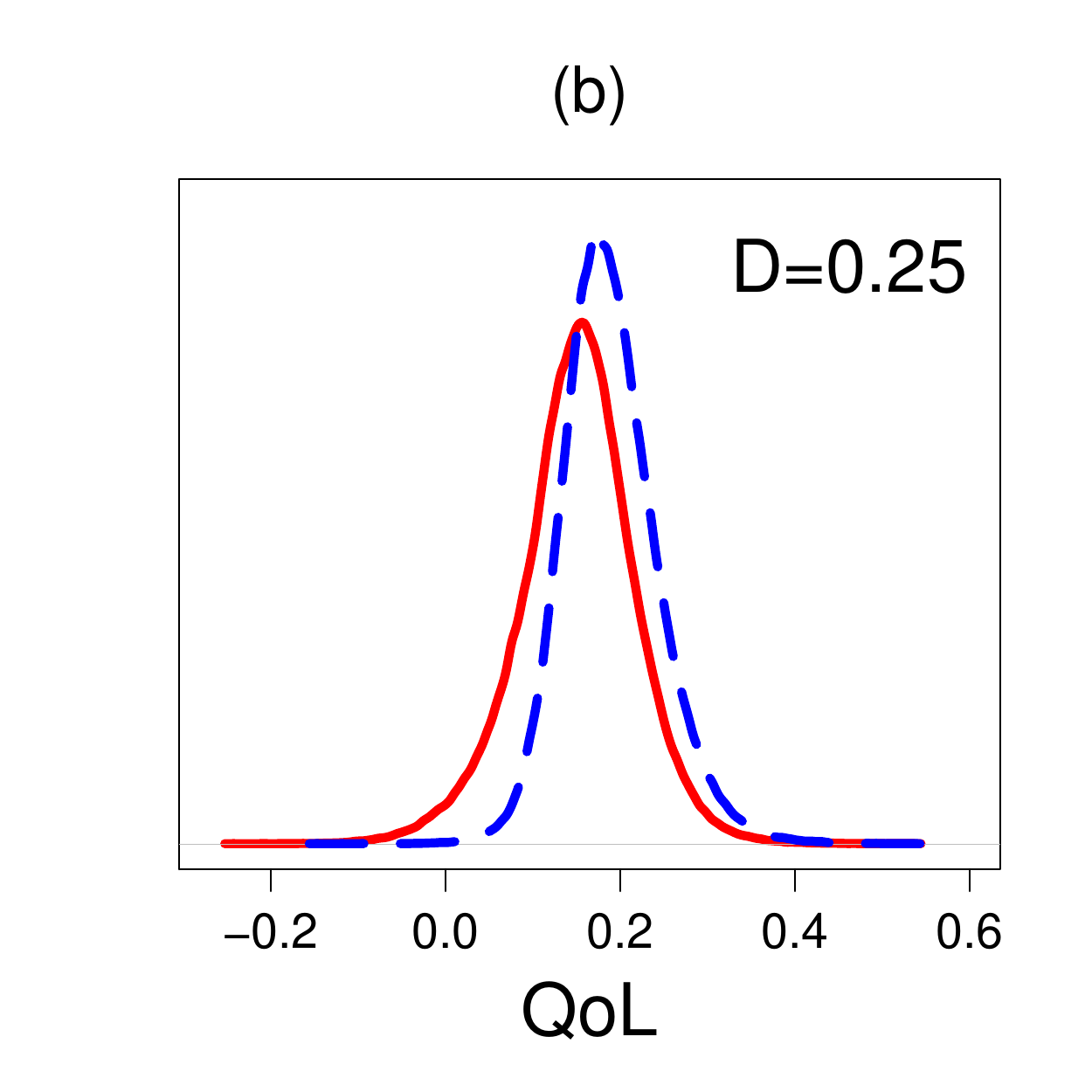}
	\includegraphics[width=.28\textwidth]{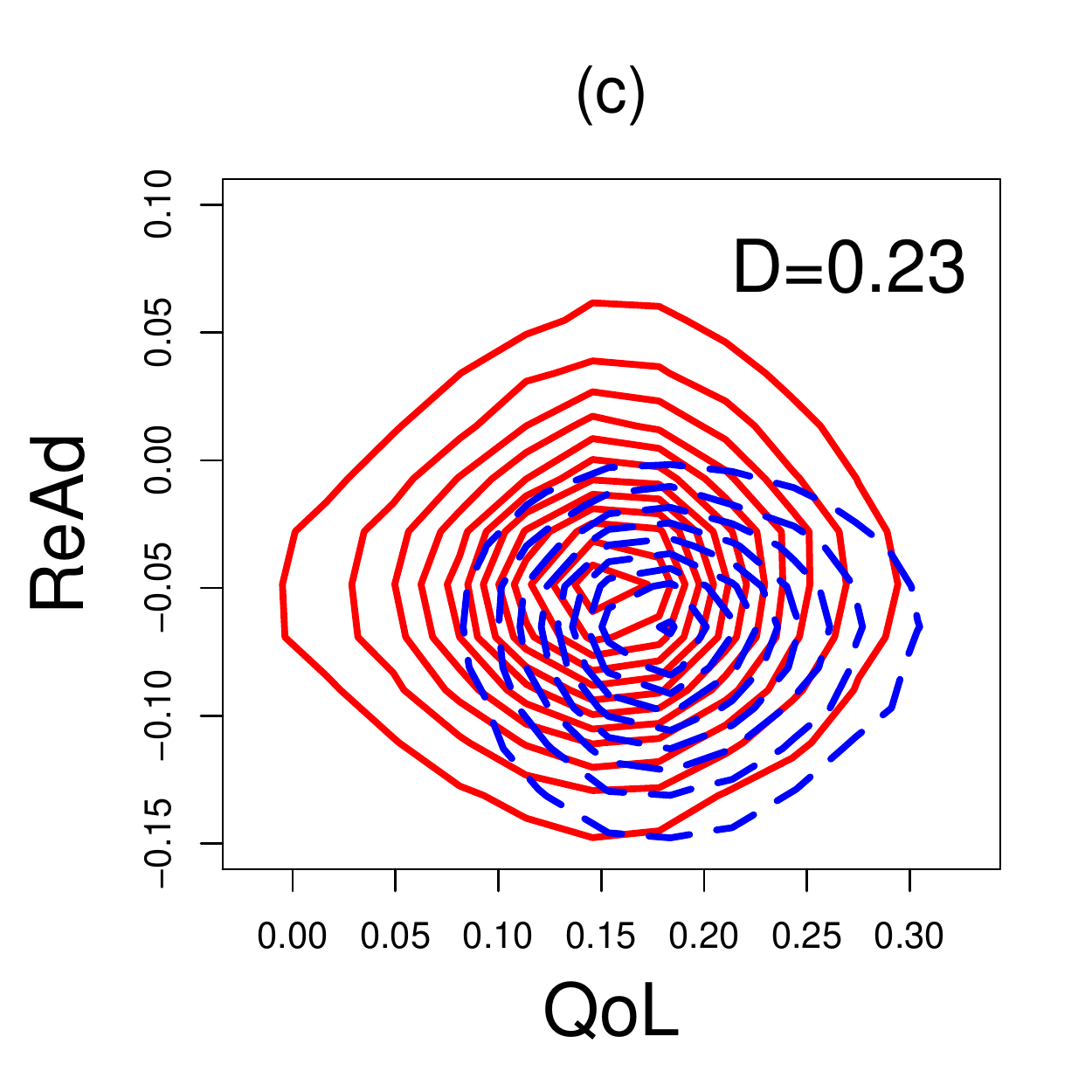} \\
	\includegraphics[width=.28\textwidth]{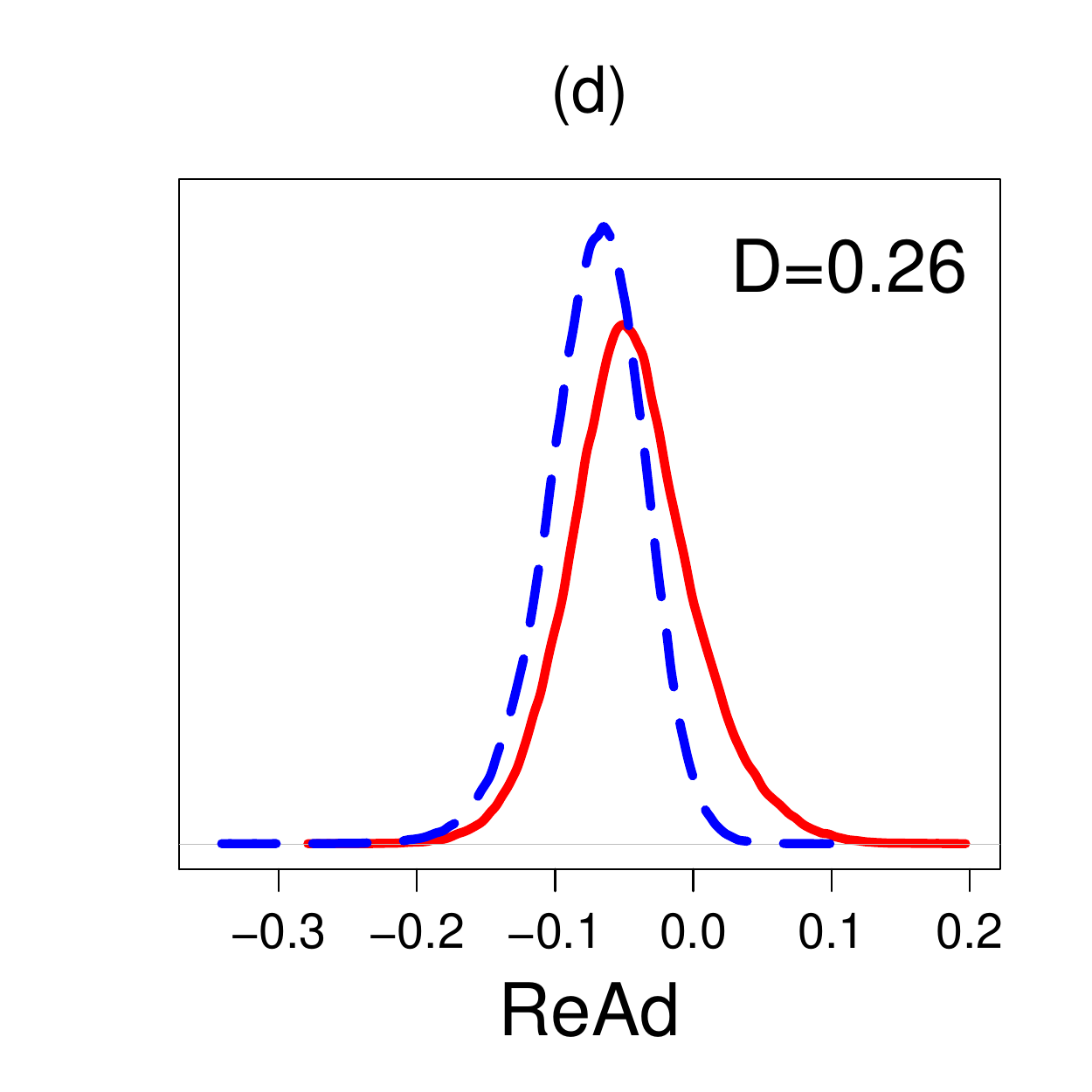} 
	\includegraphics[width=.28\textwidth]{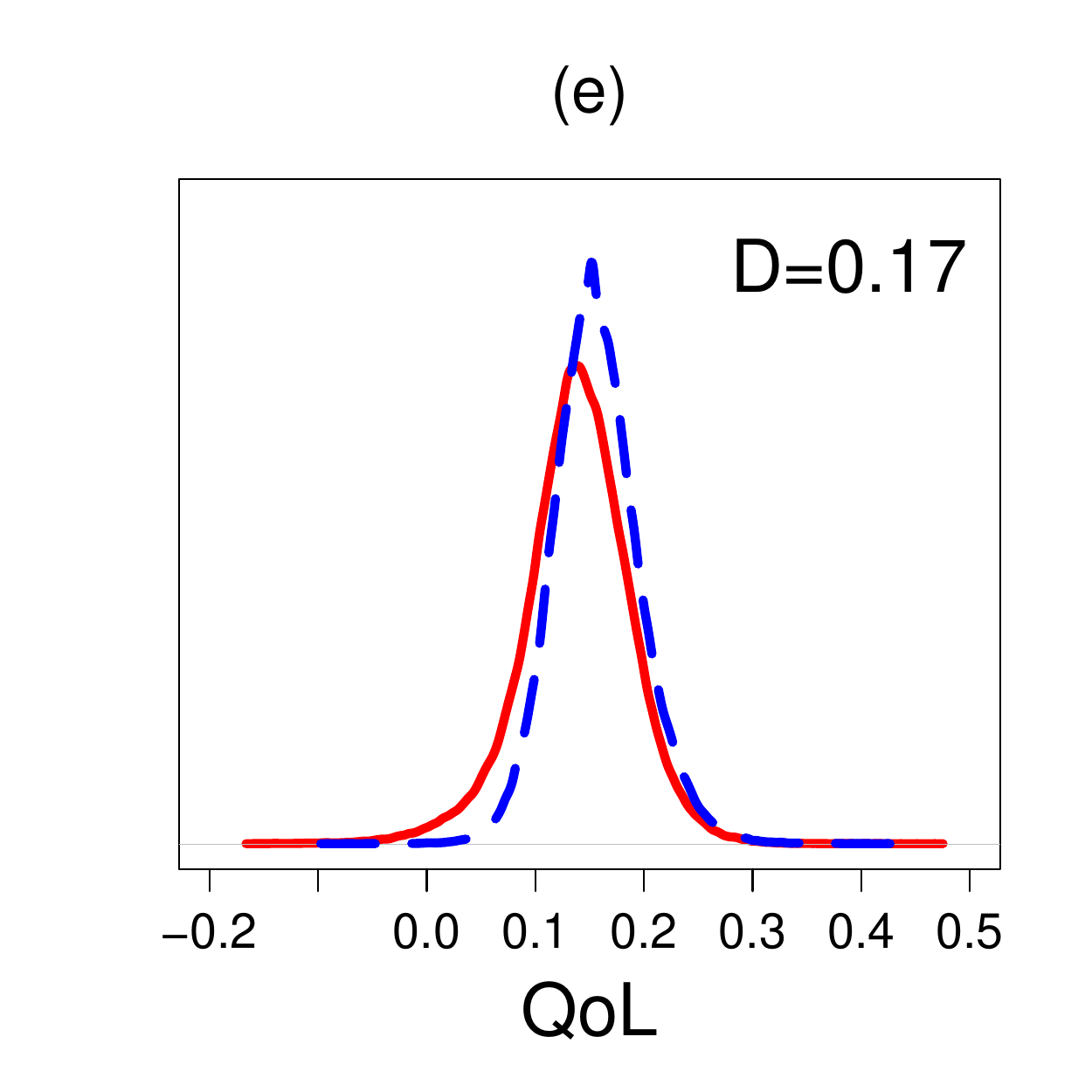} 
	\includegraphics[width=.28\textwidth]{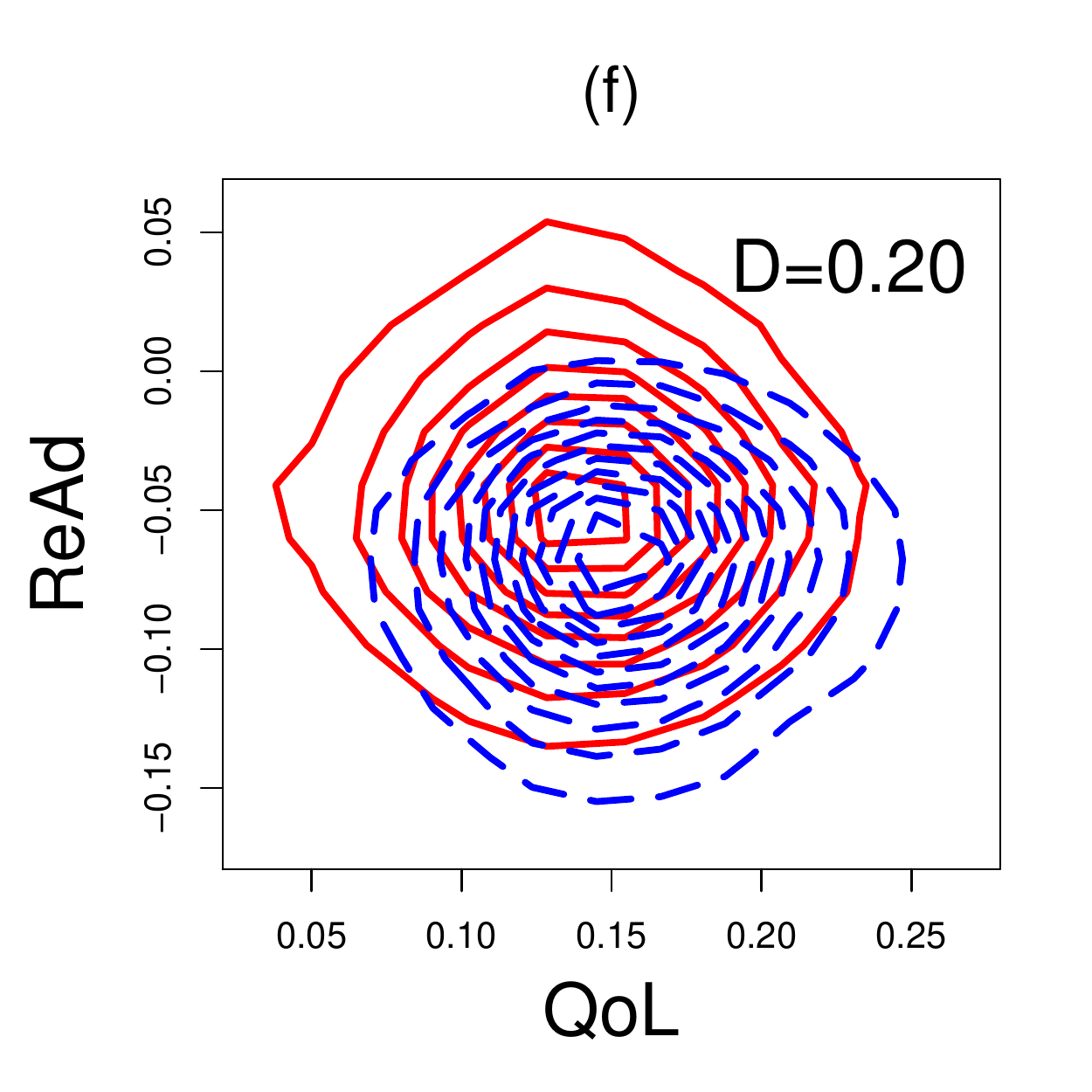}
	\caption{Panels~(a)--(c) show the results for the meta-analysis of interventions on HF patients using the \textit{published} data. Panels~(d)--(f) show the results using the \textit{updated} data.  ReAd is plotted on the log-RR scale in panels (a), (c), (d), and (f). }\label{HeartFailurePlots}
\end{figure}

In Figure~\ref{HeartFailurePlots}, we plot the posterior distributions of ReAd and QoL for the ABSORB model (solid line) and the non-bias corrected model (dashed line). In panels (a)-(c), we plot the results based on the published data, and in panels (d)-(f), we plot the results based on the updated data. For ReAd, we obtained $D = 0.25$ on the published data and $D=0.26$ on the updated data. These $D$ measures reflect the non-negligible impact from outcome reporting bias. In this case, the shift in the ReAd posterior towards the null side was enough to qualitatively change the conclusions from our meta-analysis.

For QoL, we obtained $D=0.25$ on the published data and $D=0.17$ on the updated data. By procuring more QoL outcomes from some missing studies, the updated data was less subject to ORB. This was consistent with the lower $D$ measure for QoL in our second stage analysis. Panel (b) of Figure~\ref{HeartFailureResults} and the middle two panels of Figure~\ref{HeartFailurePlots}  illustrate that the unadjusted and adjusted results for QoL were more similar to each other in the updated data than in the published data. In particular, the Jaccard index for QoL was 0.62 in the published data and 0.72 in the updated data. The Jaccard index (length of the intersection of two intervals divided by the length their union) gives a measure of consistency between the non-bias corrected and bias-corrected 95\% CIs, with a larger value indicating greater similarity. In many practical situations, it may not be possible for systematic reviewers to obtain an updated dataset. However, this case study validates that our method produces bias-corrected results that are more consistent with the unadjusted analyses when researchers \textit{are} able to mitigate some of the ORB.

Our findings have important implications for clinicians, policymakers, and HF patients. Reducing hospital readmission for HF patients has been the primary objective of these stakeholders \citep{ZiaeianFonarow2016}, and this has been the rationale for employing interventions like TM and STS. However, our results suggest that these interventions may not significantly reduce the risk of readmission. On the other hand, there seems to be a significant improvement in quality of life for HF patients who receive these interventions, compared to the patients who receive usual care. Therefore, we may conclude that TM and STS are still beneficial for the quality of life of patients, but that other approaches may be needed to significantly reduce the risk of hospital readmission.

\section{Discussion} \label{Discussion}

In this article, we have introduced a Bayesian selection model for correcting and quantifying the impact of outcome reporting bias (ABSORB) in multivariate meta-analysis. Our model enables us to not only correct the estimates of treatment effects, but also quantifies their uncertainty due to the presence of ORB. We employed the $D$ measure \eqref{Dmeasure} to quantify the \textit{impact} of ORB on the results of MMA by measuring the dissimilarity between the bias-corrected and non-bias corrected posterior densities. Our approaches were empirically evaluated through a meta-evaluation of 748 real systematic reviews from the Cochrane Database. In addition, we applied the ABSORB model to a meta-analysis on the effects of interventions on quality of life and hospital readmission for heart failure patients. Our results show that the presence of ORB can lead to qualitative differences in the conclusions from MMA. In particular, the relative risk of hospital readmission for HF patients in the intervention group shifted from a significant decrease (RR: 0.931, 95\% CI 0.862--0.993) to a statistically \textit{nonsignificant} effect (RR: 0.955, 95\% CI 0.876--1.051) once we adjusted for ORB. Furthermore, we found in our meta-evaluation that after correcting for ORB, 157 out of 748 bivariate meta-analyses from the Cochrane Database \textit{also} had a change in statistical significance for at least one outcome. Our study demonstrates the importance of accounting for ORB when conducting MMA.

In this paper, we focused on bivariate meta-analysis. However, the ABSORB model can also be extended to models with more than two outcomes. Suppose that we have $p$ outcomes of interest. When $p>2$, we can model each of the outcomes $y_{ij}$, $j = 1, \ldots, p$, exactly as we did in the bivariate case through \eqref{YgivenZ}--\eqref{latentZ}. We also model each of the correlation parameters that controls the likelihood of reporting, $\rho_j := \textrm{corr}(\epsilon_{ij}, \delta_{ij})$, and the correlations between $\epsilon_{ij}$, $\epsilon_{ij'}$, $u_{ij}$ and $u_{ij'}$ for $j \neq j'$ similarly as in \eqref{EpsilonDeltaCorrelation}--\eqref{BetweenStudyCorrelation}. While this extension of ABSORB to $p>2$ endpoints is straightforward, the potential downside is that the number of correlation parameters to estimate can be very large if $p$ is even moderately large. Thus, it may be desirable to simplify the correlation structure when $p$ is large so that the model remains parsimonious.

Another limitation when $p > 2$ is that the ABSORB model requires consideration of $2^p-1$ scenarios to completely specify its likelihood (e.g., studies with no missing endpoints, studies with only the first endpoint missing, studies with only the first two endpoints missing, etc.). While this is feasible for small $p$, it may become cumbersome if $p$ is moderately large. In the future, we plan to explore computationally efficient ways to extend the ABSORB model to handle a larger number of endpoints. This will make our model more appealing not just for MMA, but also for network meta-analysis (NMA). NMA expands the scope of a pairwise meta-analysis by simultaneously making comparisons across trials based on a common comparator (e.g., a standard treatment) \citep{Lumley2002}. NMA combines direct evidence and indirect evidence under the assumption of evidence consistency. Ignoring the impact of ORB in NMA can lead to bias in both direct evidence and indirect evidence. Thus, it is critical to develop new approaches to account for ORB in the NMA framework.

While the $D$ measure \eqref{Dmeasure} that we introduced in Section~\ref{QuantifyingORB} is a useful statistic for summarizing the \textit{sensitivity} of the results from MMA to ORB, there are several limitations to it. First, the $D$ measure does not take into account the \textit{direction} of the bias. Second, the $D$ measure does not have a variance estimate associated with it. Thus, unlike the $I^2$ statistic \citep{higgins2002quantifying} or other measures for quantifying PB \citep{LinChu2018}, there is no natural way of forming $100(1-\alpha)\%, \alpha \in (0,1)$, uncertainty intervals for the $D$ measure. One possibility is to calculate the $D$ measure on many independent, slightly perturbed datasets and to use the quantiles of the subsequent empirical distribution to obtain an interval estimate for $D$. However, this approach is also limited, and it is desirable to find more straightforward ways of obtaining \textit{interval} estimates for $D$. In the future, we hope to develop measures that not only quantify the impact of ORB, but that can also take into account both the direction of the bias and the inherent uncertainty of the measure itself.

\section*{Code}

An \textsf{R} package for implementing the model in this paper is available at \url{https://github.com/raybai07/ABSORB}.

\section*{Acknowledgments}
We acknowledge Dr. Brian Finkelman for his help in collecting the intervention studies for the case study in this paper. This work was initiated when the first listed author was a postdoctoral researcher at the University of Pennsylvania under the supervision of the last listed author.

\section*{Funding}
This research was supported in part by generous funding from the College of Arts and Sciences at the University of South Carolina (RB), National Science Foundation grant OIA-1655740 (RB), National Institutes of Health (NIH) grants 1R01LM012607, 1R01AI130460, 1R01AG073435, 1R56AG074604, 1R01LM013519, 1R56AG069880 (XL and YC), and NIH grants R01LM012982 and UL1TR002494 (HC). This work was supported partially through Patient-Centered Outcomes Research Institute (PCORI) Project Program Awards (ME-2019C3-18315 and ME-2018C3-14899). All statements in this report, including its findings and conclusions, are solely those of the authors and do not necessarily represent the views of the Patient-Centered Outcomes Research Institute (PCORI), its Board of Governors or Methodology Committee.

\bibliography{ABSORBreferences}

\begin{thebibliography}{46}
\providecommand{\natexlab}[1]{#1}
\providecommand{\url}[1]{\texttt{#1}}
\expandafter\ifx\csname urlstyle\endcsname\relax
  \providecommand{\doi}[1]{doi: #1}\else
  \providecommand{\doi}{doi: \begingroup \urlstyle{rm}\Url}\fi

\bibitem[Al-Marzouki et~al.(2008)Al-Marzouki, Roberts, Evans, and
  Marshall]{al2008selective}
S.~Al-Marzouki, I.~Roberts, S.~Evans, and T.~Marshall.
\newblock Selective reporting in clinical trials: analysis of trial protocols
  accepted by {T}he {L}ancet.
\newblock \emph{The Lancet}, 372\penalty0 (9634):\penalty0 201, 2008.

\bibitem[Bai et~al.(2020)Bai, Lin, Boland, and Chen]{BaiLinBolandChen2020}
R.~Bai, L.~Lin, M.~R. Boland, and Y.~Chen.
\newblock A robust {B}ayesian {C}opas selection model for quantifying and
  correcting publication bias.
\newblock \emph{arXiv pre-print arXiv:2005.02930}, 2020.

\bibitem[Begg and Mazumdar(1994)]{Begg1994}
C.~B. Begg and M.~Mazumdar.
\newblock Operating characteristics of a rank correlation test for publication
  bias.
\newblock \emph{Biometrics}, 50\penalty0 (4):\penalty0 1088--1101, 1994.

\bibitem[Bossuyt et~al.(2003)Bossuyt, Reitsma, E~Bruns, Gatsonis, Glasziou,
  Irwig, Lijmer, Moher, Rennie, and De~Vet]{bossuyt2003towards}
P.~M. Bossuyt, J.~B. Reitsma, D.~E~Bruns, C.~A. Gatsonis, P.~P. Glasziou, L.~M.
  Irwig, J.~G. Lijmer, D.~Moher, D.~Rennie, and H.~C.~W. De~Vet.
\newblock Towards complete and accurate reporting of studies of diagnostic
  accuracy: the {STARD} initiative.
\newblock \emph{Clinical Chemistry and Laboratory Medicine}, 41\penalty0
  (1):\penalty0 68--73, 2003.

\bibitem[Celano et~al.(2018)Celano, Villegas, Albanese, Gaggin, and
  Huffman]{Celano2018}
C.~M. Celano, A.~C. Villegas, A.~M. Albanese, H.~K. Gaggin, and J.~C. Huffman.
\newblock Depression and anxiety in heart failure: a review.
\newblock \emph{Harvard Review of Psychiatry}, 26\penalty0 (4):\penalty0
  175--184, 2018.

\bibitem[Chan et~al.(2004{\natexlab{a}})Chan, Hr{\'o}bjartsson, Haahr,
  G{\o}tzsche, and Altman]{chan2004empirical}
A.-W. Chan, A.~Hr{\'o}bjartsson, M.~T. Haahr, P.~C. G{\o}tzsche, and D.~G.
  Altman.
\newblock Empirical evidence for selective reporting of outcomes in randomized
  trials: comparison of protocols to published articles.
\newblock \emph{{JAMA}}, 291\penalty0 (20):\penalty0 2457--2465,
  2004{\natexlab{a}}.

\bibitem[Chan et~al.(2004{\natexlab{b}})Chan, Krle{\v{z}}a-Jeri{\'c}, Schmid,
  and Altman]{chan2004outcome}
A.-W. Chan, K.~Krle{\v{z}}a-Jeri{\'c}, I.~Schmid, and D.~G. Altman.
\newblock Outcome reporting bias in randomized trials funded by the {C}anadian
  {I}nstitutes of {H}ealth {R}esearch.
\newblock \emph{Canadian Medical Association Journal}, 171\penalty0
  (7):\penalty0 735--740, 2004{\natexlab{b}}.

\bibitem[Chowers et~al.(2009)Chowers, Gottesman, Leibovici, Pielmeier,
  Andreassen, and Paul]{chowers2009reporting}
M.~Y. Chowers, B.~S. Gottesman, L.~Leibovici, U.~Pielmeier, S.~Andreassen, and
  M.~Paul.
\newblock Reporting of adverse events in randomized controlled trials of highly
  active antiretroviral therapy: systematic review.
\newblock \emph{Journal of Antimicrobial Chemotherapy}, 64\penalty0
  (2):\penalty0 239--250, 2009.

\bibitem[Chu and Cole(2006)]{chu2006bivariate}
H.~Chu and S.~R. Cole.
\newblock Bivariate meta-analysis of sensitivity and specificity with sparse
  data: a generalized linear mixed model approach.
\newblock \emph{Journal of Clinical Epidemiology}, 59\penalty0 (12):\penalty0
  1331--1332, 2006.

\bibitem[Copas(1999)]{copas1999works}
J.~Copas.
\newblock What works?: selectivity models and meta-analysis.
\newblock \emph{Journal of the Royal Statistical Society: Series A (Statistics
  in Society)}, 162\penalty0 (1):\penalty0 95--109, 1999.

\bibitem[Copas and Shi(2000)]{copas2000meta}
J.~Copas and J.~Q. Shi.
\newblock Meta-analysis, funnel plots and sensitivity analysis.
\newblock \emph{Biostatistics}, 1\penalty0 (3):\penalty0 247--262, 2000.

\bibitem[Copas et~al.(2014)Copas, Dwan, Kirkham, and
  Williamson]{copas2014model}
J.~Copas, K.~Dwan, J.~Kirkham, and P.~Williamson.
\newblock A model-based correction for outcome reporting bias in meta-analysis.
\newblock \emph{Biostatistics}, 15\penalty0 (2):\penalty0 370--383, 2014.

\bibitem[Copas and Shi(2001)]{copas2001sensitivity}
J.~B. Copas and J.~Q. Shi.
\newblock A sensitivity analysis for publication bias in systematic reviews.
\newblock \emph{Statistical Methods in Medical Research}, 10\penalty0
  (4):\penalty0 251--265, 2001.

\bibitem[Egger et~al.(1997)Egger, {Davey Smith}, Schneider, and
  Minder]{Egger1997}
M.~Egger, G.~{Davey Smith}, M.~Schneider, and C.~Minder.
\newblock Bias in meta-analysis detected by a simple, graphical test.
\newblock \emph{BMJ}, 315\penalty0 (7109):\penalty0 629--634, 1997.

\bibitem[Egger et~al.(2008)Egger, Davey-Smith, and Altman]{egger2008systematic}
M.~Egger, G.~Davey-Smith, and D.~Altman.
\newblock \emph{Systematic {R}eviews in {H}ealth {C}are: {M}eta-{A}nalysis in
  {C}ontext}.
\newblock John Wiley \& Sons, 2008.

\bibitem[Frosi et~al.(2015)Frosi, Riley, Williamson, and
  Kirkham]{frosi2015multivariate}
G.~Frosi, R.~D. Riley, P.~R. Williamson, and J.~J. Kirkham.
\newblock Multivariate meta-analysis helps examine the impact of outcome
  reporting bias in {C}ochrane rheumatoid arthritis reviews.
\newblock \emph{Journal of Clinical Epidemiology}, 68\penalty0 (5):\penalty0
  542--550, 2015.

\bibitem[Guyatt et~al.(2002)Guyatt, Rennie, Meade, and Cook]{guyatt2002users}
G.~Guyatt, D.~Rennie, M.~O. Meade, and D.~J. Cook.
\newblock \emph{Users' {G}uides to the {M}edical {L}iterature: {A} {M}anual for
  {E}vidence-{B}ased {C}linical {P}ractice}, volume 706.
\newblock AMA Press Chicago, IL, 2002.

\bibitem[Hazell and Shakir(2006)]{hazell2006under}
L.~Hazell and S.~A. Shakir.
\newblock Under-reporting of adverse drug reactions.
\newblock \emph{Drug Safety}, 29\penalty0 (5):\penalty0 385--396, 2006.

\bibitem[Hemminki(1980)]{hemminki1980study}
E.~Hemminki.
\newblock Study of information submitted by drug companies to licensing
  authorities.
\newblock \emph{BMJ}, 280\penalty0 (6217):\penalty0 833--836, 1980.

\bibitem[Higgins and Thompson(2002)]{higgins2002quantifying}
J.~P.~T. Higgins and S.~G. Thompson.
\newblock Quantifying heterogeneity in a meta-analysis.
\newblock \emph{Statistics in Medicine}, 21\penalty0 (11):\penalty0 1539--1558,
  2002.

\bibitem[Higgins et~al.(2011{\natexlab{a}})Higgins, Altman, G{\o}tzsche,
  J{\"u}ni, Moher, Oxman, Savovi{\'c}, Schulz, Weeks, and
  Sterne]{higgins2011cochrane}
J.~P.~T. Higgins, D.~G. Altman, P.~C. G{\o}tzsche, P.~J{\"u}ni, D.~Moher, A.~D.
  Oxman, J.~Savovi{\'c}, K.~F. Schulz, L.~Weeks, and J.~A.~C. Sterne.
\newblock The {C}ochrane {C}ollaboration's tool for assessing risk of bias in
  randomised trials.
\newblock \emph{BMJ}, 343:\penalty0 d5928, 2011{\natexlab{a}}.

\bibitem[Higgins et~al.(2011{\natexlab{b}})Higgins, Altman, and
  Sterne]{higgins2011chapter}
J.~P.~T. Higgins, D.~G. Altman, and J.~A.~C. Sterne.
\newblock Chapter 8: assessing risk of bias in included studies.
\newblock \emph{Cochrane Handbook for Systematic Reviews of Interventions},
  2011{\natexlab{b}}.

\bibitem[Horrace(2005)]{Horrace2005}
W.~C. Horrace.
\newblock Some results on the multivariate truncated normal distribution.
\newblock \emph{Journal of Multivariate Analysis}, 94\penalty0 (1):\penalty0
  209--221, 2005.

\bibitem[Inglis et~al.(2015)Inglis, Clark, Dierckx, Prieto-Merino, and
  Cleland]{inglis2015structured}
S.~C. Inglis, R.~A. Clark, R.~Dierckx, D.~Prieto-Merino, and J.~G. Cleland.
\newblock Structured telephone support or non-invasive telemonitoring for
  patients with heart failure.
\newblock \emph{Cochrane Database of Systematic Reviews}, \penalty0 (10), 2015.

\bibitem[Jackson(2007)]{jackson2007assessing}
D.~Jackson.
\newblock Assessing the implications of publication bias for two popular
  estimates of between-study variance in meta-analysis.
\newblock \emph{Biometrics}, 63\penalty0 (1):\penalty0 187--193, 2007.

\bibitem[Jackson et~al.(2011)Jackson, Riley, and
  White]{jackson2011multivariate}
D.~Jackson, R.~Riley, and I.~R. White.
\newblock Multivariate meta-analysis: Potential and promise.
\newblock \emph{Statistics in Medicine}, 30\penalty0 (20):\penalty0 2481--2498,
  2011.

\bibitem[Kirkham et~al.(2012)Kirkham, Riley, and
  Williamson]{kirkham2012multivariate}
J.~J. Kirkham, R.~D. Riley, and P.~R. Williamson.
\newblock A multivariate meta-analysis approach for reducing the impact of
  outcome reporting bias in systematic reviews.
\newblock \emph{Statistics in Medicine}, 31\penalty0 (20):\penalty0 2179--2195,
  2012.

\bibitem[Lin and Chu(2018{\natexlab{a}})]{LinChu2018}
L.~Lin and H.~Chu.
\newblock Bayesian multivariate meta-analysis of multiple factors.
\newblock \emph{Research Synthesis Methods}, 9\penalty0 (2):\penalty0 261--272,
  2018{\natexlab{a}}.

\bibitem[Lin and Chu(2018{\natexlab{b}})]{lin2018quantifying}
L.~Lin and H.~Chu.
\newblock Quantifying publication bias in meta-analysis.
\newblock \emph{Biometrics}, 74\penalty0 (3):\penalty0 785--794,
  2018{\natexlab{b}}.

\bibitem[Lumley(2002)]{Lumley2002}
T.~Lumley.
\newblock Network meta-analysis for indirect treatment comparisons.
\newblock \emph{Statistics in Medicine}, 21\penalty0 (16):\penalty0 2313--2324,
  2002.

\bibitem[Mathieu et~al.(2009)Mathieu, Boutron, Moher, Altman, and
  Ravaud]{mathieu2009comparison}
S.~Mathieu, I.~Boutron, D.~Moher, D.~G. Altman, and P.~Ravaud.
\newblock Comparison of registered and published primary outcomes in randomized
  controlled trials.
\newblock \emph{JAMA}, 302\penalty0 (9):\penalty0 977--984, 2009.

\bibitem[Mavridis et~al.(2013)Mavridis, Sutton, Cipriani, and
  Salanti]{MavridisSuttonCiprianiSalanti2013}
D.~Mavridis, A.~Sutton, A.~Cipriani, and G.~Salanti.
\newblock A fully {B}ayesian application of the {C}opas selection model for
  publication bias extended to network meta-analysis.
\newblock \emph{Statistics in Medicine}, 32\penalty0 (1):\penalty0 51--66,
  2013.

\bibitem[Mayo-Wilson et~al.(2017)Mayo-Wilson, Fusco, Li, Hong, Canner,
  Dickersin, et~al.]{mayo2017multiple}
E.~Mayo-Wilson, N.~Fusco, T.~Li, H.~Hong, J.~Canner, K.~Dickersin, et~al.
\newblock Multiple outcomes and analyses in clinical trials create challenges
  for interpretation and research synthesis.
\newblock \emph{Journal of Clinical Epidemiology}, 2017.

\bibitem[Moher and Tsertsvadze(2006)]{moher2006systematic}
D.~Moher and A.~Tsertsvadze.
\newblock Systematic reviews: when is an update an update?
\newblock \emph{The Lancet}, 367\penalty0 (9514):\penalty0 881--883, 2006.

\bibitem[Moher et~al.(1999)Moher, Cook, Eastwood, Olkin, Rennie, Stroup, Group,
  et~al.]{moher1999improving}
D.~Moher, D.~J. Cook, S.~Eastwood, I.~Olkin, D.~Rennie, D.~F. Stroup, Q.~Group,
  et~al.
\newblock Improving the quality of reports of meta-analyses of randomised
  controlled trials: the {QUOROM} statement.
\newblock \emph{The Lancet}, 354\penalty0 (9193):\penalty0 1896--1900, 1999.

\bibitem[Moher et~al.(2009)Moher, Liberati, Tetzlaff, and
  Altman]{moher2009preferred}
D.~Moher, A.~Liberati, J.~Tetzlaff, and D.~G. Altman.
\newblock Preferred reporting items for systematic reviews and meta-analyses:
  the {PRISMA} statement.
\newblock \emph{Annals of Internal Medicine}, 151\penalty0 (4):\penalty0
  264--269, 2009.

\bibitem[Ning et~al.(2017)Ning, Chen, and Piao]{ning2017maximum}
J.~Ning, Y.~Chen, and J.~Piao.
\newblock Maximum likelihood estimation and {EM} algorithm of {C}opas-like
  selection model for publication bias correction.
\newblock \emph{Biostatistics}, 18\penalty0 (3):\penalty0 495--504, 2017.

\bibitem[Reitsma et~al.(2005)Reitsma, Glas, Rutjes, Scholten, Bossuyt, and
  Zwinderman]{reitsma2005bivariate}
J.~B. Reitsma, A.~S. Glas, A.~W.~S. Rutjes, R.~J. P.~M. Scholten, P.~M.
  Bossuyt, and A.~H. Zwinderman.
\newblock Bivariate analysis of sensitivity and specificity produces
  informative summary measures in diagnostic reviews.
\newblock \emph{Journal of Clinical Epidemiology}, 58\penalty0 (10):\penalty0
  982--990, 2005.

\bibitem[Riley(2009)]{riley2009multivariate}
R.~D. Riley.
\newblock Multivariate meta-analysis: the effect of ignoring within-study
  correlation.
\newblock \emph{Journal of the Royal Statistical Society: Series A (Statistics
  in Society)}, 172\penalty0 (4):\penalty0 789--811, 2009.

\bibitem[Riley et~al.(2007{\natexlab{a}})Riley, Abrams, Lambert, Sutton, and
  Thompson]{riley2007evaluation}
R.~D. Riley, K.~R. Abrams, P.~C. Lambert, A.~J. Sutton, and J.~R. Thompson.
\newblock An evaluation of bivariate random-effects meta-analysis for the joint
  synthesis of two correlated outcomes.
\newblock \emph{Statistics in Medicine}, 26\penalty0 (1):\penalty0 78--97,
  2007{\natexlab{a}}.

\bibitem[Riley et~al.(2007{\natexlab{b}})Riley, Abrams, Sutton, Lambert, and
  Thompson]{riley2007bivariate}
R.~D. Riley, K.~R. Abrams, A.~J. Sutton, P.~C. Lambert, and J.~R. Thompson.
\newblock Bivariate random-effects meta-analysis and the estimation of
  between-study correlation.
\newblock \emph{BMC Medical Research Methodology}, 7\penalty0 (1):\penalty0 3,
  2007{\natexlab{b}}.

\bibitem[Riley et~al.(2007{\natexlab{c}})Riley, Thompson, and
  Abrams]{RileyThompsonAbrams2007}
R.~D. Riley, J.~R. Thompson, and K.~R. Abrams.
\newblock {An alternative model for bivariate random-effects meta-analysis when
  the within-study correlations are unknown}.
\newblock \emph{Biostatistics}, 9\penalty0 (1):\penalty0 172--186,
  2007{\natexlab{c}}.

\bibitem[Sterne et~al.(2016)Sterne, Egger, and Moher]{sterne2016chapter}
J.~A.~C. Sterne, M.~Egger, and D.~Moher.
\newblock Chapter 10: Addressing reporting biases. {C}ochrane {H}andbook of
  {S}ystematic {R}eviews of {I}ntervention {V}ersion 5.1. 0 [updated {M}arch
  2011].
\newblock \emph{Cochrane Handbook for Systematic Reviews of Interventions},
  5\penalty0 (0), 2016.

\bibitem[Stewart et~al.(2015)Stewart, Clarke, Rovers, Riley, Simmonds, Stewart,
  and Tierney]{stewart2015preferred}
L.~A. Stewart, M.~Clarke, M.~Rovers, R.~D. Riley, M.~Simmonds, G.~Stewart, and
  J.~F. Tierney.
\newblock Preferred reporting items for a systematic review and meta-analysis
  of individual participant data: the {PRISMA-IPD} statement.
\newblock \emph{JAMA}, 313\penalty0 (16):\penalty0 1657--1665, 2015.

\bibitem[Vera-Badillo et~al.(2013)Vera-Badillo, Shapiro, Ocana, Amir, and
  Tannock]{vera2013bias}
F.~E. Vera-Badillo, R.~Shapiro, A.~Ocana, E.~Amir, and I.~F. Tannock.
\newblock Bias in reporting of end points of efficacy and toxicity in
  randomized, clinical trials for women with breast cancer.
\newblock \emph{Annals of Oncology}, 24\penalty0 (5):\penalty0 1238--1244,
  2013.

\bibitem[Ziaeian and Fonarow(2016)]{ZiaeianFonarow2016}
B.~Ziaeian and G.~C. Fonarow.
\newblock The prevention of hospital readmissions in heart failure.
\newblock \emph{Progress in Cardiovascular Diseases}, 58\penalty0 (4):\penalty0
  379--385, 2016.

\end{thebibliography}

\pagebreak

\begin{appendix}

\counterwithin{figure}{section}
\counterwithin{table}{section}

\section{Additional Details and Results for Our Meta-Evaluation of the Cochrane Database} \label{AdditionalResults}

\subsection{Determining the 748 Meta-Analyses to Include} \label{StudyEligibility}

In order to arrive at the 748 reviews from the Cochrane database for our meta-evaluation in Section~\ref{meta-meta}, we first extracted 5652 available reviews out of 13,505 reviews from the Cochrane Database before January 16, 2020. We applied the following selection rules to select reviews with multiple continuous or binary outcomes:
\begin{itemize}
\item[1.] The review must contain at least one comparison that includes at least two \textit{distinct} outcomes. For example, some reviews reported the same outcome twice from a sensitivity analysis.
\item[2.] For the meta-analysis of these two distinct outcomes, there should be at least five studies reporting \textit{both} outcomes, and they must contain at least 10 studies in total (reporting \textit{either} a single outcome or both outcomes).
\end{itemize}
These two rules were defined so that we could understand how the ABSORB model and the $D$ measure \eqref{Dmeasure} perform when there are a relatively small or medium number of studies, which is more common in practice. A review could contain several outcome pairs that satisfied both rules 1 and 2, and in that case, we selected the pair that had the largest total number of studies. If there were several such pairs, we further selected the one that had the largest number of shared studies. 

Next, the bivariate observations were formed by matching study names. We only used studies whose effect size was estimable; the meta-analyses where sub-group analysis entered as separate studies were deleted. We also excluded reviews where the non-missing values for both endpoints were identical or nearly identical. This suggested that the two endpoints were not true bivariate observations, but rather, they were a single outcome that was reported twice from a sensitivity analysis. To further exclude reviews where the two endpoints were really the same outcome (e.g. $y_1$ and $y_2$ were the same outcome with rounding differences), we also calculated the Pearson correlation coefficient for the non-missing pairs $(y_1, y_2)$ in each review and removed the reviews with a correlation greater than 0.99. Finally, several reviews where one endpoint contained all or almost all zeroes were excluded from the analysis. Once we had performed all these quality controls, we were left with a final sample of 748 meta-analyses with which to perform our meta-evaluation.

\begin{table}[ht!]
\centering
\caption{The overall quantiles for $D_1$, $D_2$, and $D_{12}$ from all 748 meta-analyses in our meta-evaluation} 
\medskip
  \resizebox{.66\textwidth}{!}{  
\begin{tabular}{||c | c c c||c | c c c||} 
 \hline
 Quantile & $D_1$ & $D_2$ & $D_{12}$ & Quantile & $D_1$ & $D_2$ & $D_{12}$ \\ [0.5ex] 
 \hline\hline
1\% & 0.02 & 0.02 & 0.02 & 51\% & 0.10 & 0.11 & 0.10 \\ 
  2\% & 0.02 & 0.02 & 0.02 & 52\% & 0.11 & 0.11 & 0.10 \\ 
  3\% & 0.02 & 0.02 & 0.02 & 53\% & 0.11 & 0.11 & 0.10 \\ 
  4\% & 0.03 & 0.03 & 0.02 & 54\% & 0.11 & 0.11 & 0.10 \\ 
  5\% & 0.03 & 0.03 & 0.03 & 55\% & 0.11 & 0.11 & 0.10 \\ 
  6\% & 0.03 & 0.03 & 0.03 & 56\% & 0.11 & 0.12 & 0.10 \\ 
  7\% & 0.03 & 0.04 & 0.03 & 57\% & 0.11 & 0.12 & 0.11 \\ 
  8\% & 0.03 & 0.04 & 0.03 & 58\% & 0.12 & 0.12 & 0.11 \\ 
  9\% & 0.04 & 0.04 & 0.03 & 59\% & 0.12 & 0.12 & 0.11 \\ 
  10\% & 0.04 & 0.04 & 0.03 & 60\% & 0.12 & 0.12 & 0.11 \\ 
  11\% & 0.04 & 0.04 & 0.04 & 61\% & 0.12 & 0.13 & 0.11 \\ 
  12\% & 0.04 & 0.04 & 0.04 & 62\% & 0.12 & 0.13 & 0.11 \\ 
  13\% & 0.04 & 0.05 & 0.04 & 63\% & 0.13 & 0.13 & 0.12 \\ 
  14\% & 0.04 & 0.05 & 0.04 & 64\% & 0.13 & 0.13 & 0.12 \\ 
  15\% & 0.05 & 0.05 & 0.04 & 65\% & 0.13 & 0.13 & 0.12 \\ 
  16\% & 0.05 & 0.05 & 0.04 & 66\% & 0.13 & 0.14 & 0.12 \\ 
  17\% & 0.05 & 0.05 & 0.05 & 67\% & 0.14 & 0.14 & 0.13 \\ 
  18\% & 0.05 & 0.05 & 0.05 & 68\% & 0.14 & 0.14 & 0.13 \\ 
  19\% & 0.05 & 0.05 & 0.05 & 69\% & 0.14 & 0.15 & 0.13 \\ 
  20\% & 0.05 & 0.05 & 0.05 & 70\% & 0.15 & 0.15 & 0.13 \\ 
  21\% & 0.06 & 0.05 & 0.05 & 71\% & 0.15 & 0.15 & 0.14 \\ 
  22\% & 0.06 & 0.05 & 0.05 & 72\% & 0.15 & 0.16 & 0.14 \\ 
  23\% & 0.06 & 0.06 & 0.05 & 73\% & 0.16 & 0.16 & 0.14 \\ 
  24\% & 0.06 & 0.06 & 0.05 & 74\% & 0.16 & 0.17 & 0.15 \\ 
  25\% & 0.06 & 0.06 & 0.05 & 75\% & 0.17 & 0.17 & 0.15 \\ 
  26\% & 0.06 & 0.06 & 0.06 & 76\% & 0.17 & 0.17 & 0.15 \\ 
  27\% & 0.06 & 0.06 & 0.06 & 77\% & 0.18 & 0.18 & 0.16 \\ 
  28\% & 0.06 & 0.06 & 0.06 & 78\% & 0.19 & 0.18 & 0.16 \\ 
  29\% & 0.07 & 0.07 & 0.06 & 79\% & 0.20 & 0.19 & 0.17 \\ 
  30\% & 0.07 & 0.07 & 0.06 & 80\% & 0.20 & 0.19 & 0.17 \\ 
  31\% & 0.07 & 0.07 & 0.06 & 81\% & 0.21 & 0.20 & 0.18 \\ 
  32\% & 0.07 & 0.07 & 0.07 & 82\% & 0.22 & 0.21 & 0.18 \\ 
  33\% & 0.07 & 0.07 & 0.07 & 83\% & 0.22 & 0.22 & 0.19 \\ 
  34\% & 0.07 & 0.07 & 0.07 & 84\% & 0.23 & 0.23 & 0.19 \\ 
  35\% & 0.07 & 0.08 & 0.07 & 85\% & 0.23 & 0.24 & 0.20 \\ 
  36\% & 0.08 & 0.08 & 0.07 & 86\% & 0.24 & 0.25 & 0.21 \\ 
  37\% & 0.08 & 0.08 & 0.07 & 87\% & 0.25 & 0.26 & 0.21 \\ 
  38\% & 0.08 & 0.08 & 0.07 & 88\% & 0.27 & 0.27 & 0.23 \\ 
  39\% & 0.08 & 0.08 & 0.08 & 89\% & 0.28 & 0.28 & 0.23 \\ 
  40\% & 0.08 & 0.09 & 0.08 & 90\% & 0.29 & 0.29 & 0.24 \\ 
  41\% & 0.08 & 0.09 & 0.08 & 91\% & 0.31 & 0.30 & 0.24 \\ 
  42\% & 0.09 & 0.09 & 0.08 & 92\% & 0.32 & 0.32 & 0.26 \\ 
  43\% & 0.09 & 0.09 & 0.08 & 93\% & 0.34 & 0.33 & 0.27 \\ 
  44\% & 0.09 & 0.09 & 0.08 & 94\% & 0.37 & 0.35 & 0.29 \\ 
  45\% & 0.09 & 0.09 & 0.09 & 95\% & 0.41 & 0.37 & 0.30 \\ 
  46\% & 0.09 & 0.09 & 0.09 & 96\% & 0.45 & 0.38 & 0.33 \\ 
  47\% & 0.10 & 0.10 & 0.09 & 97\% & 0.53 & 0.42 & 0.36 \\ 
  48\% & 0.10 & 0.10 & 0.09 & 98\% & 0.66 & 0.47 & 0.42 \\ 
  49\% & 0.10 & 0.10 & 0.09 & 99\% & 0.76 & 0.53 & 0.46 \\ 
  50\% & 0.10 & 0.10 & 0.09 & 100\% & 0.99 & 0.81 & 0.72 \\[1ex]  \hline
  \end{tabular}} \label{Dtable1}
\end{table}

\begin{table}[ht!]
\centering
\caption{The quantiles for $D_1$, $D_2$, and $D_{12}$ from only the meta-analyses that had $D$ measure greater than 0.10 in our meta-evaluation}
\medskip
\begin{tabular}{||c | c c c||} 
 \hline
 Quantile & $D_1$ & $D_2$ & $D_{12}$ \\ [0.5ex] 
 \hline\hline
10\% & 0.11 & 0.11 & 0.11 \\ 
  20\% & 0.12 & 0.12 & 0.11 \\ 
  30\% & 0.13 & 0.13 & 0.13 \\ 
  40\% & 0.15 & 0.15 & 0.14 \\ 
  50\% & 0.17 & 0.17 & 0.16 \\ 
  60\% & 0.20 & 0.19 & 0.18 \\ 
  70\% & 0.23 & 0.23 & 0.21 \\ 
  80\% & 0.29 & 0.29 & 0.24 \\ 
  90\% & 0.41 & 0.36 & 0.31 \\[1ex] 
 \hline
 \end{tabular} \label{Dtable2}
\end{table}

\subsection{Empirical Quantiles for Our Meta-Evaluation of the Cochrane Database}  \label{AdditionalMetaEvaluationResults}

	In Tables \ref{Dtable1}-\ref{Dtable2}, we provide the quantiles of the empirical distribution of the $D$ measure (3.32) from our meta-evaluation of 748 meta-analyses from the Cochrane Database of Systematic Reviews. We denote the $D$ measure for the first outcome $\mu_1$ by $D_1$, the $D$ measure for the second outcome $\mu_2$ by $D_2$, and the $D$ measure for the joint outcomes $(\mu_1, \mu_2)'$ by $D_{12}$.


Table~\ref{Dtable1} reports the overall quantiles for the $D$ measure. Based on our meta-evaluation, we determined that $D > 0.10$ might represent non-negligible impact from ORB. Table~\ref{Dtable2} reports the quantiles for the empirical distribution for \textit{only} the $D$ measures that were greater than 0.10. We used the overall quantiles (Table~\ref{Dtable1}) to determine ranges for negligible impact from ORB. For the meta-analyses with \textit{non}-negligible impact, we used the empirical quantiles of \textit{only} the $D$ measures greater than 0.10 (Table~\ref{Dtable2}) to determine ranges for ``moderate,'' ``substantial,'' and ``severe'' impact from ORB.

\section{Sampling from the ABSORB Likelihood} \label{Sampling}

The main challenge with implementing the ABSORB model is sampling from the truncated densities \eqref{JointYZbothoutcomes}, \eqref{jointYZfirstoutcomeonly}, and \eqref{jointYZsecondoutcomeonly} in the full likelihood \eqref{ABSORBLikelihood}. However, we can use the sparse structure of our model to approximately sample from these three densities.

We begin with sampling from \eqref{JointYZbothoutcomes}. Since the latent variables $(z_{i1}, z_{i2})$ are uncorrelated, one can use the arguments in \cite{MavridisSuttonCiprianiSalanti2013} to show that $z_{i1}$ is marginally distributed as $\mathcal{N}(\gamma_{01} + \gamma_{11}/s_{i1}, 1) \mathbb{I}_{[z_{i1} > 0]}$ and $z_{i2}$ is marginally distributed as $\mathcal{N}(\gamma_{02} + \gamma_{12} / s_{i2}, 1) \mathbb{I}_{[z_{i2} > 0]}$. By Theorem 6 of \cite{Horrace2005}, the joint distribution of $(z_{i1}, z_{i2})$ in \eqref{JointYZbothoutcomes} is also a bivariate truncated normal distribution,
\begin{align} \label{jointZs}
    \begin{pmatrix} z_{i1} \\ z_{i2} \end{pmatrix} \sim \mathcal{N} \left( \begin{pmatrix} \gamma_{01} + \gamma_{11} / s_{i1} \\ \gamma_{02} + \gamma_{12} / s_{i12} \end{pmatrix}, \begin{pmatrix} 1 & 0 \\ 0 & 1 \end{pmatrix}  \right) \mathbb{I}_{[z_{i1} > 0 \cap z_{i2} > 0]}.
\end{align}
Therefore, we can approximately sample $(y_{i1}, y_{i2}, z_{i1}, z_{i2})$ from \eqref{JointYZbothoutcomes} by first sampling $z_{i1} \sim \mathcal{N}(\gamma_{01} + \gamma_{11}/s_{i1}, 1) \mathbb{I}_{[z_{i1} > 0]}$ and $z_{i2} \sim \mathcal{N}(\gamma_{02} + \gamma_{12} / s_{i2}, 1) \mathbb{I}_{[z_{i2} > 0]}$. Then using basic properties about conditional distributions for the multivariate normal distribution and \eqref{jointZs}, we sample $(y_{i1}, y_{i2})$ conditionally on $(z_{i1}, z_{i2})$ as
\begin{align} \label{jointYsconditionalonZs}
    \begin{pmatrix} y_{i1} \\ y_{i2} \end{pmatrix} \sim \mathcal{N} \left( \begin{pmatrix} \theta_{i1} + \rho_1 s_{i1} ( z_{i1} - \gamma_{01} - \gamma_{11} / s_{i1}) \\ \theta_{i2} + \rho_2 s_{i2} (z_{i2} - \gamma_{02} - \gamma_{12} / s_{i2})  \end{pmatrix}, \begin{pmatrix} s_{i1}^2 (1-\rho_1^2) & \rho_\text{W} s_{i1} s_{i2} \\ \rho_\text{W} s_{i1} s_{i2} & s_{i2}^2 (1-\rho_2^2) \end{pmatrix} \right).
\end{align}
Similarly, to approximately sample $(y_{i1}, z_{i1}, z_{i2})$ from \eqref{jointYZfirstoutcomeonly}, we first sample $z_{i1} \sim \mathcal{N}(\gamma_{01} + \gamma_{11}/s_{i1}, 1) \mathbb{I}_{[z_{i1}>0]}$ and $z_{i2} \sim \mathcal{N}(\gamma_{02}+\gamma_{12}/\widehat{s}_{i2}, 1) \mathbb{I}_{[z_{i2} < 0]}$ (where recall that we have replaced the missing $s_{i2}$'s with their estimates $\widehat{s}_{i2}$). Then conditionally on $(z_{i1}, z_{i2})$, we sample $y_{i1}$ as $y_{i1} \mid (z_{i1}, z_{i2} ) \sim \mathcal{N} ( \widetilde{\theta}_{i1} + \rho_1 s_{i1} (z_{i1} - \gamma_{01} - \gamma_{11} / s_{i1}), s_{i1}^2(1-\rho_1^2) )$. Finally, to approximately sample $(y_{i2}, z_{i1}, z_{i2})$ from \eqref{jointYZsecondoutcomeonly}, we first sample $z_{i1} \sim \mathcal{N}(\gamma_{01} + \gamma_{11}/\widehat{s}_{i1}, 1) \mathbb{I}_{[z_{i1}<0]}$ (where we have replaced the missing $s_{i1}$'s with their estimates $\widehat{s}_{i1}$) and $z_{i2} \sim \mathcal{N}(\gamma_{02}+\gamma_{12}/ s_{i2}, 1) \mathbb{I}_{[z_{i2} > 0]}$. Conditionally on $(z_{i1}, z_{i2})$, we sample $y_{i2}$ as $y_{i2} \mid (z_{i1}, z_{i2} ) \sim \mathcal{N} ( \check{\theta}_{i2} + \rho_2 s_{i2} (z_{i2} - \gamma_{02} - \gamma_{12} / s_{i2}), s_{i2}^2(1-\rho_2^2) )$. 

Using the approximate sampling schemes above for the individual components of the ABSORB likelihood \eqref{ABSORBLikelihood}, we are able to specify the full likelihood and the prior distributions on the parameters in our model. We may then use any MCMC software to estimate the marginal posterior distributions for the model parameters.

\section{Simulation Studies} \label{Simulations}

In this section, we evaluate the ABSORB model for MMA in a variety of settings when ORB is present. As competitors to ABSORB, we considered other meta-analyses performed under missing completely at random (MCAR), missing at random (MAR), and selection bias assumptions. Under the MCAR assumption, we simply removed any studies with missing endpoints. Meanwhile, under the MAR assumption, we imputed the missing endpoints and standard errors using multiple imputation. We also compared the MMA approaches to \textit{univariate} meta-analyses (UMA) for each endpoint $\mu_1$ and $\mu_2$ separately. One of these UMA approaches was the Copas selection model \citep{copas1999works, copas2000meta, copas2001sensitivity}, which adjusts for publication bias in UMA under the assumption of selection bias.  

Specifically, in addition to ABSORB, we also considered: 1) MMA where studies with missing endpoints were removed (MMA-rem), 2) MMA where missing endpoints and standard errors were imputed (MMA-imp), 3) UMA where studies with missing endpoints were removed (UMA-rem), 4) UMA where missing endpoints and standard errors were imputed (UMA-imp), and 5) the Copas selection model applied to the non-missing values for each endpoint (Copas). For MMA-imp and UMA-imp, we used the \textsf{R} package \texttt{mice} to impute missing values with multiple imputation. Once we had only complete data, we fit either MMA with the \texttt{riley} function or UMA with the \texttt{uvmeta} function in \textsf{R} package \texttt{metamisc}. For the Copas selection model, we used the \texttt{copas} function in the \textsf{R} package \texttt{metasens}. 

For each of these six models, we generated the data according to the selection mechanism in \eqref{YgivenZ}-\eqref{ZeroCorrelations}. We fixed $(\mu_1, \mu_2, \gamma_{01}, \gamma_{11}, \gamma_{02}, \gamma_{02}, \rho_\text{W}, \rho_\text{B}) = (0.3, -0.3, -1, 0.6, -1, 0.6, 0.5, 0.5)$ in our simulation studies. However, we varied the parameters $(\tau_1, \tau_2, \rho_1, \rho_2)$ and the percentage of missing endpoints to determine how the models would perform under varying degrees of between-study heterogeneity and missingness. The within-study standard errors ($s_{i1}, s_{i2}), i = 1, \ldots, n$, were generated from a $\mathcal{U}(0.2, 0.8)$ distributions, and the study sizes were generated uniformly from $\{ 20, 21, \ldots, 100 \}$. We considered four simulation settings, with $n= 25$ and $n=50$:
\begin{itemize}
    \item Experiment 1: Moderate heterogeneity ($\tau_1 = 0.5, \tau_2 = 0.5, \rho_1 = 0.4, \rho_2 = 0.4$) with 20\% missing endpoints for $y_1$ and 20\% missing endpoints for $y_2$;
    \item Experiment 2: Strong heterogeneity ($\tau_1=1, \tau_2=1, \rho_1=0.4, \rho_2=0.4$) with 20\% missing endpoints for $y_1$ and 20\% missing endpoints for $y_2$;
    \item Experiment 3: Moderate heterogeneity ($\tau_1=0.5, \tau_2=0.5, \rho_1=0.5, \rho_2=0.7$) with 20\% missing endpoints for $y_1$ and 40\% missing endpoints for $y_2$; 
    \item Experiment 4: Mixed heterogeneity ($\tau_1 = 0.8, \tau_2 = 0.4, \rho_1 = 0.7, \rho_2 = 0.3$) with 32\% missing endpoints for $y_1$ and 12\% missing endpoints for $y_2$.
\end{itemize}
We repeated each of these experiments 1000 times. We computed the following statistics for each individual endpoint: the bias ($\widehat{\mu}_j - \mu_j, j = 1, 2$), the standard error (SE), and the coverage probability (CP), i.e., the proportion of experiments where the 95\% posterior credible interval (for ABSORB) or the 95\% confidence interval (for MMA-rem, MMA-imp, UMA-imp, UMA-rem, Copas) contained the true $\mu_j, j = 1, 2$. The ABSORB model was fit by running three MCMC chains of 50,000 iterations each, discarding the first 10,000 samples as burn-in, and combining the remaining samples. For ABSORB, we used the posterior mean as the point estimate, and for all the other approaches, we used the MLEs as point estimates. For MMA-rem, MMA-imp, UMA-rem, and UMA-rem, we used the \textsf{R} package \texttt{metasens} to obtain asymptotic SE and asymptotic 95\% CIs. For Copas, we used the parametric bootstrap with 100 bootstrap samples to compute the SE and construct bootstrap confidence intervals.

\begin{table}\centering 
\caption{Bias, standard error (SE), and coverage probability (CP) results for our simulation studies, averaged across 1000 replications. The approaches with the lowest average bias, the lowest average SE, and the highest CP for each endpoint $\mu_1$ and $\mu_2$ are in bold.}
\medskip

  \resizebox{.6\textwidth}{!}{  
\begin{tabular}{llccccccc}
\multicolumn{9}{l}{Experiment 1 (moderate heterogeneity, 20\% missing $y_1$, 20\% missing $y_2$)}  \\
\midrule
\phantom{abc} & \phantom{abc} & \multicolumn{3}{c}{$n = 25$} & \phantom{abc}& \multicolumn{3}{c}{$n=50$}  \\
\cmidrule{3-5} \cmidrule{7-9}  
& & Bias & SE & CP && Bias & SE & CP  \\ \midrule
 ABSORB & $\mu_1$ & \textbf{0.032} & 0.208 & \textbf{0.97} && \textbf{0.005} & 0.208 & \textbf{0.98} \\ 
 & $\mu_2$ & \textbf{0.006} & 0.208 & \textbf{0.98} && \textbf{0.002} & 0.165 & \textbf{0.97}  \\ 
 MMA-rem & $\mu_1$ & 0.167 & 0.165 & 0.79 && 0.131 & 0.114 & 0.81 \\
 & $\mu_2$ & 0.060 & 0.137 & 0.94 && 0.052 & 0.097 & 0.94 \\ 
 MMA-imp & $\mu_1$ & 0.122 & 0.131 & 0.79 && 0.080 & 0.092 & 0.81  \\ 
 & $\mu_2$ & 0.039 & 0.118 & 0.91 && 0.020 & 0.084 & 0.92 \\ 
 UMA-rem & $\mu_1$ & 0.123 & 0.141 & 0.88 && 0.088 & 0.100 & 0.84 \\
 & $\mu_2$ & 0.043 & 0.122 & 0.93 && 0.023 & 0.087 & 0.95 \\ 
 UMA-imp & $\mu_1$ & 0.128 & \textbf{0.125} & 0.81 && 0.089 & \textbf{0.088} & 0.73  \\
 & $\mu_2$ & 0.046 & \textbf{0.108} & 0.90 && 0.020 & \textbf{0.077} & 0.88  \\
 Copas & $\mu_1$ & 0.053 & 0.187 & 0.88 && 0.071 & 0.140 & 0.93  \\
 & $\mu_2$ & 0.020 & 0.153 & 0.93 && 0.003 & 0.115 & 0.96  \\ 
\bottomrule
\end{tabular}}

\medskip

  \resizebox{.6\textwidth}{!}{  
\begin{tabular}{llccccccc}
\multicolumn{9}{l}{Experiment 2 (strong heterogeneity, 20\% missing $y_1$, 20\% missing $y_2$)}  \\
\midrule
\phantom{abc} & \phantom{abc} & \multicolumn{3}{c}{$n = 25$} & \phantom{abc}& \multicolumn{3}{c}{$n=50$}  \\
\cmidrule{3-5} \cmidrule{7-9}  
& & Bias & SE & CP && Bias & SE & CP  \\ \midrule
 ABSORB & $\mu_1$ & \textbf{0.053} & 0.303 & \textbf{0.94} && \textbf{0.005} & 0.165 & \textbf{1.00} \\ 
 & $\mu_2$ & 0.034 & 0.303 & \textbf{0.99} && \textbf{0.002} & 0.165 & \textbf{0.97} \\ 
 MMA-rem & $\mu_1$ & 0.167 & 0.165 & 0.79 && 0.131 & 0.114 & 0.81 \\
 & $\mu_2$ & 0.060 & 0.137 & 0.94 && 0.052 & 0.097 & 0.94 \\ 
 MMA-imp & $\mu_1$ & 0.122 & 0.131 & 0.79 && 0.080 & 0.092 & 0.81  \\ 
 & $\mu_2$ & 0.039 & \textbf{0.118} & 0.91 && 0.020 & \textbf{0.084} & 0.92 \\ 
 UMA-rem & $\mu_1$ & 0.123 & 0.141 & 0.88 && 0.088 & 0.100 & 0.84 \\
 & $\mu_2$ & 0.043 & 0.122 & 0.93 && 0.023 & 0.087 & 0.92 \\ 
 UMA-imp & $\mu_1$ & 0.128 & \textbf{0.125} & 0.81 && 0.089 & \textbf{0.088} & 0.73 \\
 & $\mu_2$ & 0.292 & 0.126 & 0.38 && 0.052 & 0.097 & 0.94  \\
 Copas & $\mu_1$ & 0.093 & 0.187 & 0.88 && 0.047 & 0.206 & 0.93  \\
 & $\mu_2$ & \textbf{0.020} & 0.153 & 0.93 && 0.036 & 0.169 & 0.91  \\ 
\bottomrule
\end{tabular}}

\medskip

  \resizebox{.6\textwidth}{!}{  
\begin{tabular}{llccccccc}
\multicolumn{9}{l}{Experiment 3 (moderate heterogeneity, 20\% missing $y_1$, 40\% missing $y_2$)}  \\
\midrule
\phantom{abc} & \phantom{abc} & \multicolumn{3}{c}{$n = 25$} & \phantom{abc}& \multicolumn{3}{c}{$n=50$}  \\
\cmidrule{3-5} \cmidrule{7-9}  
& & Bias & SE & CP && Bias & SE & CP  \\ \midrule
ABSORB & $\mu_1$ & \textbf{0.049} & 0.236 & \textbf{0.98} && \textbf{0.029} & 0.182 & \textbf{1.00} \\ 
 & $\mu_2$ & \textbf{0.007} & 0.236 & \textbf{0.97} && \textbf{0.021} & 0.182 & \textbf{0.94}  \\ 
MMA-rem & $\mu_1$ & 0.337 & 0.259 & 0.71 && 0.347 & 0.175 & 0.47 \\
 & $\mu_2$ & 0.169 & 0.137 & 0.81 && 0.150 & 0.092 & 0.65 \\ 
MMA-imp & $\mu_1$ & 0.139 & 0.150 & 0.82 && 0.142 & 0.106 & 0.69  \\ 
 & $\mu_2$ & 0.078 & 0.102 & 0.76 && 0.062 & 0.070 & 0.68 \\ 
UMA-rem & $\mu_1$ & 0.140 & 0.160 & 0.91 && 0.132 & 0.116 & 0.86 \\
 & $\mu_2$ & 0.060 & 0.110 & 0.88 && 0.047 & 0.076 & 0.82 \\ 
UMA-imp & $\mu_1$ & 0.144 & \textbf{0.142} & 0.79 && 0.124 & \textbf{0.104} & 0.75  \\
 & $\mu_2$ & 0.071 & \textbf{0.085} & 0.71 && 0.049 & \textbf{0.059} & 0.71  \\
Copas & $\mu_1$ & 0.078 & 0.209 & 0.92 && 0.045 & 0.162 & 0.94  \\
 & $\mu_2$ & 0.024 & 0.152 & 0.91 && 0.039 & 0.111 & 0.90  \\ 
\bottomrule
\end{tabular}}

\medskip

  \resizebox{.6\textwidth}{!}{  
\begin{tabular}{llccccccc}
\multicolumn{9}{l}{Experiment 4 (mixed heterogeneity, 32\% missing $y_1$, 12\% missing $y_2$)}  \\
\midrule
\phantom{abc} & \phantom{abc} & \multicolumn{3}{c}{$n = 25$} & \phantom{abc}& \multicolumn{3}{c}{$n=50$}  \\
\cmidrule{3-5} \cmidrule{7-9}  
& & Bias & SE & CP && Bias & SE & CP  \\ \midrule
ABSORB & $\mu_1$ & \textbf{0.043} & 0.270 & \textbf{0.98} && \textbf{0.004} & 0.220 & \textbf{0.99} \\ 
 & $\mu_2$ & \textbf{0.065} & 0.270 & \textbf{0.94} && \textbf{0.045} & 0.220 & \textbf{0.95}  \\ 
MMA-rem & $\mu_1$ & 0.125 & 0.220 & 0.94 && 0.121 & 0.156 & 0.83 \\
 & $\mu_2$ & 0.137 & 0.121 & 0.81 && 0.128 & 0.079 & 0.68 \\
MMA-imp & $\mu_1$ & 0.083 & 0.165 & 0.89 && 0.083 & 0.119 & 0.78  \\ 
 & $\mu_2$ & 0.062 & 0.104 & 0.84 && 0.073 & 0.072 & 0.73 \\ 
UMA-rem & $\mu_1$ & 0.091 & 0.196 & 0.94 && 0.086 & 0.140 & 0.87 \\
 & $\mu_2$ & 0.071 & 0.101 & 0.85 && 0.071 & 0.071 & 0.81 \\ 
UMA-imp & $\mu_1$ & 0.074 & \textbf{0.157} & 0.82 && 0.089 & \textbf{0.115} & 0.77  \\
 & $\mu_2$ & 0.072 & \textbf{0.095} & 0.82 && 0.071 & \textbf{0.066} & 0.73  \\
Copas & $\mu_1$ & 0.061 & 0.252 & 0.93 && 0.005 & 0.204 & 0.98 \\
 & $\mu_2$ & 0.075 & 0.125 & 0.89 && 0.056 & 0.097 & 0.89  \\ 
\bottomrule
\end{tabular}}
\label{SimulationsTable}
\end{table}

Table~\ref{SimulationsTable} shows the average bias, SE, and CP averaged across all 1000 replications. We see that in all but one of scenarios considered, the ABSORB model had the smallest average bias and the highest CP. In the high heterogeneity scenario (Experiment 2) with $n=25$, ABSORB had the second lowest average bias for $\mu_2$, while Copas had the lowest. On the other hand, the SE tended to be larger on average for ABSORB. However, the approaches that had lower SE also tended to be \textit{overcertain}, with greater bias and worse coverage of the true parameters. Additionally, in the presence of ORB, it is not unreasonable to expect that there should be somewhat greater uncertainty about the true parameter values. 

	Our results for the ABSORB model are especially encouraging from the perspective of quantifying uncertainty. Across various degrees of between-study heterogeneity and missingness, ABSORB consistently had coverage close to the nominal level. Since practitioners typically use 95\% CIs to draw conclusions about treatment effects (e.g., determining whether effects are statistically significant), it is critical for MMA models to have robust coverage. ABSORB offers such reliable uncertainty quantification in the presence of ORB.
	
In contrast, MMA-imp, MMA-rem, UMA-imp, and UMA-rem not only had greater bias than ABSORB but they also typically had coverage \textit{below} the nominal level, especially when there was a higher degree of missingness for one endpoint (Experiments 3 and 4). This suggests that approaches based on the MAR and MCAR assumptions do not quantify uncertainty as well when there is ORB. Meanwhile, Copas performed the second best overall in terms of average bias and CP. Our results suggest that selection models which explicitly model the selection bias (i.e. ABSORB and Copas) outperform approaches that remove or impute missing outcomes when there is ORB.

\section{Extension of ABSORB to Account for Publication Bias} \label{ABSORBISM}

In addition to modeling ORB, it may also be of interest to explicitly model the completely missing studies that do not report \textit{either} $y_1$ or $y_2$ because of publication bias. Since neither outcome is observed in this case, incorporating these missing studies into the model can be very challenging. Nevertheless, if we know the \textit{number} of missing studies (say $K$), we can augment the ABSORB likelihood \eqref{ABSORBLikelihood} to include these $K$ missing studies. By modeling all $n+K$ studies, we can correct for \textit{both} ORB \textit{and} PB, and thus, potentially improve the efficiency of our estimates. We call this extension to our model from Section~\ref{ABSORB} the ABSORB-ISM model, i.e., ABSORB \textbf{I}ncluding \textbf{S}tudies that are \textbf{M}issing.

For the $n$ studies that have reported \textit{at least} one outcome $y_1$ or $y_2$, it is permissible to estimate the missing standard errors $s_{i1}$ and $s_{i2}$ using the estimation approach of \cite{copas2014model}, as we did in Section~\ref{LikelihoodImplementation}. However, for the $K$ \textit{completely} missing studies, this method is no longer applicable since we do not know the missing studies' sample sizes.  Thus, we adopt a Bayesian approach for estimating these missing studies' standard errors. 

In order to emphasize that these standard errors are for the $K$ studies that are \textit{completely} missing, we denote them as $\widetilde{s}_{i1}$ and $\widetilde{s}_{i2}$, respectively. Meanwhile, $s_{i1}$ and $s_{i2}$ refer to \textit{observed} standard errors in the $n$ \textit{published} studies. In general, it is very difficult to place ``noninformative'' priors on $(\widetilde{s}_{i1}, \widetilde{s}_{i2})$. Instead, we use the observed data to construct \textit{informative} priors for $(\widetilde{s}_{i1}, \widetilde{s}_{i2})$. For each of the $K$ missing studies, we place the following priors on $\widetilde{s}_{i1}$ and $\widetilde{s}_{i2}$:
\begin{equation} \label{missingstandarderrorpriors}
	\widetilde{s}_{ij} \sim \mathcal{U} ( \min_{i} s_{ij}, \max_{i} s_{ij}) \hspace{.5cm} i=n+1, \ldots, K, \hspace{.2cm} j = 1, 2,
\end{equation}
that is, we assume that the standard errors for completely missing studies lie in between the minimum of the observed $s_{ij}$'s and the maximum of the observed $s_{ij}$'s for each $j = 1, 2$. 

With the priors \eqref{missingstandarderrorpriors} on $(\widetilde{s}_{i1}, \widetilde{s}_{i2})$, we proceed to model each of the missing studies with the truncated normal density, 
\begin{align} \label{jointZbothoutcomesmissing}
	\begin{pmatrix} z_{i1} \\ z_{i2} \end{pmatrix} \sim \mathcal{N} \left( \begin{pmatrix} \gamma_{01}+ \gamma_{11} / \widetilde{s}_{i1} \\ \gamma_{02} + \gamma_{12} / \widetilde{s}_{i2} \end{pmatrix}, \begin{pmatrix} 1 & 0 \\ 0 & 1 \end{pmatrix} \right) \mathbb{I}_{[ z_{i1} < 0 \cap z_{i2} < 0]},
\end{align}
where the truncation in \eqref{jointZbothoutcomesmissing} indicates that $y_{i1}$ and $y_{i2}$ are not reported because their corresponding latent variables $z_{i1}$ and $z_{i2}$ are \textit{both} less than zero. With \eqref{jointZbothoutcomesmissing}, we can then define the likelihood for the $K$ missing studies as
\begin{align} \label{LikelihoodMissingStudies}
	L_4 ( \bm{\Xi} \mid \bm{y}_{n+1}, \ldots, \bm{y}_{n+K} ) = \prod_{i=n+1}^{n+K} f( z_{i1}, z_{i2} \mid \gamma_{01}, \gamma_{11}, \gamma_{02}, \gamma_{12}, \widetilde{s}_{i1}, \widetilde{s}_{i2}),
\end{align} 
where $f( z_{i1}, z_{i2} \mid \cdot )$ is the pdf of the truncated normal distribution in \eqref{jointZbothoutcomesmissing}. The complete likelihood function for all $n+K$ studies is then
\begin{align} \label{ABSORBAugmentedLikelihood}
	L_{\textrm{extended}}( \bm{\Xi} \mid \bm{y}_1, \ldots, \bm{y}_{n+K} ) = L ( \bm{\Xi} \mid \bm{y}_1, \ldots, \bm{y}_n )  L_4 ( \bm{\Xi} \mid \bm{y}_{n+1}, \ldots, \bm{y}_{n+K}), 
\end{align}
where $L(\bm{\Xi} \mid \bm{y}_1, \ldots, \bm{y}_n)$ is the likelihood for the $n$ studies with at least one reported outcome in \eqref{ABSORBLikelihood} and $L_4 (\bm{\Xi})$ is the likelihood for the missing studies in \eqref{LikelihoodMissingStudies}. Posterior inference under the augmented model \eqref{ABSORBAugmentedLikelihood} proceeds as usual by placing the priors \eqref{muprior}-\eqref{rhoprior} and \eqref{missingstandarderrorpriors} on the model parameters.

In practice, the number of studies $K$ that do not report either of the two outcomes may not be known. For these scenarios, we should use the regular ABSORB model from Section~3, which \textit{only} uses the data from the $n$ available studies that report at least one outcome. However, if we \textit{do} know the number of missing studies, then we can use the ABSORB-ISM model with the augmented likelihood \eqref{ABSORBAugmentedLikelihood} in order to correct for both ORB \textit{and} PB. In Section~\ref{AdditionalHFResults}, we illustrate one such application of the ABSORB-ISM model for the HF patients meta-analysis from Section~\ref{Application}.

\section{Additional Results for the Heart Failure Patients Meta-Analysis}  \label{AdditionalHFResults}

\subsection{Trace Plots for the ABSORB Model} \label{TracePlots}

\begin{figure}[t]
\centering
\includegraphics[width=.9\textwidth]{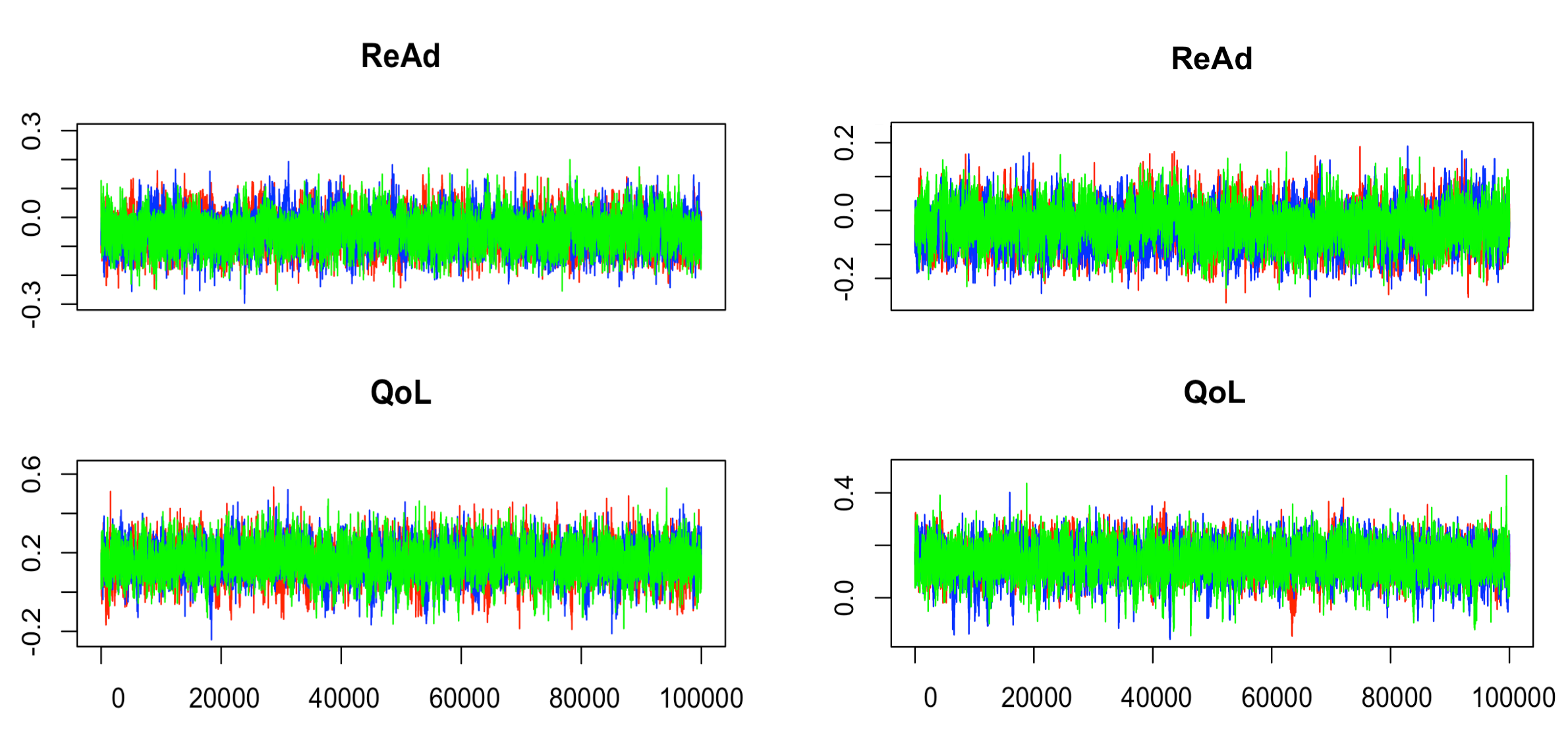}
\caption{The trace plots for the ABSORB model fit to the published data (left two panels) and the updated data (right two panels) for the meta-analysis of interventions on HF patients.} \label{app:trace-plots-combined}
\end{figure}

Here, we show the trace plots from the ABSORB models that were fit to the published and updated data that we analyzed in Section~\ref{Application}. The results for the ABSORB model fit to the published data are given in the left two panels of Figure~\ref{app:trace-plots-combined}, while the model fit to the updated data are given in the right two panels. These plots confirm that the three MCMC chains that we ran for the ABSORB models mixed very well and converged to the stationary posterior distributions for $\mu_1$ and $\mu_2$ within 100,000 iterations.

\subsection{ABSORB-ISM for Meta-analysis of the Effects of Interventions for Heart Failure Patients} \label{ABSORBISMApplication}

In this section, we revisit the meta-analysis from Sections \ref{MotivatingData} and \ref{Application} using the ABSORB-ISM model from Appendix~\ref{ABSORBISM}. As discussed in this case study, we had \textit{a priori} knowledge that among the 45 studies, there were $K=4$ studies that did not publish either ReAd or QoL. After querying corresponding authors for the missing outcomes, we had an updated dataset with only $K=1$ study that did not report either ReAd or QoL.

\begin{figure}[t!]
\centering
\hspace{.4cm} \includegraphics[width=.5\linewidth]{Table3_legend.png} \\
\includegraphics[width=.45\textwidth]{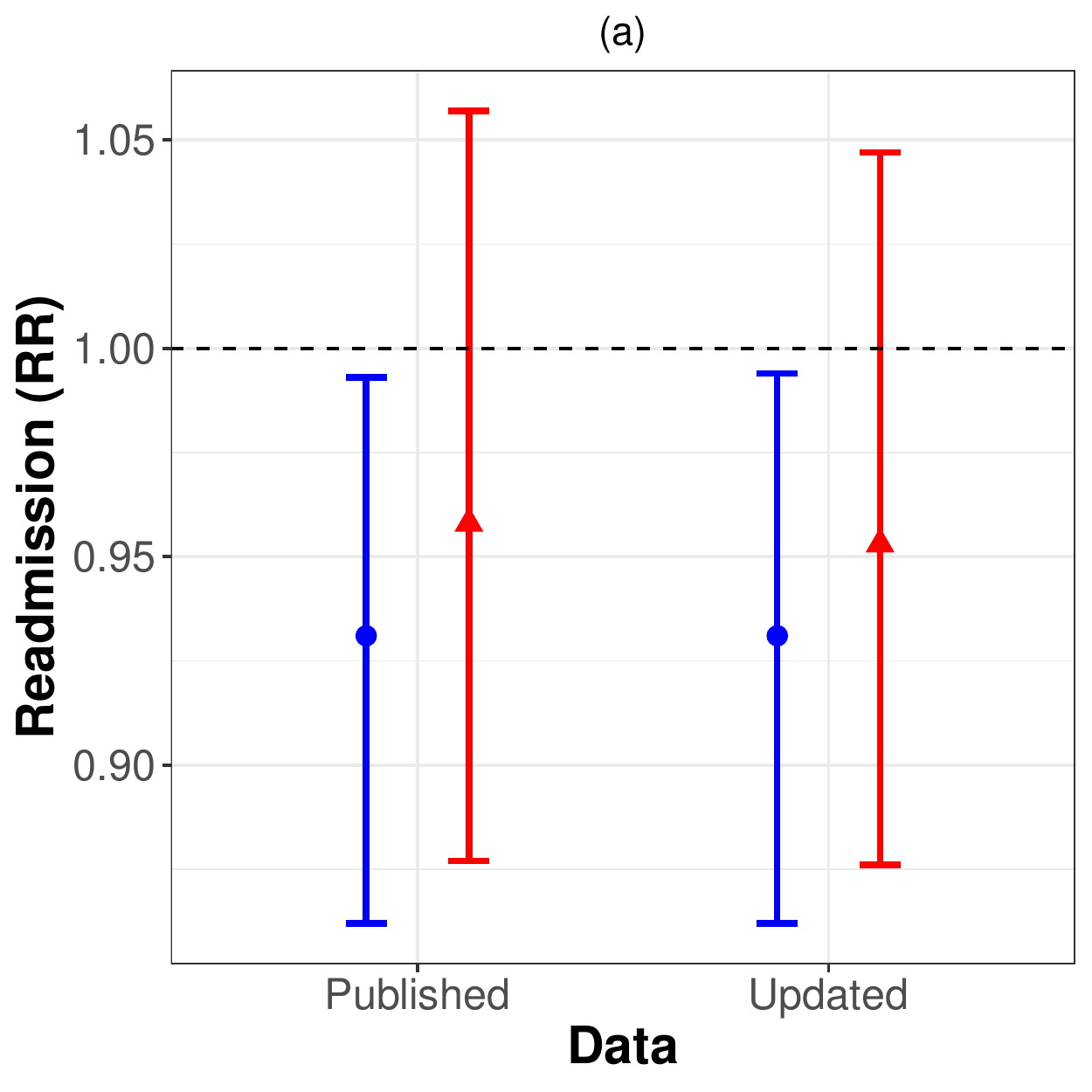}
\includegraphics[width=.45\textwidth]{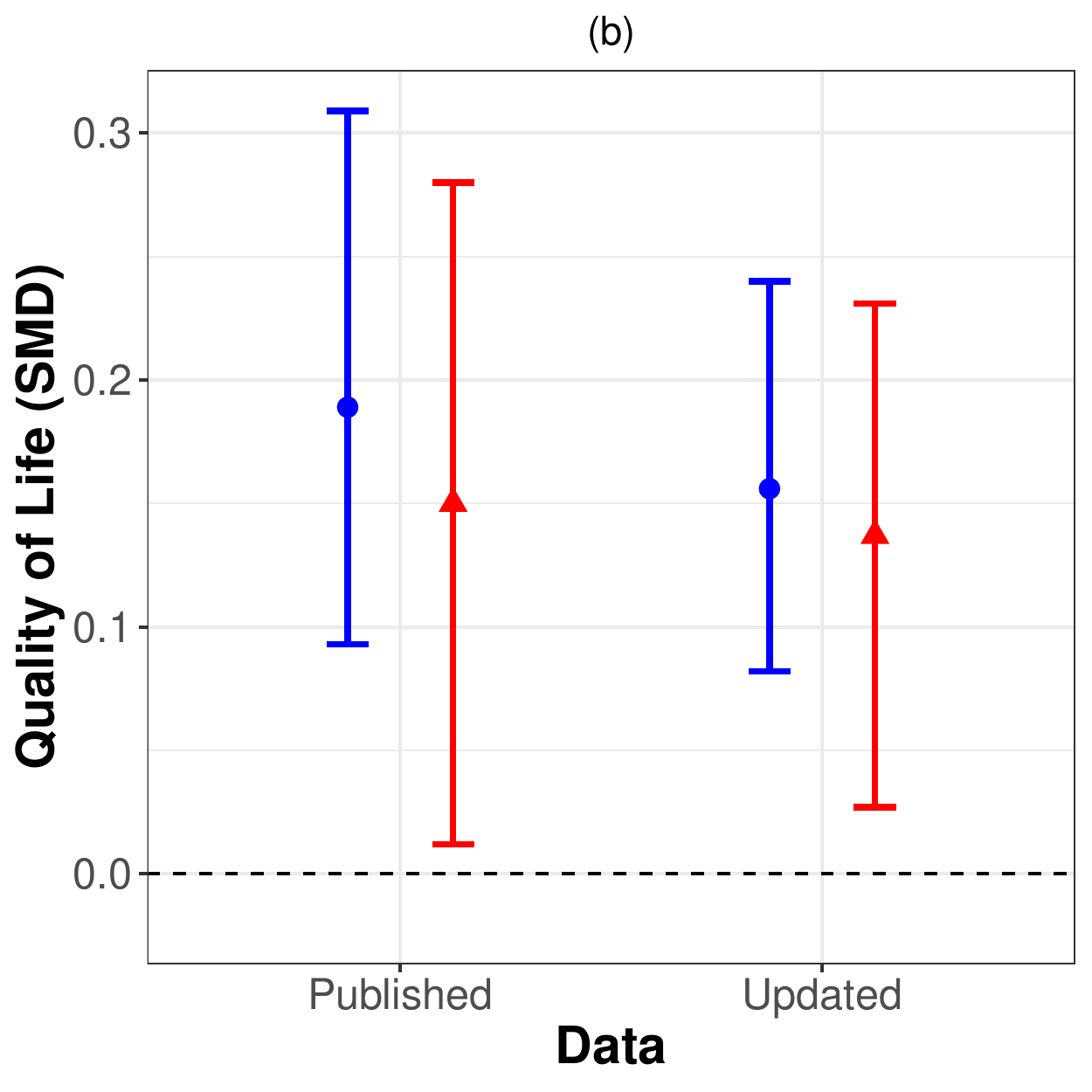}
\caption{Plots of the posterior means and 95\% posterior credible intervals for our case study on interventions for HF patients under the non-bias corrected and ABSORB-ISM models. Panel~(a) plots the results for ReAd and panel~(b) plots the results for QoL.} \label{HeartFailureResultsABSORBISM}
\end{figure}

With the ABSORB-ISM model, we are able to adjust for both ORB and PB simultaneously and thus we can include all 45 studies in our model. We report our results for the ABSORB-ISM model in Figure~\ref{HeartFailureResultsABSORBISM}. The clinical conclusions based on the 95\% CIs for ABSORB-ISM were essentially the same as those that we obtained for the regular ABSORB model in Section~\ref{Application}. One interesting thing to note is that in the published data, the ABSORB-ISM model had a slightly greater shift towards the null side in the mean effect for ReAd (0.958), compared to the mean for ReAd effect (0.955) under the ABSORB model. In addition, the CIs for ReAd were slightly larger for ABSORB-ISM rather than ABSORB. 

\begin{figure}[ht]
\centering
\hspace{.4cm} \includegraphics[width=.55\linewidth]{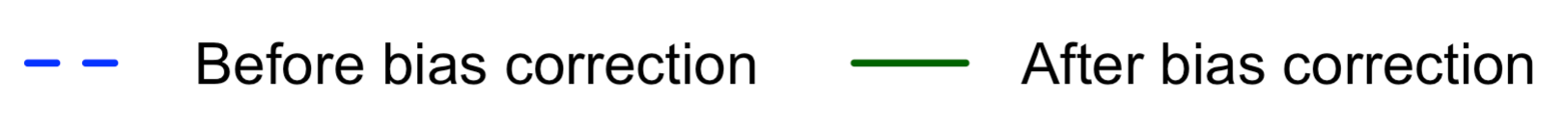} \\
\includegraphics[width=.28\textwidth]{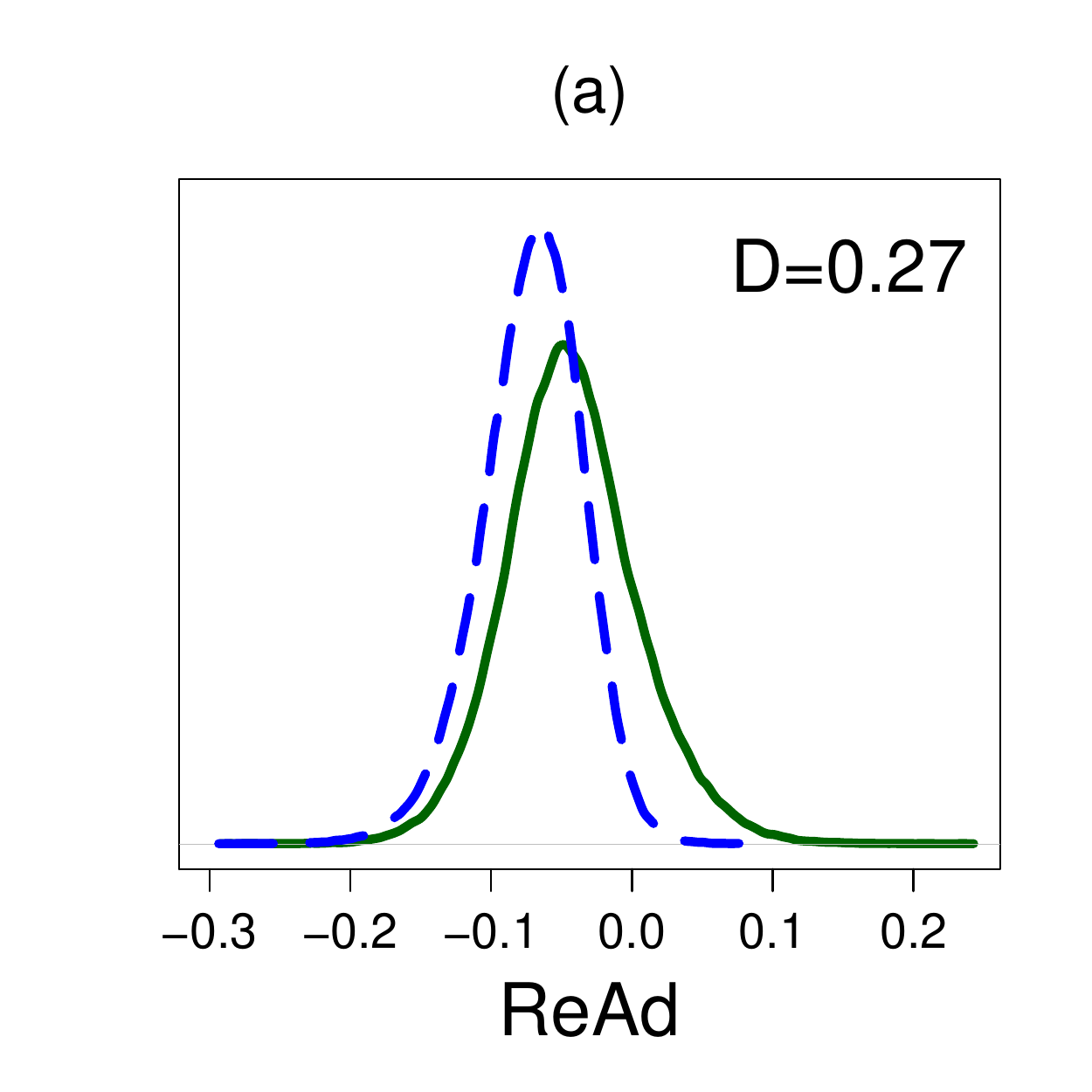}
\includegraphics[width=.28\textwidth]{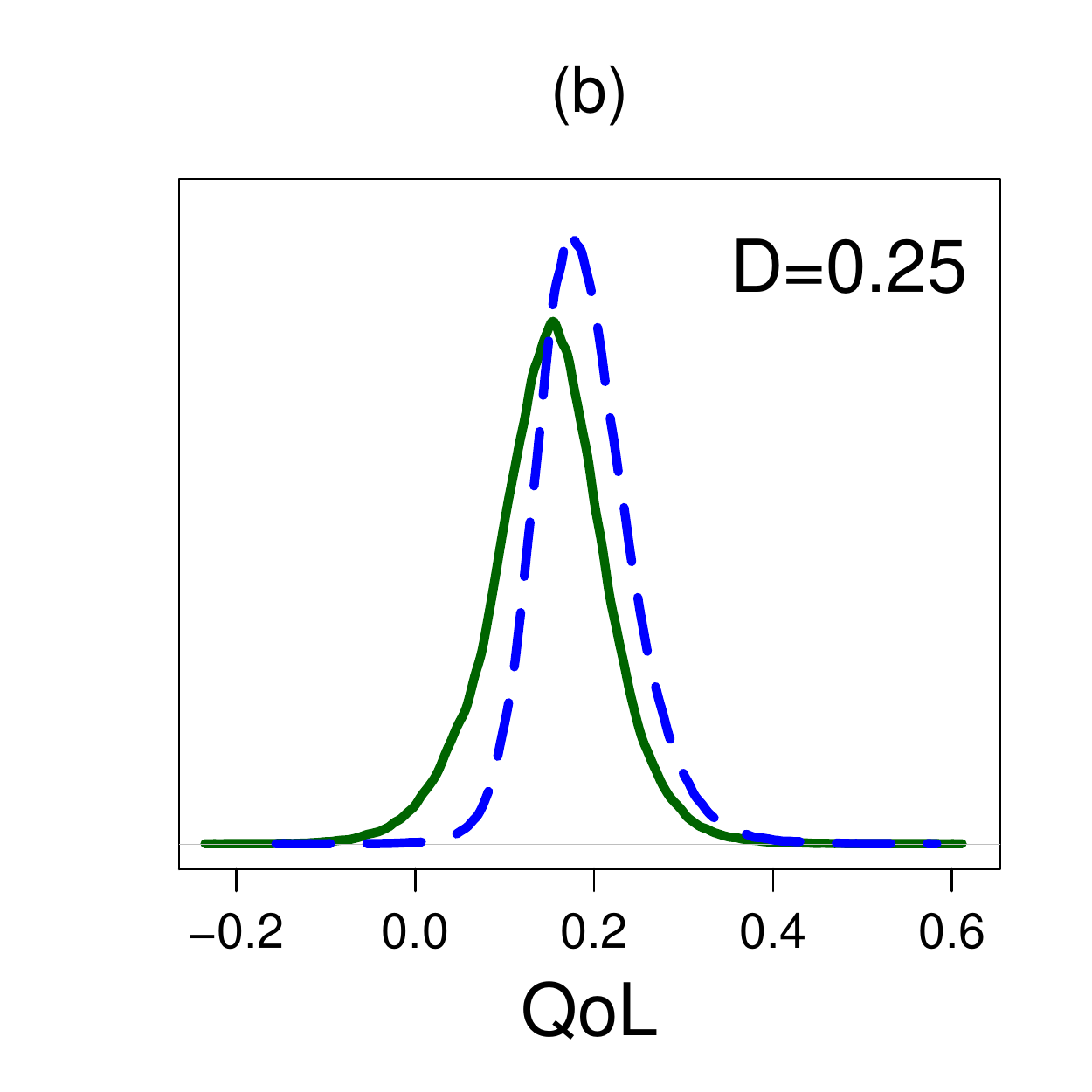}
 \includegraphics[width=.28\textwidth]{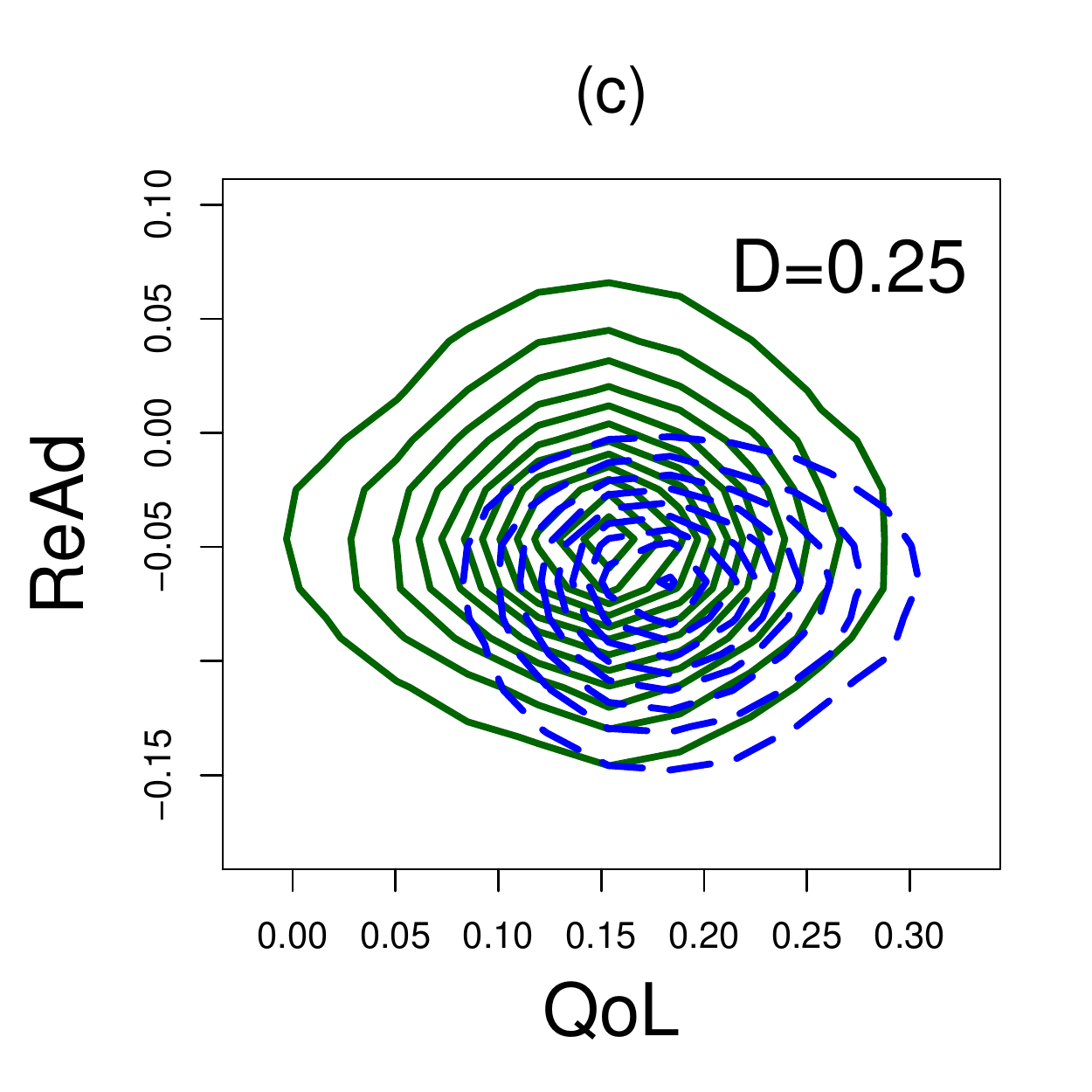} \\
\includegraphics[width=.28\textwidth]{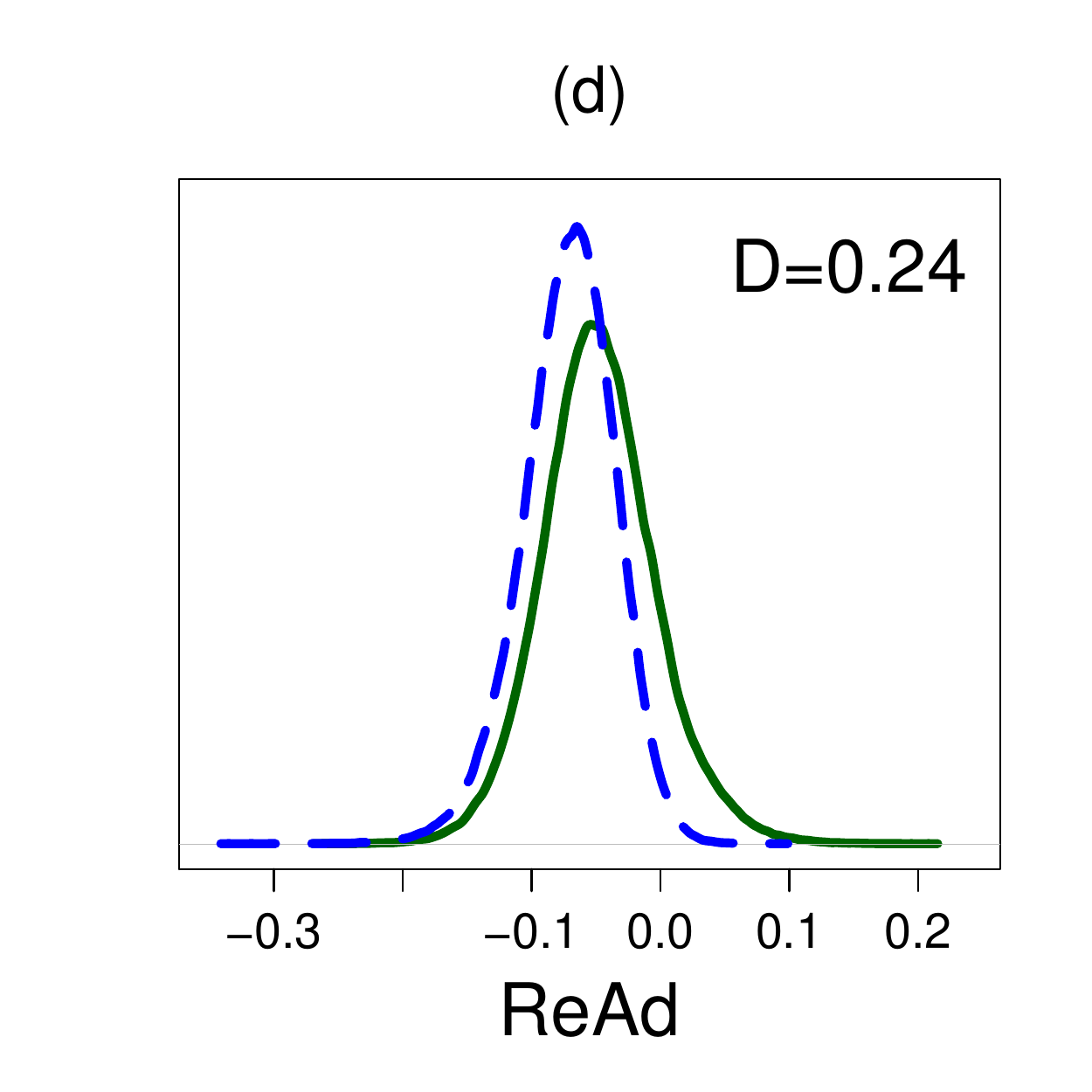} 
\includegraphics[width=.28\textwidth]{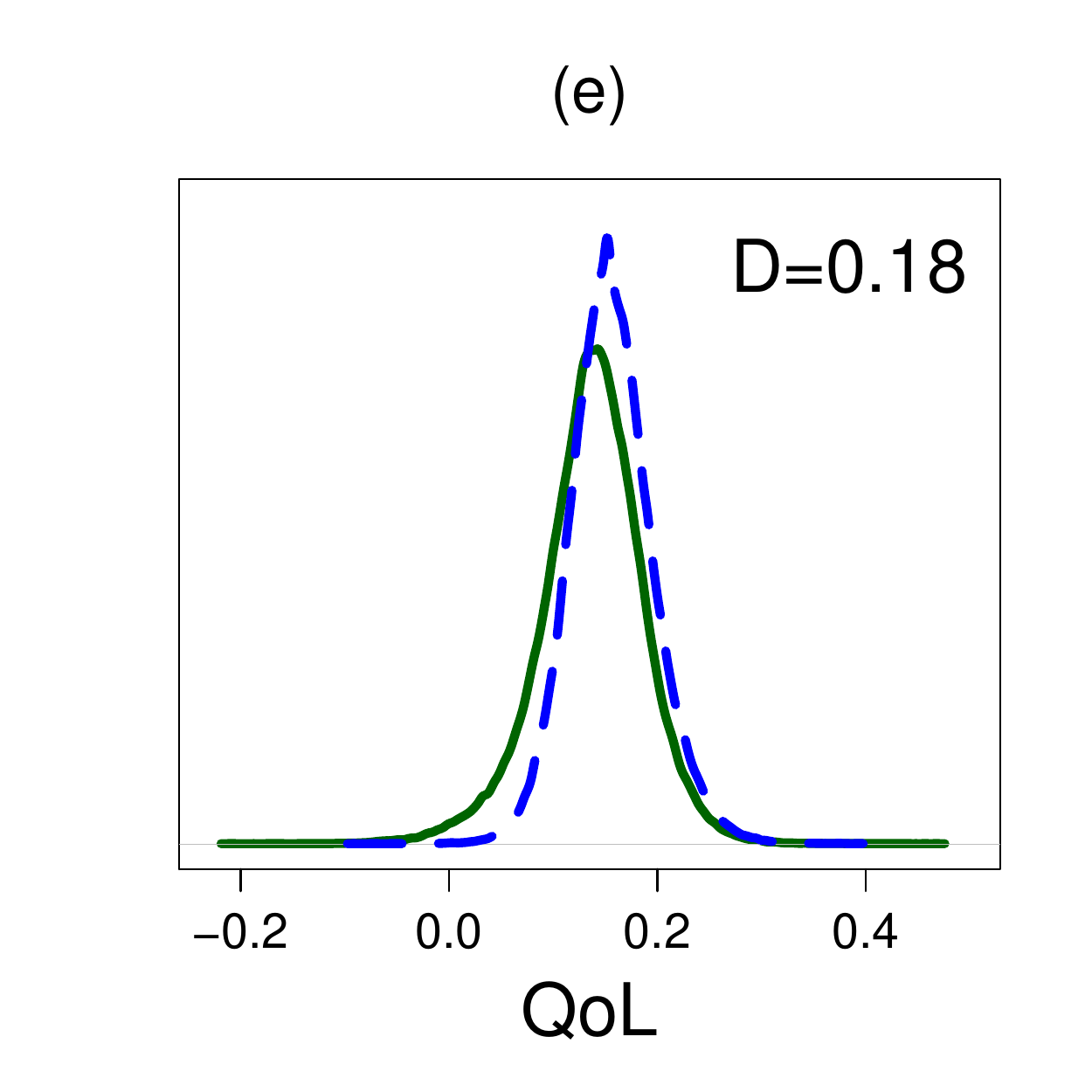} 
 \includegraphics[width=.28\textwidth]{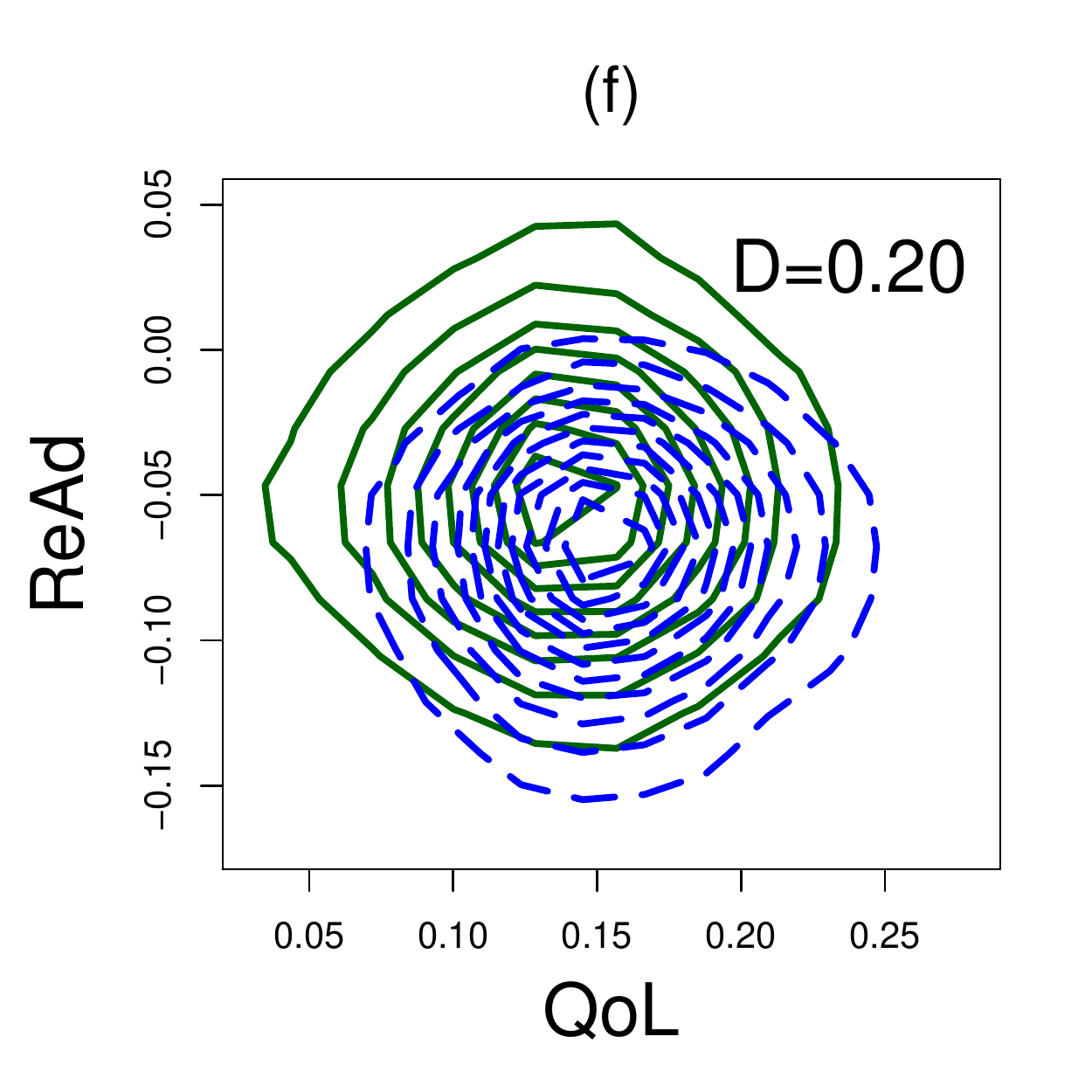}
\caption{Panels~(a)--(c) show the results for the ABSORB-ISM model using the \textit{published} data. Panels~(d)--(f) show the ABSORB-ISM results using the \textit{updated} data.  ReAd is plotted on the log-RR scale in panels (a), (c), (d), and (f).} \label{HeartFailureABSORBISMPlots}
\end{figure}

In Figure~\ref{HeartFailureABSORBISMPlots}, we plot the bias-corrected posteriors against the non-bias corrected posteriors (dashed line) for the published data in the panels (a)-(c) and the updated data in panels (d)-(f). Similarly as with the ABSORB model, the $D$ measure indicated that the impact from ORB became more negligible once we had more data points for QoL, decreasing from $D=0.25$ for the published data to $D=0.18$ under the updated data. Moreover, the Jaccard index for the 95\% CIs for QoL was 0.63 in the published data and 0.70 in the updated data, indicating greater agreement between the non-bias corrected and bias-corrected posterior credible intervals in the updated data. On the other hand, there was non-negligible impact from ORB under the ABSORB-ISM model for ReAd in both the published data ($D=0.27$) and the updated data ($D = 0.24$). 

In this particular example, including the completely missing studies in our model did not result in any qualitative changes in the conclusions from the regular ABSORB model. This may be because the completely missing studies did not contribute much additional information to the estimation of the effects of intervention on ReAd and QoL. However, it is conceivable that in other scenarios, including missing studies due to publication bias (provided that the number of missing studies is known) with the ABSORB-ISM model can improve the efficiency of our estimates. 

\end{appendix}

\end{document}